\documentclass[fleqn,usenatbib]{mnras}

\usepackage{newtxtext,newtxmath}

\usepackage[T1]{fontenc}

\DeclareRobustCommand{\VAN}[3]{#2}
\let\VANthebibliography\thebibliography
\def\thebibliography{\DeclareRobustCommand{\VAN}[3]{##3}\VANthebibliography}

\usepackage{graphicx}	
\usepackage{amsmath}	

\usepackage{pdflscape}

\newcommand{\oiii}{[O~{\sc iii}]}

\newcommand{\ha}{H$\alpha$}
\newcommand{\hb}{H$\beta$}
\newcommand{\nii}{[N~{\sc ii}]}
\newcommand{\sii}{[S~{\sc ii}]}
\newcommand{\fesc}{$f_{\rm{esc}}$}
\newcommand{\xion}{$\xi_{\rm{ion}}$}
\newcommand{\xionnofesc}{$\xi_{\rm{ion},0}$}
\newcommand{\ndot}{$\dot{n}_{\rm{ion}}$}
\newcommand{\Ndot}{$\dot{N}_{\rm{ion}}$}

\title[What drives $\xi_{\rm{ion}}$?]{Low-mass bursty galaxies in JADES efficiently produce ionising photons and could represent the main drivers of reionisation}

\author[C. Simmonds et al.]{
C. Simmonds,$^{1,2}$\thanks{E-mail: cs2210@cam.ac.uk}
S. Tacchella,$^{1,2}$
K. Hainline,$^{3}$
B.~D. Johnson,$^{4}$
W. McClymont,$^{1,2}$
B. Robertson,$^{6}$
\newauthor
A. Saxena,$^{7,8}$
F. Sun,$^{3}$
C. Witten,$^{1,5}$ 
W.~M. Baker,$^{1,2}$
R. Bhatawdekar,$^{9}$
K. Boyett,$^{10,11}$
A.~J. Bunker,$^{7}$
\newauthor
S. Charlot,$^{12}$
E. Curtis-Lake,$^{13}$
E. Egami,$^{3}$
D.~J. Eisenstein,$^{4}$
R. Hausen,$^{14}$
R. Maiolino,$^{1,2,8}$
\newauthor
M.~V. Maseda,$^{15}$
J. Scholtz,$^{1,2}$
C.~C. Williams,$^{16}$
C. Willot$^{17}$
and J. Witstok$^{1,2}$
\\
$^{1}$The Kavli Institute for Cosmology (KICC), University of Cambridge, Madingley Road, Cambridge, CB3 0HA, UK\\
$^{2}$Cavendish Laboratory, University of Cambridge, 19 JJ Thomson Avenue, Cambridge, CB3 0HE, UK\\
$^{3}$Steward Observatory, University of Arizona, 933 N. Cherry Avenue, Tucson, AZ 85721, USA\\
$^{4}$Center for Astrophysics $|$ Harvard \& Smithsonian, 60 Garden St., Cambridge MA 02138 USA\\
$^{5}$Institute of Astronomy, University of Cambridge, Madingley Road, Cambridge, CB3 0HA, UK\\
$^{6}$Department of Astronomy and Astrophysics University of California, Santa Cruz, 1156 High Street, Santa Cruz CA 96054, USA\\
$^{7}$Department of Physics, University of Oxford, Denys Wilkinson Building, Keble Road, Oxford OX1 3RH, UK\\
$^{8}$Department of Physics and Astronomy, University College London, Gower Street, London WC1E 6BT, UK\\
$^{9}$European Space Agency (ESA), European Space Astronomy Centre (ESAC), Camino Bajo del Castillo s/n, 28692 Villanueva de la Cañada, Madrid, Spain\\
$^{10}$School of Physics, University of Melbourne, Parkville 3010, VIC, Australia\\
$^{11}$ARC Centre of Excellence for All Sky Astrophysics in 3 Dimensions (ASTRO 3D), Canberra 2611, Australia\\
$^{12}$Sorbonne Universit\'e, CNRS, UMR 7095, Institut d'Astrophysique de Paris, 98 bis bd Arago, 75014 Paris, France\\
$^{13}$Centre for Astrophysics Research, Department of Physics, Astronomy and Mathematics, University of Hertfordshire, Hatfield AL10 9AB, UK\\
$^{14}$Department of Physics and Astronomy, The Johns Hopkins University, 3400 N. Charles St., Baltimore, MD 21218, USA\\
$^{15}$Department of Astronomy, University of Wisconsin-Madison, 475 N. Charter St., Madison, WI 53706, USA\\
$^{16}$NSF’s National Optical-Infrared Astronomy Research Laboratory, 950 North
Cherry Avenue, Tucson, AZ 85719, USA \\
$^{17}$NRC Herzberg, 5071 West Saanich Rd, Victoria, BC V9E 2E7, Canada\\
}

\date{Accepted XXX. Received YYY; in original form ZZZ}

\begin{document}
\label{firstpage}
\pagerange{\pageref{firstpage}--\pageref{lastpage}}
\maketitle

\begin{abstract} %
We study galaxies in JADES Deep to study the evolution of the ionising photon production efficiency, \xion\/, observed to increase with redshift.
We estimate \xion\ for a sample of 677 galaxies at $z \sim 4 - 9$ using NIRCam photometry. Specifically, combinations of the medium and wide bands F335M-F356W and F410M-F444W to constrain emission lines that trace \xion\/: \ha\ and \oiii\/. Additionally, we use the spectral energy distribution fitting code \texttt{Prospector} to fit all available photometry and infer galaxy properties.
The flux measurements obtained via photometry are consistent with FRESCO and NIRSpec-derived fluxes. Moreover, the emission-line-inferred measurements are in tight agreement with the \texttt{Prospector} estimates. We also confirm the observed \xion\ trend with redshift and M$_{\rm{UV}}$, and find: $\log \xi_{\rm{ion}} (z,\text{M}_{\rm{UV}}) = (0.05 \pm 0.02)z + (0.11 \pm 0.02) \text{M}_{\rm{UV}} + (27.33 \pm 0.37)$. 
We use \texttt{Prospector} to investigate correlations of \xion\ with other galaxy properties. We see a clear correlation between \xion\ and burstiness in the star formation history of galaxies, given by the ratio of recent to older star formation, where burstiness is more prevalent at lower stellar masses. We also convolve our \xion\ relations with luminosity functions from the literature, and constant escape fractions of 10 and 20\%, to place constraints on the cosmic ionising photon budget. 
By combining our results, we find that if our sample is representative of the faint low-mass galaxy population, galaxies with bursty star formation are efficient enough in producing ionising photons and could be responsible for the reionisation of the Universe.
\end{abstract}




\section{Introduction}
\label{INTRO}

The Epoch of Reionisation (EoR) describes one of the Universe's major phase changes, during which the intergalactic medium (IGM) became transparent to Lyman Continuum (LyC; E $\geq$ 13.6 eV) radiation. Observations place the end of this epoch at $z \sim 6$ \citep{Becker2001,Fan2006,Yang2020}, with some studies favouring a later reionisation closer to $z \sim 5$ \citep{Keating2020,Bosman2022}. It is widely believed that young massive stars in galaxies are the main drivers of this transition, due to their copious production of LyC photons that escape the interstellar medium (ISM), and eventually ionise the IGM \citep{Hassan2018,Rosdahl2018,Trebitsch2020}. However, there is a debate whether faint, low-mass galaxies or bright, massive galaxies dominate the photon budget of reionisation \citep{Finkelstein2019,Naidu2020,Robertson2022}. In particular, the mass of galaxies has been seen to correlate with both the production efficiency and escape of ionising photons \citep{Paardekooper2015}, both key factors to understand the EoR. Moreover, the contribution of Active Galactic Nuclei (AGN) to this budget might be more important than previously believed \citep[AGN + host galaxy $> 10$\%; ][]{Maiolino2023}. For galaxies to be the main sources of reionisation, adopting canonical values of ionising photon production efficiencies, relatively high average escape fractions are necessary \citep[\fesc\/ = 10-20\%; ][]{Ouchi2009,Robertson2013,Robertson2015,Finkelstein2019,Naidu2020}. High \fesc\ values have been observed in some galaxies \citep[e.g. ][]{Borthakur2014,Bian2017,Vanzella2018,Izotov2021}, but usually not in large samples \citep{Leitet2013,Leitherer2016,Steidel2018,Flury2022a}. Another important quantity to measure is the ionising photon production efficiency (\xion\/), which is a measure of the production rate of ionising photons over the non-ionising ultra-violet (UV) luminosity density. Promisingly, by gaining observational access to the early Universe (up to $z \sim 9$), studies have found that as we go to higher redshifts, \xion\ increases \citep[e.g. ][]{Bouwens2016,Faisst2019,Endsley2021,Stefanon2022,Tang2023,Simmonds2023,Atek2023}. An increase of \xion\ implies that lower \fesc\ values are required in galaxies, in order for them to be responsible for the reionisation of the Universe.

Current constraints place the mean redshift of reionisation somewhere between $z = 7.8 - 8.8$ \citep{Planck2016}. Since the launch and deployment of the James Webb Space Telescope \citep[JWST; ][]{Gardner2023}, we have an unprecedented view of the Universe deep into the EoR. Moreover, by using deep photometry taken with the Near-Infrared Camera \citep[NIRCam; ][]{Rieke2023instrument}, we can gain insight into the rest-frame optical properties of large and statistically significant samples of galaxies at this epoch.  In particular, there are three important ingredients that contribute to our overall understanding of the ionising photon budget of the Universe: (1) a prescription for the \fesc\ of the population, (2) an appropriate luminosity density function, $\rho_{\rm{UV}}$, describing how many objects per unit volume of a certain UV luminosity exist as a function of redshift \citep[for example ][]{Bouwens2021}, and (3) \xion\/. Until recently, it was common practice to set (1) and (3) as constants \cite[e.g. ][]{Boyett2022}. However, the launch of JWST has given us unprecedented access to the rest-frame optical regime at high redshift, providing enough additional constraints on the stellar population to better infer \xion\ across the population. Therefore, studies shedding light on how \fesc\ and/or \xion\ evolve with galaxy properties, especially at high redshift, are of utmost relevance to the field.

In \cite{Simmonds2023}, JWST Extragalactic Medium Band Survey \cite[JEMS; ][]{Williams2023} photometry was used to estimate \xion\ for a sample of 30 Lyman-$\alpha$ emitters (LAE) at $z \sim 6$. In this work we use deep NIRCam imaging \citep{Rieke2023} to create a sample of 677 galaxies at $z \sim 4 - 9$, with photometric redshifts provided by the template-fitting code \texttt{EAZY} \citep{Brammer2008}. We use two filter pair combinations: F335M-F356W, and F410M-F444W, to estimate \ha\ and/or \oiii\ emission line fluxes, which can be used to infer \xion\/. To test the reliability of our derived fluxes, we compare (when available) our measurements to those obtained by First Reionisation Epoch Spectroscopic Complete Survey  \citep[FRESCO; ][PI: Oesch]{Oesch2023} grism spectra. In addition, we compare our fluxes and ionising photon production efficiencies to NIRSpec measurements \citep{Saxena2023}. Simultaneously, we use the Spectral Energy Distribution (SED) fitting code \texttt{Prospector} \citep{Johnson2019,Johnson2021} to infer galaxy properties such as star formation rates (SFRs) and histories (SFHs), both closely related to the production of ionising photons through star formation. Finally, we investigate how our findings affect the cosmic ionising photon budget, and make conclusions about which kind of galaxies could be the main sources responsible for the reionisation of the Universe.

The structure of this paper is the following. In $\S$~\ref{DATA} we present the data used in this work, along with the sample selection criteria. In  $\S$~\ref{OBS_METHODS} we explain the two observational methods used to estimate \xion\/ (through \ha\ and \oiii\/$_{\lambda 5007}$), and how the respective fluxes were measured from photometry. In $\S$~\ref{PROSPECTOR} we present our \texttt{Prospector} fitting method. Our \xion\ constraints are given in $\S$~\ref{CONSTRAINTS}, followed by a discussion in  $\S$~\ref{DISCUSSION}, and brief conclusions in $\S$~\ref{CONLCUSIONS}.

Throughout this work we assume $\Omega_0 = 0.315$ and $H_0 = 67.4$ km s$^{-1}$ Mpc$^{-1}$, following \cite{Planck2020}.

\section{Data and selection criteria}
\label{DATA}
In this section we describe the data and selection criteria applied to build a sample for which we can infer \xion\ through emission line fluxes, specifically \ha\ and \oiii\/$_{\lambda 5007}$. We caution the reader that by making this choice we are introducing a bias towards galaxies with strong emission lines, which will be discussed later.

\subsection{Data}
We make use of the NIRCam Deep imaging \citep{Rieke2023} released by the JWST Advanced Deep Extragalactic Survey \cite[JADES; ][]{Eisenstein2023}. This data covers an area of $\sim 25$ arcmin$^2$ overlapping with the Hubble Ultra Deep Field \citep[HUDF; ][]{Beckwith2006}, and portions of the Great Origins Deeps Survey South \citep[GOODS-S; ][]{Giavalisco2004}. The images were taken by a combination of 9 medium and wide-band infrared filters: F090W, F115W, F150W, F200W, F277W, F335M, F356W, F410M, and F444W. When in an overlapping region, some galaxies also have JEMS photometry, adding 5 more medium filters: F182M, F210M, F430M, F460M and F480M. This exquisite data set is ideal to estimate photometric redshifts (photo-z) with great accuracy. In this work we use photo-z inferred by the template-fitting code \texttt{EAZY}, as described in 
 \cite{Hainline2023} and \cite{Rieke2023}. \footnote{For a visual comparison between the inferred photometric and spectroscopic redshifts we refer the reader to Figure 13 of \cite{Rieke2023}.}

Regarding the photometric catalogue, the source detection and photometry leverage both the JEMS NIRCam medium band and JADES NIRCam broad and medium band imaging. Detection is performed using the \texttt{photutils} \citep{bradley2022a} Software package, identifying sources with contiguous regions of the SNR mosaic with signal $>3\sigma$ and five or more contiguous pixels. We also use {\tt photutils} to perform circular aperture photometry with filter-dependent aperture corrections based on empirical point-spread-functions measured from stars in the mosaic. The details of the catalogue generation and photometry will be presented in Robertson et al., (in prep). In this work we adopt a circular aperture of diameter $0.3''$ throughout, and impose a floor error of 5\% in each band. 

Finally, when available, we compare our photometry-derived emission-line fluxes to those obtained through an independent reduction of the spectra taken with the FRESCO program \citep[][]{Oesch2023}, which will be presented in Sun et al. (in prep), and to NIRSpec measurements provided in \cite{Saxena2023}. 

   \begin{figure}
        \centering
   \includegraphics[width=0.5\textwidth]{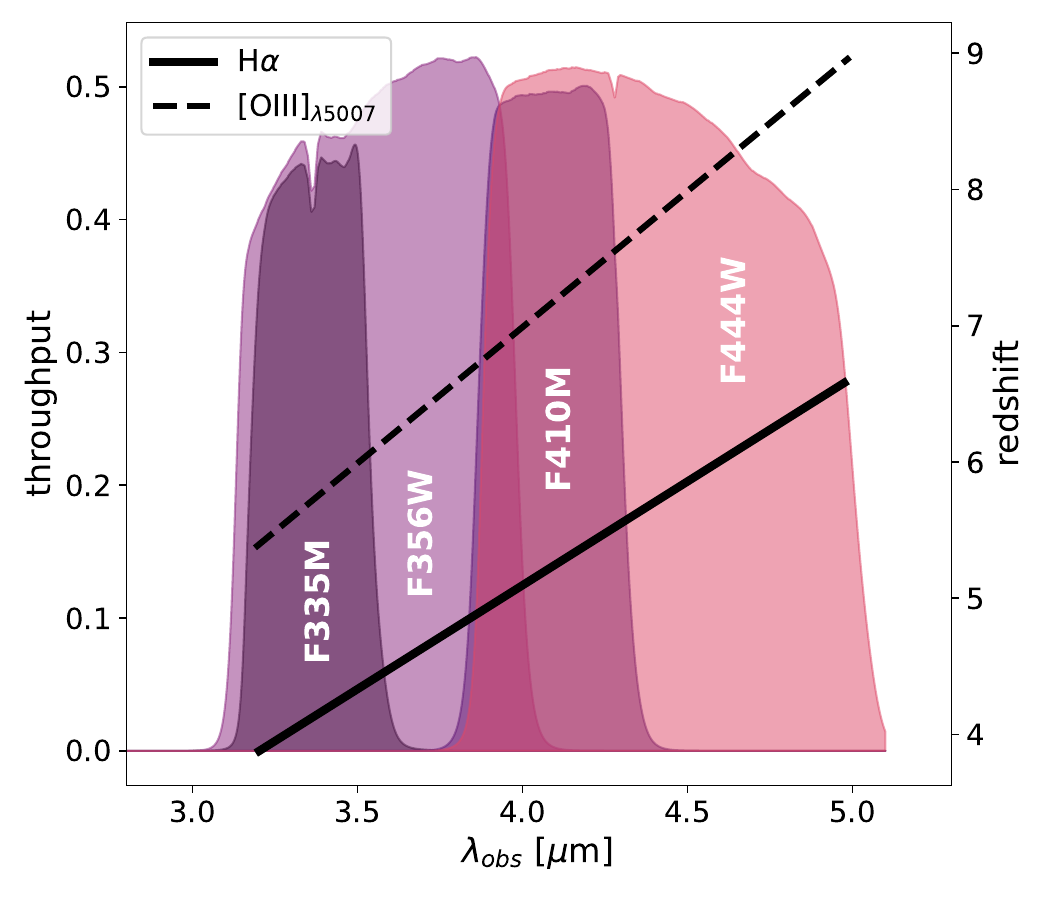}
   \caption{Medium and wide band NIRCam filters used to estimate \ha\ and/or \oiii\ emission line fluxes in this work. From left to right: F335M, F356W, F410M, F444W. The left y-axis indicates their throughput. The black lines show the observed wavelength of \ha\ (filled) and \oiii\/$_{\lambda 5007}$ (dashed) with redshift (right axis). Depending on the redshift, the emission lines can be estimated by combining the medium and wide band filter pairs: F335M-F356W and/or F410M-F444W.}
              \label{fig:filters}%
    \end{figure}

\subsection{Sample selection criteria}
The focus of this work is to constrain \xion\ for a large sample of emission line galaxies, thought to have had a significant role in reionisation \citep[e.g. ][]{Rinaldi2023arXiv,Rinaldi2023}. As is discussed in Section~\ref{OBS_METHODS}, this requires \ha\ and/or \oiii\/$_{\lambda 5007}$ in emission. The combination of broad and medium photometric bands is powerful to estimate emission lines when spectra are not available \citep[e.g. ][]{Bunker1995,Stark2013,Faisst2016}. Therefore, we select galaxies where the desired emission lines fall on one (or more) of the following filters: F335M, F356W, F410M or F444W. Figure~\ref{fig:filters} shows the throughput and wavelengths of these filters, as well as the redshift evolution of the observed wavelength of \ha\ and \oiii\/$_{\lambda 5007}$. As shown in the right vertical axis, this constrains the sample to $z = 3.9 - 9.0$. We note that the medium bands from the JEMS survey cover a smaller region in the sky, therefore, we use them (when available) to feed our SED-fitting routine, but not for estimating emission line fluxes. 

We apply this redshift cut to galaxies based on their photo-z. Furthermore, in order to be able to detect emission lines, we impose a conservative minimum flux difference between medium and wide bands to ensure a 5$\sigma$ line detection, as follows: 
\begin{itemize}
    \item  |F335M - F356W| $\geq$ 10 nJy
    \item  |F410M - F444W| $\geq$ 10 nJy
\end{itemize}
Where the excess in flux in a given band (depending on redshift) is assumed to be dominated by either \ha\ or \oiii\/ (i.e. neglecting \nii\/, \hb\/ and \sii\/ contamination).  A visual inspection was then performed on all the SEDs that satisfied this condition.

\begin{figure}
        \centering
   \includegraphics[width=0.5\textwidth]{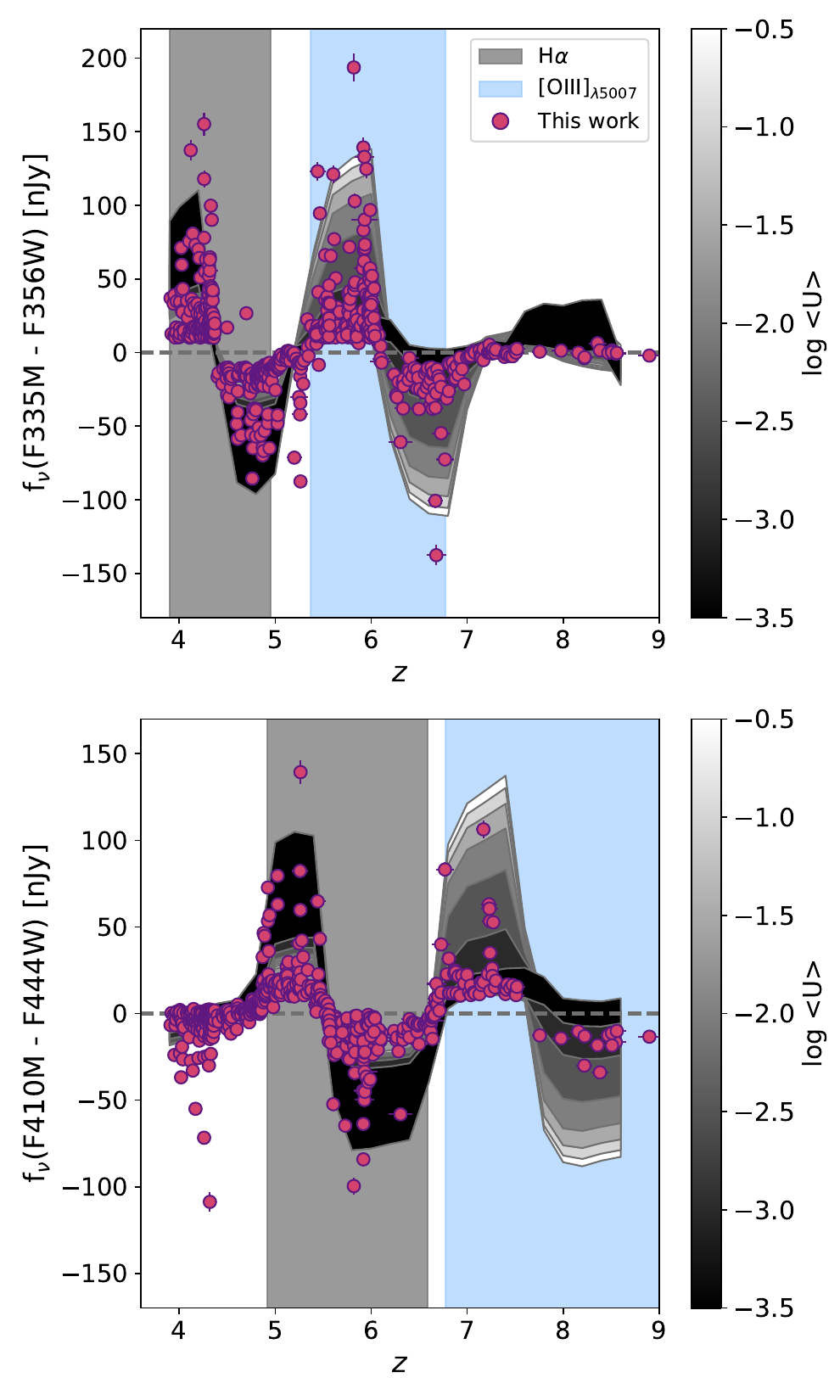}
   \caption{Expected shape for each filter pair flux differences, along with the corresponding emission lines, shown as vertical grey (\ha\/) and light blue (\oiii\/$_[\lambda 5007]$) bands. The purple circles represent the sample analysed in this work, they overall follow the idealised Cloudy models, shown as shaded areas colour-coded by ionisation parameter (log$\langle U \rangle$). This agreement corroborates the reliability of the photo-z inferred using \texttt{EAZY}. Fluxes are in units of nJy. \textsl{Top panel:} F335M - F356W. \textsl{Bottom panel:} F410M - F444W.}
              \label{fig:cloudy_obs}%
    \end{figure}

Once the sample has been constructed, we compare our flux excesses to a grid of simple Cloudy \citep{Ferland2017} photoionisation models, using stellar populations from the Binary Population and Spectral Synthesis version 2.2.1 \citep[BPASS; ][]{Eldridge2017} as intrinsic SEDs. The models were run to convergence assuming a constant SFH. The stellar and nebular parameters were varied to cover a broad range of metallicities (Z = 0.001, 0.006, 0.014 and 0.030; in this convention Z$_{\odot}$ = 0.014), ages ($3\times 10^6$,$5\times 10^6$,$10^7$ and $5\times 10^7$ years), ionisation parameters (log$\langle U \rangle = -3.5$ to -0.5 in steps of 0.5), and densities (log $\rho$/[cm$^{-3}$] = 0, 1, 2 and 3). The net transmitted SEDs (that include nebular emission) were then redshifted between $z = 3.9$ and $9.0$, in steps of 0.1, and photometry was simulated in the filters of interest (F335M, F356W, F410M and F444W) using the code \texttt{Bagpipes} \citep{Carnall2018}. Figure~\ref{fig:cloudy_obs} shows the results of this Cloudy exercise, for visualisation purposes the models are shown as shaded areas colour-coded by log$\langle U \rangle$, and represent the shape expected in each filter pair, as a function of redshift. The regions where the respective emission lines dominate either filter pair are highlighted as vertical bands. As expected, there is a strong dependency of \oiii\/$_{\lambda 5007}$ emission with log$\langle U \rangle$. The final sample, composed of 677 galaxies in the redshift range $z = 3.9 - 8.9$ is shown as purple circles.

   \begin{figure*} 
        \centering
   \includegraphics[width=0.85\textwidth,trim={0 1.3cm 0 0}]{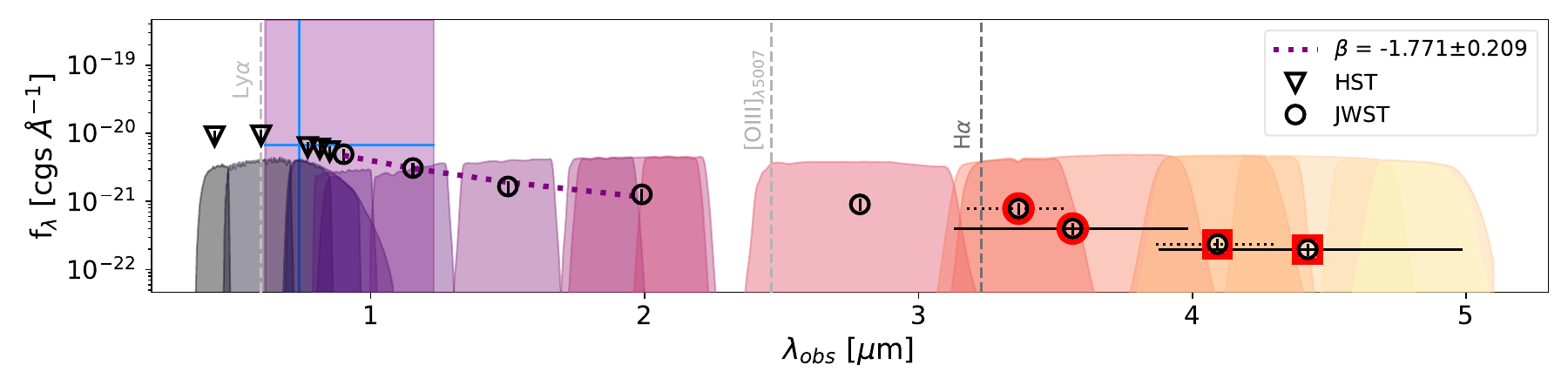}
   \includegraphics[width=0.85\textwidth,trim={0 1.3cm 0 0}]{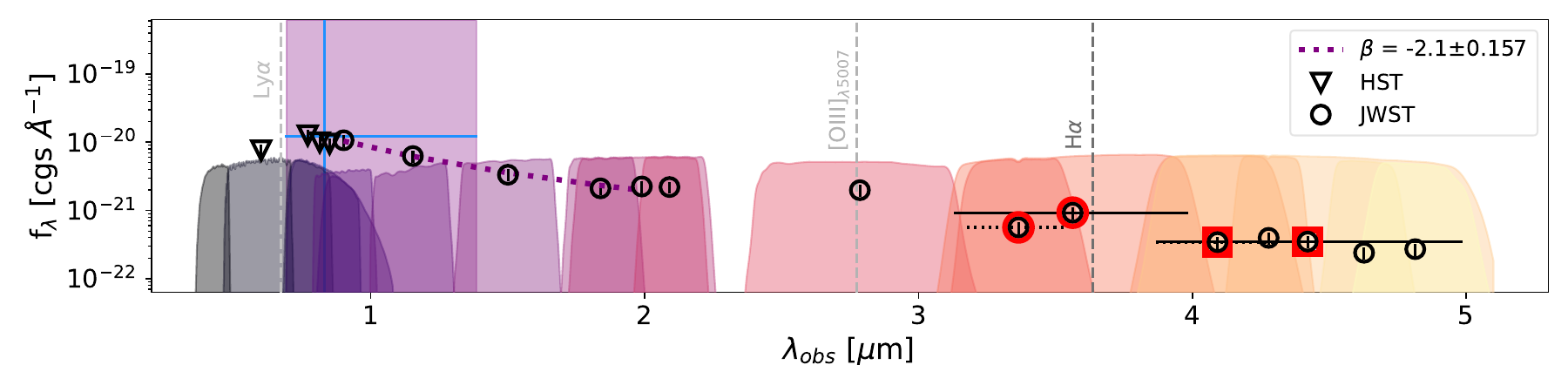}
   \includegraphics[width=0.85\textwidth,trim={0 1.3cm 0 0}]{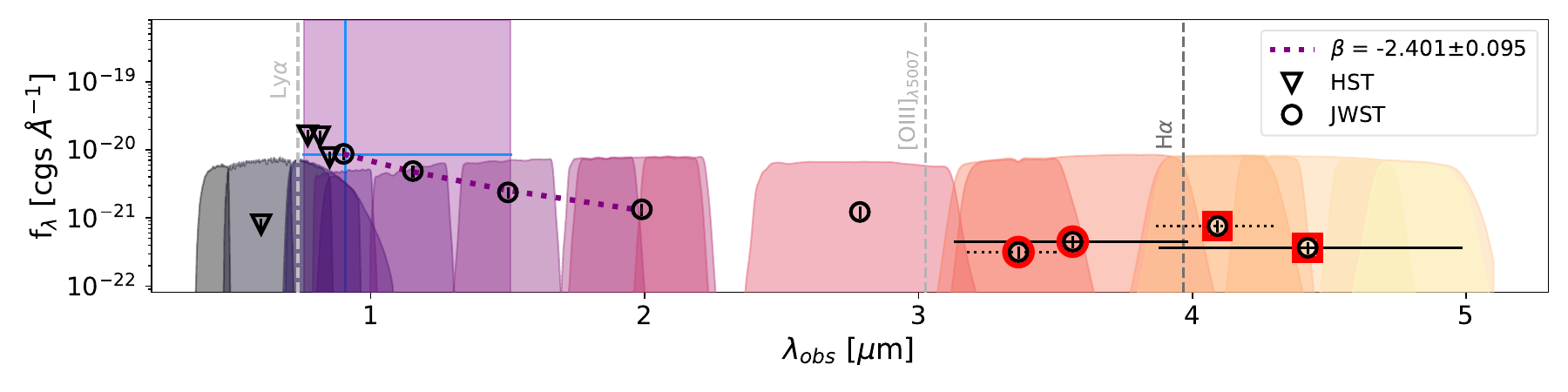}
   \includegraphics[width=0.85\textwidth]{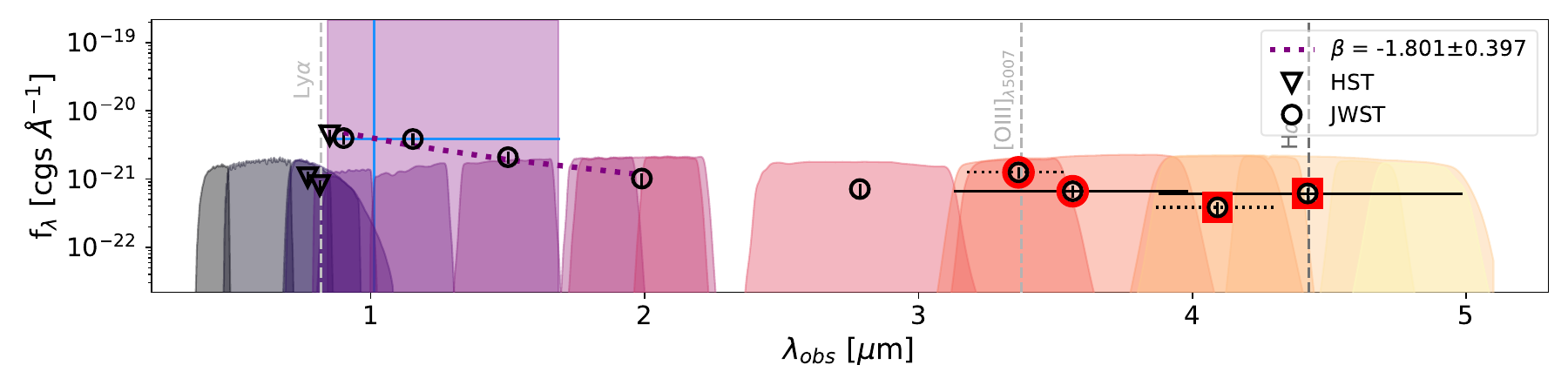}
   \caption{Representative example SEDs of each \ha\ redshift bin, with redshift increasing downward. From top to bottom: JADES-GS+53.20925-27.75711 ($z = 3.92 \pm 0.04$), JADES-GS+53.11398-27.80615 ($z = 4.54 \pm 0.05$), JADES-GS+53.15638-27.80966 ($z = 5.04 \pm 0.08$), and JADES-GS+53.15825-27.74091 ($z = 5.74 \pm 0.06$). The coloured curves show the transmission of the filters used in this work. Specifically, the HST/ACS bands: F435W, F606W, F775W, F814W, and F850LP. Followed by the JADES NIRCam bands: F090W, F115W, F150W, F200W, F277W, F335M, F356W, F410M, and F444W. Finally (when available), JEMS medium band photometry in the bands: F182M, F210M, F430M, F460M, and F480M. HST fluxes are shown as triangles, while the circles show JWST NIRCam photometry. The photometry of the filter-pairs of interest are highlighted in red (circles for F335M-F356W, squares for F410M-F444W). The purple vertical band marks the rest-frame 1250 - 2500 \AA\ region. The $\beta$ slope is given in the legend of each panel, and corresponds to the purple dashed line. Finally, the blue cross shows the observed wavelength and flux corresponding to rest-frame 1500 \AA\/. For every redshift bin the \ha\ line falls dominantly on either F335M, F356W, F410M or F444W. The detection and flux measurement of  \oiii\/$_{\lambda 5007}$ is performed in an analogous manner.} 
              \label{fig:SEDs_example_Ha}%
    \end{figure*}

    \begin{figure}
        \centering
   \includegraphics[width=0.5\textwidth]{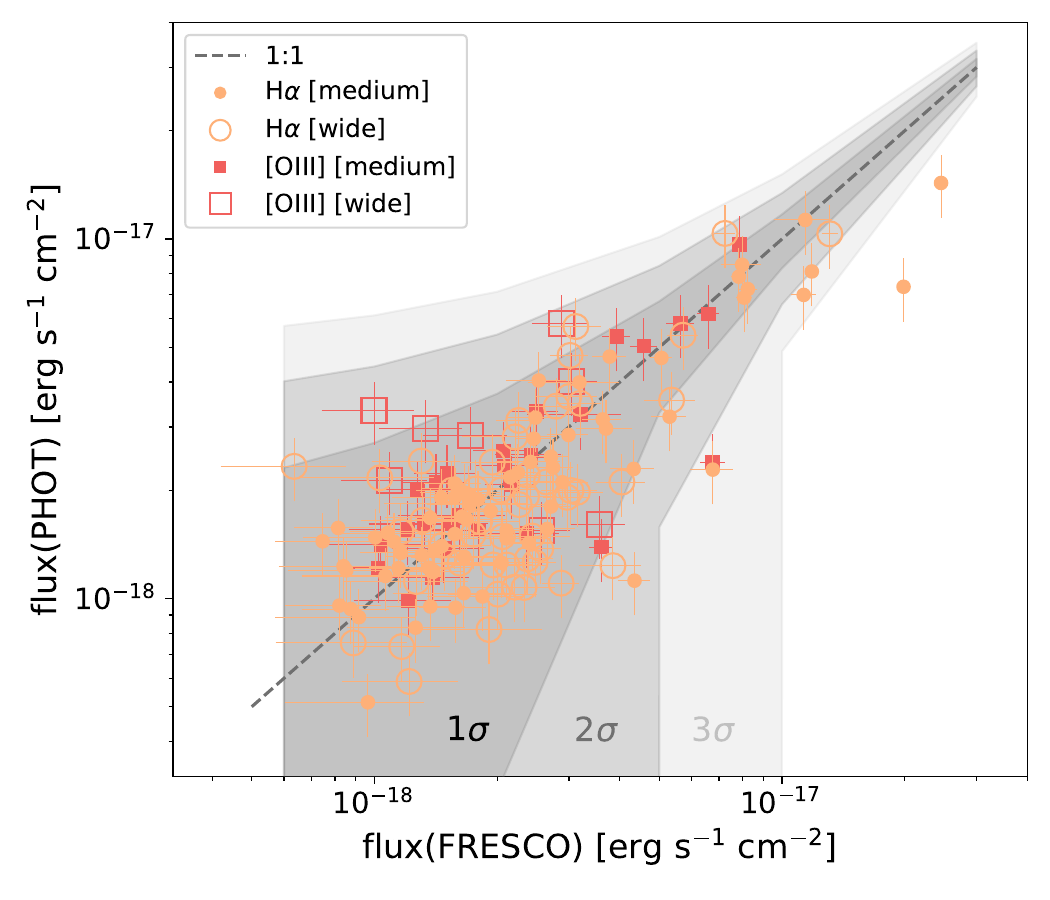}
   \caption{Comparison of fluxes obtained via photometry with those obtained through FRESCO grism spectra (when available). There are 122 overlapping cases with \ha\ fluxes (orange circles), and 36 with \oiii\/$_{\lambda 5007}$ fluxes (red squares). The filled and open symbols indicate if the emission line falls predominantly in a medium (F335M or F410M) or wide (F356W or F444W) band. The shaded areas show 1, 2 and 3$\sigma$, respectively. We find the values obtained by photometry in this work are in significant agreement with those measured with grism spectra, suggesting there is minimal contamination from other emission lines (\nii\/, \sii\/, \hb\/). The fluxes that fall in wide bands contain more continuum and noise, resulting in a larger scatter. We note the background subtraction in grism spectra can lead to an underestimation of emission line fluxes. In a similar manner, our photometrically-derived fluxes have a fixed aperture of diameter $0.3''$, and might not capture the total flux of a source.}
              \label{fig:fresco_comparison}%
    \end{figure}

\begin{table*}
    \small
    \centering
    \begin{tabular}{ccccc}
    \hline
    \noalign{\smallskip}
    Name & $z$ & $\log$ \xionnofesc\ (\ha\/) & $\log$ \xion\ (\oiii\/) & $\log$ \xion\ (Prospector) \\
        &    &  [Hz erg$^{-1}$]  & [Hz erg$^{-1}$]  & [Hz erg$^{-1}$] \\
    \noalign{\smallskip}
    \hline
    \noalign{\smallskip}
    JADES-GS+53.11634-27.81272 & 3.9094 $\pm$ 0.0476 & 25.26 $_{-0.30}^{+0.30}$ &     -     & 25.48 $_{-0.06}^{+0.03}$ \\ 
JADES-GS+53.20925-27.75711 & 3.9196 $\pm$ 0.0379 & 25.52 $_{-0.30}^{+0.30}$ &     -     & 25.52 $_{-0.04}^{+0.04}$ \\ 
JADES-GS+53.12549-27.78044 & 3.9413 $\pm$ 0.0462 & 25.17 $_{-0.30}^{+0.30}$ &     -     & 25.35 $_{-0.03}^{+0.05}$ \\ 
JADES-GS+53.18436-27.80581 & 3.9496 $\pm$ 0.0402 & 25.43 $_{-0.30}^{+0.30}$ &     -     & 25.19 $_{-0.14}^{+0.17}$ \\ 
JADES-GS+53.16268-27.73611 & 3.9496 $\pm$ 0.0399 & 25.31 $_{-0.30}^{+0.30}$ &     -     & 25.37 $_{-0.06}^{+0.06}$ \\ 
JADES-GS+53.15123-27.79826 & 3.9595 $\pm$ 0.0446 & 25.24 $_{-0.30}^{+0.30}$ &     -     & 25.01 $_{-0.15}^{+0.18}$ \\ 
JADES-GS+53.15282-27.79549 & 3.9688 $\pm$ 0.0417 & 25.43 $_{-0.30}^{+0.30}$ &     -     & 25.71 $_{-0.04}^{+0.03}$ \\ 
JADES-GS+53.19804-27.76002 & 3.9696 $\pm$ 0.0421 & 25.25 $_{-0.30}^{+0.30}$ &     -     & 25.39 $_{-0.05}^{+0.04}$ \\ 
JADES-GS+53.13905-27.75893 & 3.9740 $\pm$ 0.0515 & 25.39 $_{-0.30}^{+0.30}$ &     -     & 25.65 $_{-0.01}^{+0.03}$ \\ 
JADES-GS+53.15186-27.75258 & 3.9992 $\pm$ 0.0561 & 25.34 $_{-0.30}^{+0.30}$ &     -     & 25.58 $_{-0.04}^{+0.07}$ \\ 

JADES-GS+53.12644-27.79200 & 5.3746 $\pm$ 0.0581 & 25.47 $_{-0.30}^{+0.30}$ & 25.55 $_{-0.29}^{+0.32}$ & 25.77 $_{-0.04}^{+0.07}$ \\ 
JADES-GS+53.12775-27.78098 & 5.3751 $\pm$ 0.0650 & 25.38 $_{-0.30}^{+0.30}$ & 25.45 $_{-0.29}^{+0.32}$ & 25.52 $_{-0.04}^{+0.05}$ \\ 
JADES-GS+53.16729-27.75273 & 5.3785 $\pm$ 0.0617 & 25.28 $_{-0.30}^{+0.30}$ & 25.21 $_{-0.29}^{+0.31}$ & 25.45 $_{-0.07}^{+0.05}$ \\ 
JADES-GS+53.14381-27.80835 & 5.4189 $\pm$ 0.0763 & 25.27 $_{-0.30}^{+0.30}$ & 25.60 $_{-0.29}^{+0.32}$ & 25.72 $_{-0.05}^{+0.04}$ \\ 
JADES-GS+53.10726-27.81102 & 5.4290 $\pm$ 0.0729 & 25.90 $_{-0.30}^{+0.30}$ & 25.61 $_{-0.29}^{+0.32}$ & 25.88 $_{-0.03}^{+0.08}$ \\ 
JADES-GS+53.14676-27.79738 & 5.4294 $\pm$ 0.0578 & 25.36 $_{-0.30}^{+0.30}$ & 25.52 $_{-0.29}^{+0.32}$ & 25.71 $_{-0.05}^{+0.03}$ \\ 
JADES-GS+53.12301-27.79661 & 5.4393 $\pm$ 0.0601 & 25.57 $_{-0.30}^{+0.30}$ & 25.50 $_{-0.29}^{+0.32}$ & 25.80 $_{-0.02}^{+0.02}$ \\ 
JADES-GS+53.12247-27.79653 & 5.4418 $\pm$ 0.0666 & 25.49 $_{-0.30}^{+0.30}$ & 25.44 $_{-0.29}^{+0.32}$ & 25.67 $_{-0.05}^{+0.05}$ \\ 
JADES-GS+53.16407-27.79972 & 5.4422 $\pm$ 0.0826 & 25.40 $_{-0.30}^{+0.30}$ & 25.45 $_{-0.29}^{+0.32}$ & 25.05 $_{-0.11}^{+0.11}$ \\ 
JADES-GS+53.12874-27.79788 & 5.4430 $\pm$ 0.0572 & 25.41 $_{-0.30}^{+0.30}$ & 25.44 $_{-0.29}^{+0.32}$ & 25.65 $_{-0.05}^{+0.06}$ \\ 

JADES-GS+53.19106-27.79732 & 7.2634 $\pm$ 0.0745 &     -     & 25.44 $_{-0.29}^{+0.32}$ & 25.57 $_{-0.16}^{+0.11}$ \\ 
JADES-GS+53.17976-27.77465 & 7.2717 $\pm$ 0.0805 &     -     & 25.57 $_{-0.29}^{+0.32}$ & 25.63 $_{-0.03}^{+0.07}$ \\ 
JADES-GS+53.16579-27.82179 & 7.2890 $\pm$ 0.0873 &     -     & 25.63 $_{-0.29}^{+0.32}$ & 25.73 $_{-0.05}^{+0.15}$ \\ 
JADES-GS+53.18334-27.79050 & 7.2994 $\pm$ 0.0665 &     -     & 25.76 $_{-0.28}^{+0.32}$ & 25.71 $_{-0.05}^{+0.09}$ \\ 
JADES-GS+53.13219-27.78578 & 7.3890 $\pm$ 0.0851 &     -     & 25.74 $_{-0.28}^{+0.32}$ & 25.73 $_{-0.10}^{+0.18}$ \\ 
JADES-GS+53.16638-27.81237 & 7.3992 $\pm$ 0.0762 &     -     & 25.68 $_{-0.29}^{+0.32}$ & 25.80 $_{-0.08}^{+0.02}$ \\ 
JADES-GS+53.18405-27.79783 & 7.4091 $\pm$ 0.0826 &     -     & 25.60 $_{-0.29}^{+0.32}$ & 25.71 $_{-0.05}^{+0.06}$ \\ 
JADES-GS+53.18536-27.77319 & 7.4095 $\pm$ 0.0590 &     -     & 25.57 $_{-0.29}^{+0.32}$ & 25.62 $_{-0.15}^{+0.28}$ \\ 
JADES-GS+53.18393-27.79999 & 7.4193 $\pm$ 0.0719 &     -     & 25.45 $_{-0.29}^{+0.32}$ & 25.62 $_{-0.10}^{+0.06}$ \\ 
JADES-GS+53.18301-27.78946 & 7.4494 $\pm$ 0.0672 &     -     & 25.63 $_{-0.29}^{+0.32}$ & 25.91 $_{-0.17}^{+0.03}$ \\ 

    \end{tabular}
    \caption{Table excerpt showing a selection of galaxies in our sample. Depending on the redshift and the detection of emission lines, galaxies can have \xion\ estimations from \ha\/, \oiii\/$_{\lambda 5007}$, or both. \textsl{Column 1:} JADES identifier, composed of the coordinates of the centroid rounded to the fifth decimal place, in units of degrees. \textsl{Column 2:} photometric redshift from inferred using the template-fitting code \texttt{EAZY}. \textsl{Columns 3, 4 and 5:} logarithm of the ionising photon production efficiency estimations in units of Hz erg$^{-1}$. Columns 4 and 5 have values obtained through photometry using the \ha\ and \oiii\ methods, the uniformity in their errors arises from flooring the photometric uncertainties. Column 6 values where provided by \texttt{Prospector}.}
    \label{table:excerpt}
\end{table*}

    \begin{figure*}
        \centering
   \includegraphics[width=1\textwidth,trim={0 0.5cm 0 0}]{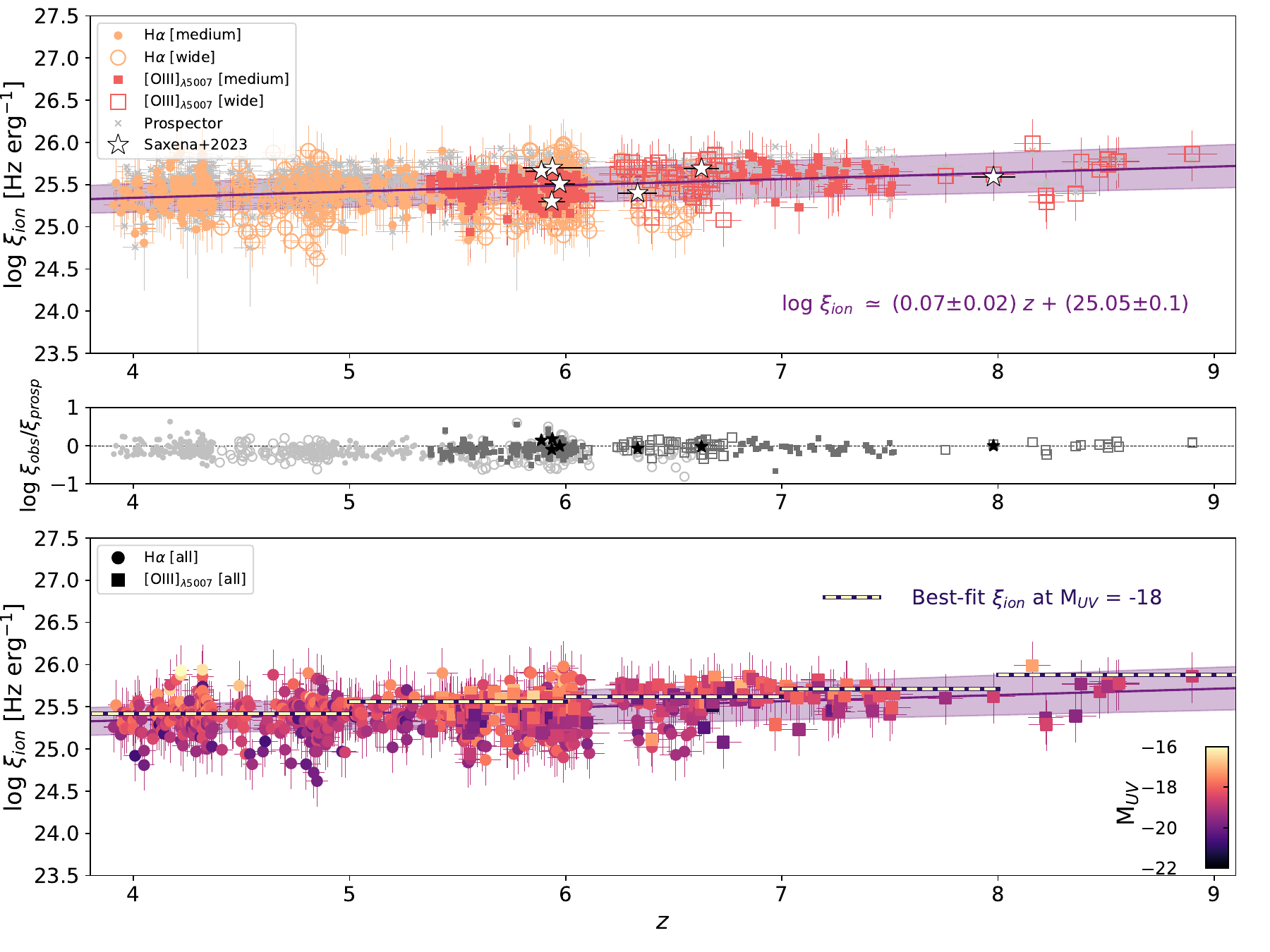}
   \caption{\xion\ values inferred through \ha\/ and \oiii\/$_{\lambda 5007}$ emission lines, as well as through through SED fitting. For comparison, we include NIRSpec measurements for 7 galaxies from \citet{Saxena2023} that overlap with our sample (white stars), most of which were derived from \ha\ fluxes. We note that for the \ha\ and \oiii\/$_{\lambda 5007}$ results, an SMC dust attenuation curve was assumed, and remind the reader that the \ha\ method in addition assumes an escape fraction of zero. \textsl{Top panel:} \xion\ versus redshift for the entire sample (677 galaxies). The symbols and colours of the values estimated by photometry are the same as in Figure~\ref{fig:fresco_comparison}. The \texttt{Prospector} estimations are shown in grey. The line represents the best fit to the photometrically-inferred results. As expected, there is more scatter when the emission lines fall on wide bands (either F356W or F444W), due to more noise and continuum being introduced. \textsl{Middle panel:} residuals between the values inferred through photometrically-estimated emission lines and via \texttt{Prospector}. The symbols are the same as in the upper panel, light grey corresponds to comparisons with the \ha\ method, while dark grey corresponds to comparisons with the \oiii\/$_{\lambda 5007}$ method. We find a good agreement between all methods, and confirm an increased \xion\ with redshift given by log \xion\ $= (0.07\pm 0.02)z + 25.05\pm 0.11$, consistent with literature. \textsl{Bottom panel:} same as top panel but only showing the photometrically-estimated \xion\/ values, and colour-coded by M$_{\rm{UV}}$. The dashed horizontal lines represent the intercepts of the best-fit relations shown in Figure~\ref{fig:xion_MUV} for a fixed M$_{\rm{UV}}$ of -18. It can be seen that at fixed M$_{\rm{UV}}$, \xion\ evolves with redshift.} 
              \label{fig:xion_redshift}%
    \end{figure*}

    \begin{figure}
        \centering
   \includegraphics[width=0.5\textwidth]{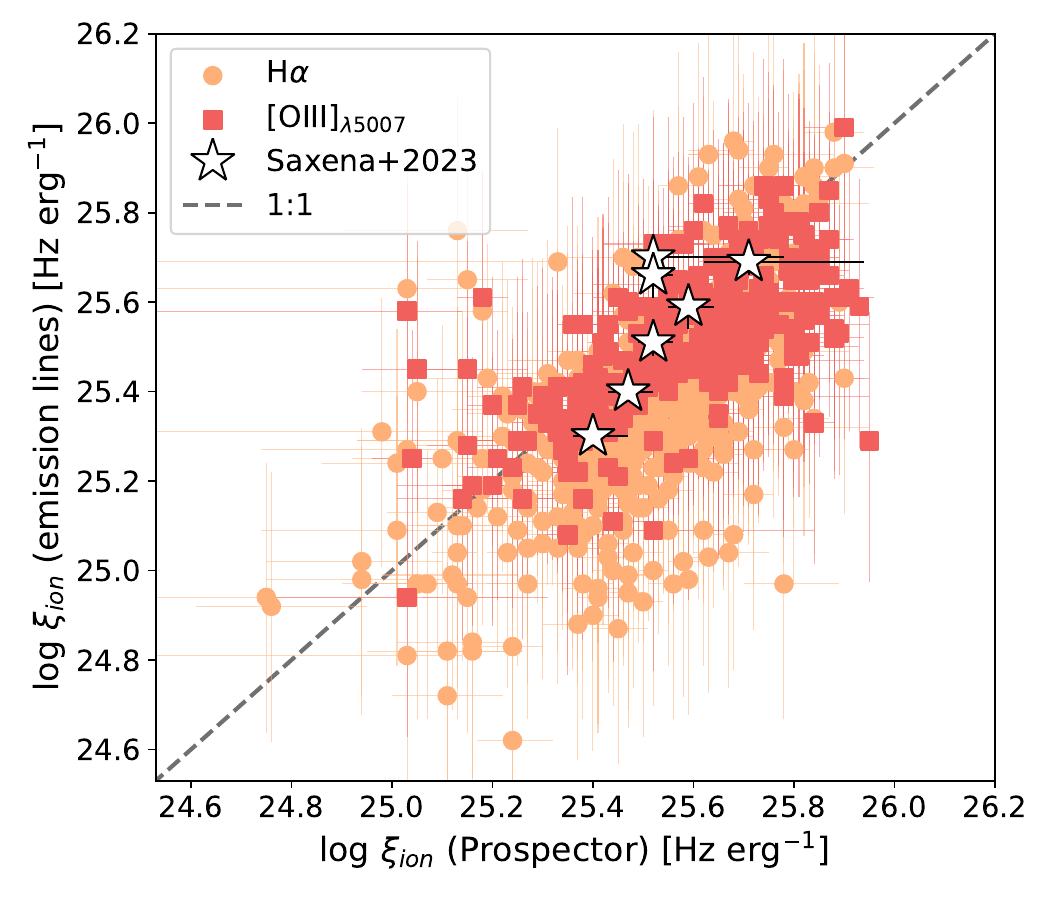}
   \caption{Comparison of \xion\ inferred through \ha\ fluxes (orange circles) and \oiii\/$_{\lambda 5007}$ EWs (red squares), with the values inferred by \texttt{Prospector}. We have also included the measurements from \citet{Saxena2023} as stars. The values scatter around the 1:1 relation, shown as a dashed grey line with 1,2 and 3$\sigma$ shaded areas.}
              \label{fig:xi_comparison}%
    \end{figure}

\section{Using photometry to constrain the ionising photon production efficiency of galaxies}
\label{OBS_METHODS}
To estimate \xion\ photometrically we use two methods, both of which rely on emission lines measurements, particularly \ha\ and \oiii\/$_{\lambda 5007}$. We now briefly present them, along with a description of how they were applied in this work. We remind the reader that all errors in photometric points were floored to 5\% in these calculations.

\subsection{\ha\ as proxy for ionising radiation production}

If we assume no ionising photons escape from a galaxy (\fesc\ = 0) and Case B recombination, the dust-corrected \ha\ luminosity is directly related to the amount of ionising photons (\ndot\/) that are being emitted. Adopting a temperature of $10^{4}$ K and an electron density of $100\, \text{cm}^{-3}$, these quantities are related by:
\begin{equation}
\label{eq:nion}
    \dot{n}_{\rm{ion}} = 7.28 \times 10^{11} \text{L(\ha\/)}
\end{equation}
as given in \cite{Osterbrock2006}, where \ndot\ is in units of photon s$^{-1}$, and L(\ha\/) in erg s$^{-1}$. This equation has a slight dependence on temperature and metallicity \citep{Charlot2001}, but for the purpose of this work this has been ignored. We note that the Case B recombination assumption yields a conservative estimation of the amount of ionising photons being produced, and non-zero escape fractions would lead to a boost in the derived \ndot\ values. Additionally, instead of ionising the surrounding gas or escaping, a significant amount of ionising photons could be absorbed by dust \citep[$\sim 30$\%; ][]{Tacchella2022stellar_pops}, resulting in a lack of nebular emission lines.

To estimate the ionising photon production efficiency per UV luminosity assuming Case B recombination, \xionnofesc\/ (the zero subscript indicates \fesc\/ = 0), we insert \ndot\ into the following equation:
\begin{equation}
    \xi_{\rm{ion,0}} = \frac{\dot{n}_{\text{ion}}}{\text{L}_{\rm{UV}}}
\end{equation}
where L$_{\rm{UV}}$ is the observed monochromatic luminosity in units of erg s$^{-1}$ Hz$^{-1}$, measured at the rest-frame wavelength of $\lambda$ = 1500 \AA\/.

\subsection*{Measurements from photometry} 
We define four redshift bins to estimate \ha\ fluxes, based the expected wavelength of \ha\/, as follows: 
\begin{enumerate}
    \item $3.90 \leq z \leq 4.37$: f(\ha\/) falls in F335M
    \item $4.37 < z < 4.92$: f(\ha\/) falls in F356W but outside F335M
    \item $4.92 \leq z \leq 5.61$: f(\ha\/) falls in F410M
    \item $5.61 < z \leq 6.59$: f(\ha\/) falls in F444W but outside F410M
\end{enumerate}

Where we assume the excess flux in the filter containing \ha\ is dominated by \ha\ emission, reasonable at high redshifts \citep[e.g. ][]{Cameron2023}. To obtain L$_{\rm{UV}}$ we fit a straight line in logarithmic space using the \texttt{curve\_fit} function in \texttt{SciPy} \citep{Virtanen2020}, between rest-frame 1250 and 2500 \AA\/, in the form $f_\lambda \propto \lambda^{\beta}$, where $\beta$ is the rest-frame UV continuum slope \citep[$\beta$; ][]{Calzetti1994}. We use all the available photometry in this region for each redshift bin.

Figure~\ref{fig:SEDs_example_Ha} shows a representative example SED of each \ha\ redshift bin. The identifier and redshift of each galaxy are given in the caption. The expected wavelengths of \ha\ and \oiii\/$_{\lambda 5007}$ are shown as vertical lines, it can be seen that \ha\ falls primarily in a different filter as redshift increases. The photometry of the four NIRCam filters of interest are highlighted with red edges. The $\beta$ slope is given in the legend, and corresponds to the purple dashed line. Finally, the blue cross shows the observed wavelength and flux corresponding to rest-frame 1500 \AA\/.

Once \ha\ fluxes have been calculated, they must be corrected for dust attenuation. This is not trivial for our sample since this parameter is not well understood at high redshifts \citep{Gallerani2010,Ma2019}, and we do not have measurements for Balmer line ratios. Moreover, the geometry and effect of dust attenuation in early galaxies is highly uncertain \citep{Bowler2018,Bowler2022}. Nevertheless, it has been shown that a steep attenuation curve, such as seen in the Small Magellanic Cloud \cite[SMC; ][]{Prevot1984,Gordon1998}, is appropriate for young high-redshift galaxies \citep{Shivaei2020}. Thus, we apply an average SMC attenuation curve \citep{Gordon2003} to our \ha\ and UV measurements, using $\beta$ to infer the nebular continuum colour excess E(B-V) , given by E(B-V) = $(\beta + 2.616)\times\frac{1}{11.259}$ \citep[][; adopting SMC attenuation]{Reddy2018}.

We note that in redshift bins (ii) and (iv), \ha\ falls in the wide band filter, and thus, more noise is introduced. In addition, they could be affected by \nii\ and/or \sii\ contamination. As discussed in \cite{Simmonds2023}, this contamination is not expected to be significant at high redshift \citep[see also; ][]{Maiolino2019, Onodera2020,Sugahara2022,Cameron2023}.

\newpage
\subsection{\oiii\ equivalent widths as proxy for ionising radiation production}

The previous method has some limitations, such as the assumption of Case B recombination, and the high dependence on the attenuation curve adopted. Moreover, at $z \gtrsim 6$, \ha\ is redshifted to wavelengths challenging to observe. To circumvent these limitations, an alternative method that depends on \oiii\/$_{\lambda 5007}$ instead was proposed in \cite{Chevallard2018}, granting access to higher redshifts (up to $z \sim 9.5$). Strong \oiii\ emission is indicative of intense ionisation conditions, such as those found at the early Universe. In brief, they use 10 local analogues ($z \sim 0$) to high-redshift galaxies and derive an empirical relation between \xion\ and \oiii\/$_{\lambda 5007}$ equivalent widths (EWs). \cite{Tang2019} conducted a similar project but with a larger sample and at higher redshift ($z \sim 2$). Since their sample is closer in parameter space to ours, we follow Equation 4 of their work, 
\begin{equation}
    \log \xi_{\rm{ion}} = (0.73 \pm 0.08) \times \log (\text{EW [OIII]}_{\lambda 5007}) + (23.45 \pm 0.23)
\end{equation}
assuming an SMC attenuation law. 

\subsection*{Measurements from photometry}
As in the \ha\ case, we define four redshift bins to estimate \oiii\/$_{\lambda 5007}$ fluxes, as follows: 
\begin{enumerate}
    \item $5.37 \leq z \leq 6.15$: f(\oiii\/$_{\lambda 5007}$) falls in F335M
    \item $6.15 < z < 6.77$: f(\oiii\/$_{\lambda 5007}$) falls in F356W but outside F335M
    \item $6.77 \leq z \leq 7.55$: f(\oiii\/$_{\lambda 5007}$) falls in F410M
    \item $7.55 < z \leq 9.00$: f(\oiii\/$_{\lambda 5007}$) falls in F444W but outside F410M
\end{enumerate}

Our data allows us to estimate \oiii\/$_{\lambda\lambda 4959+5007}$ fluxes, therefore, to obtain \oiii\/$_{\lambda 5007}$ we adopt the standard ratio between the components of the \oiii\ doublet: \oiii\/$_{\lambda 5007} = 0.75 \times$ \oiii\/$_{\lambda\lambda 4959,5007}$. Unless stated differently, all \oiii\ fluxes in this work hereafter represent \oiii\/$_{\lambda 5007}$. The EWs are then the division between the \oiii\ line fluxes and the local continuum. The latter was estimated following two prescriptions depending if the line falls on the medium or the wide band of each filter pair (F335M-F356W or F410M-F444W). If the line falls in the medium band, then the wide band also includes it, so the local continuum is measured from the corresponding wide band minus the line contribution. On the other hand, if the line falls in the wide band, then the corresponding medium band is assumed to represent the continuum. The differential dust attenuation between continuum and nebular emission is uncertain at high redshifts. Here we assume a ratio of $1.3$ between the reddening affecting emission lines and continuum, appropriate at $z \sim 1$ \citep{Pannella2015}, but caution that this value can be closer to $2$ for galaxies with low metallicities \citep{Shivaei2020}. Adopting the latter would systematically increase our \oiii\/ EWs and consequently, our inferred \xion\ measurements. We note that \oiii\ might suffer from \hb\ contamination, below we investigate the importance of this contamination by comparing our fluxes to those measured in FRESCO grism spectra.

\subsection{Flux comparisons to FRESCO grism spectra}
To investigate the importance of contamination from other emission lines in our \ha\ and \oiii\ fluxes, as well as to test the simplistic approach to measuring the fluxes, we compare our measurements (when available) with those obtained through an independent reduction of FRESCO grism spectra. The detailed FRESCO grism line flux measurements and validation for a larger sample of $z = 5 - 9$ galaxies will be presented in a forthcoming paper (Sun et al. in prep). 
We find 122 (36) overlapping galaxies with \ha\ (\oiii\/) flux measurements. Figure~\ref{fig:fresco_comparison} shows the results for both \ha\ (circles) and \oiii\ (squares). The filled and open symbols in each case denote if the emission line falls in a medium or wide band, respectively. We find that most measurements are within 3$\sigma$ of a 1:1 relation, confirming that our approach, while simplistic, is overall acceptable. Moreover, it indicates that if other lines are contaminating our flux estimations (\nii\ or \sii\ in the case of \ha\/, \hb\ in the case of \oiii\/), then the contribution is not significant on average.  We draw attention to the limitations of estimating emission-line fluxes using these two methods: grism spectra can potentially be affected by background subtraction, while aperture photometry can neglect some flux in extended sources. Both cases would lead to an underestimation in the measurement of emission line fluxes.

\section{SED fitting with \texttt{Prospector}}
\label{PROSPECTOR}

We use the galaxy SED fitting code \texttt{Prospector} \citep{Johnson2019,Johnson2021} to study our sample, and compare to our \xion\ estimations. This code uses photometry and/or spectroscopy as an input in order to infer stellar population parameters, from UV to IR wavelengths. In this work we use photometry from the HST ACS bands: F435W ($\lambda_{\rm{eff}} =  0.432\text{ }\mu$m), F606W ($\lambda_{\rm{eff}} = 0.578\text{ }\mu$m), F775W ($\lambda_{\rm{eff}} = 0.762\text{ }\mu$m), F814W ($\lambda_{\rm{eff}} = 0.803\text{ }\mu$m), F850LP ($\lambda_{\rm{eff}} = 0.912\text{ }\mu$m). In addition, we use the JADES NIRCam photometry from: F090W ($\lambda_{\rm{eff}} =  0.898\text{ }\mu$m), F115W ($\lambda_{\rm{eff}} =  1.143\text{ }\mu$m), F150W ($\lambda_{\rm{eff}} =  1.487\text{ }\mu$m), F200W ($\lambda_{\rm{eff}} =  1.968\text{ }\mu$m), F277W ($\lambda_{\rm{eff}} = 2.786\text{ }\mu$m), F335M ($\lambda_{\rm{eff}} =  3.365\text{ }\mu$m), F356W ($\lambda_{\rm{eff}} =  3.563\text{ }\mu$m), F410M ($\lambda_{\rm{eff}} =  4.092\text{ }\mu$m), and F444W ($\lambda_{\rm{eff}} =  4.421\text{ }\mu$m). Finally, when available, we include JEMS photometry: F182M ($\lambda_{\rm{eff}} = 1.829\text{ }\mu$m), F210M ($\lambda_{\rm{eff}} = 2.091\text{ }\mu$m), F430M ($\lambda_{\rm{eff}} = 4.287\text{ }\mu$m), F460M ($\lambda_{\rm{eff}} = 4.627\text{ }\mu$m), and F480M ($\lambda_{\rm{eff}} = 4.814\text{ }\mu$m). The same circular aperture of diameter $0.30''$ is used to extract the HST, JADES and JEMS convolved photometry. All photometry has been aperture corrected.

For the redshift, we adopt a normal distribution using the \texttt{EAZY} photo-z as a mean, with the sigma given by the photo-z errors. We vary the dust attenuation and stellar population properties following \cite{Tacchella2022}. In particular, we use a two component dust model described in  \cite{Conroy2009}. This model accounts for the differential effect of dust on young stars ($< 10$ Myr) and nebular emission lines, through a variable dust index. We adopt a Chabrier initial mass function \citep{Chabrier2003}, with mass cutoffs of 0.1 and 100 M$_{\odot}$, respectively, allowing the stellar metallicity to explore a range between 0.01 - 1 Z$_{\odot}$, and include nebular emission. The continuum and emission properties of the SEDs are provided by the Flexible Stellar Population Synthesis (FSPS) code \citep{Byler2017}, based on Cloudy models \citep[v.13.03; ][]{Ferland2013}. This earlier version of Cloudy introduces an upper limit on the permitted ionisation parameters (log$\langle U \rangle$$_{\rm{max}} = -1.0$). Due to the stochastic nature of the IGM absorption, we set a flexible IGM model based on a scaling of the Madau model \citep{Madau1995}, with the scaling left as a free parameter with a clipped normal prior ($\mu = 1.0, \sigma = 0.03$, in a range [0.0, 2.0]). Last but not least, we use a non-parametric SFH \citep[continuity SFH; ][]{Leja2019}. This model describes the SFH as six different SFR bins, the ratios and amplitudes between them are in turn, controlled by the bursty-continuity prior \citep{Tacchella2022bursty}.

In this work we use \texttt{Prospector} to calculate \xion\ for our entire sample, as well as to infer galaxy properties. The latter can be found in Appendix~\ref{Appendix:Prospector}. \texttt{Prospector} has the ability to reconstruct the full SED of galaxies, therefore, \xion\ is calculated from direct integration of the spectra, allowing to marginalise over most of the assumptions made for the most direct observational estimates from the emission line excess presented in Section~\ref{OBS_METHODS}.

    \begin{figure*}
        \centering
   \includegraphics[width=1\textwidth]{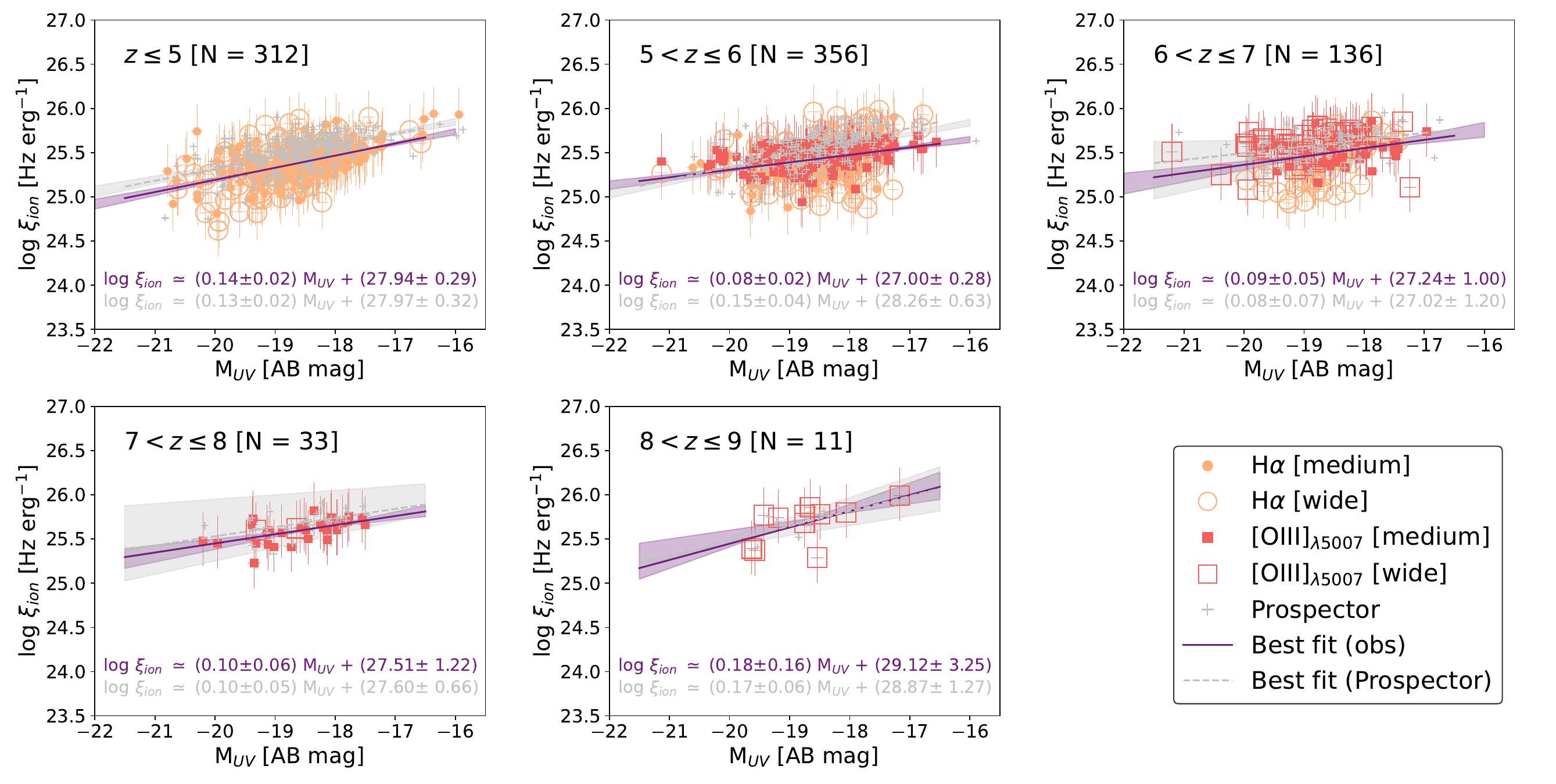}
   \caption{Dependence of \xion\ on UV magnitude, separated in redshift bins. The symbols and colours are the same as in Figures~\ref{fig:fresco_comparison} and ~\ref{fig:xion_MUV}. The number of galaxies in each redshift bin is indicated in the top left corner of each panel. The filled (dashed) line is the best fit to the data obtained via emission lines  \texttt{Prospector}). We find that fainter galaxies are more efficient at producing ionising radiation, as expected from previous studies. A version of this figure with \ndot\ can be found in Appendix~\ref{Appendix:nion}.}
              \label{fig:xion_MUV}%
    \end{figure*}

    \begin{figure}
        \centering
   \includegraphics[width=0.5\textwidth]{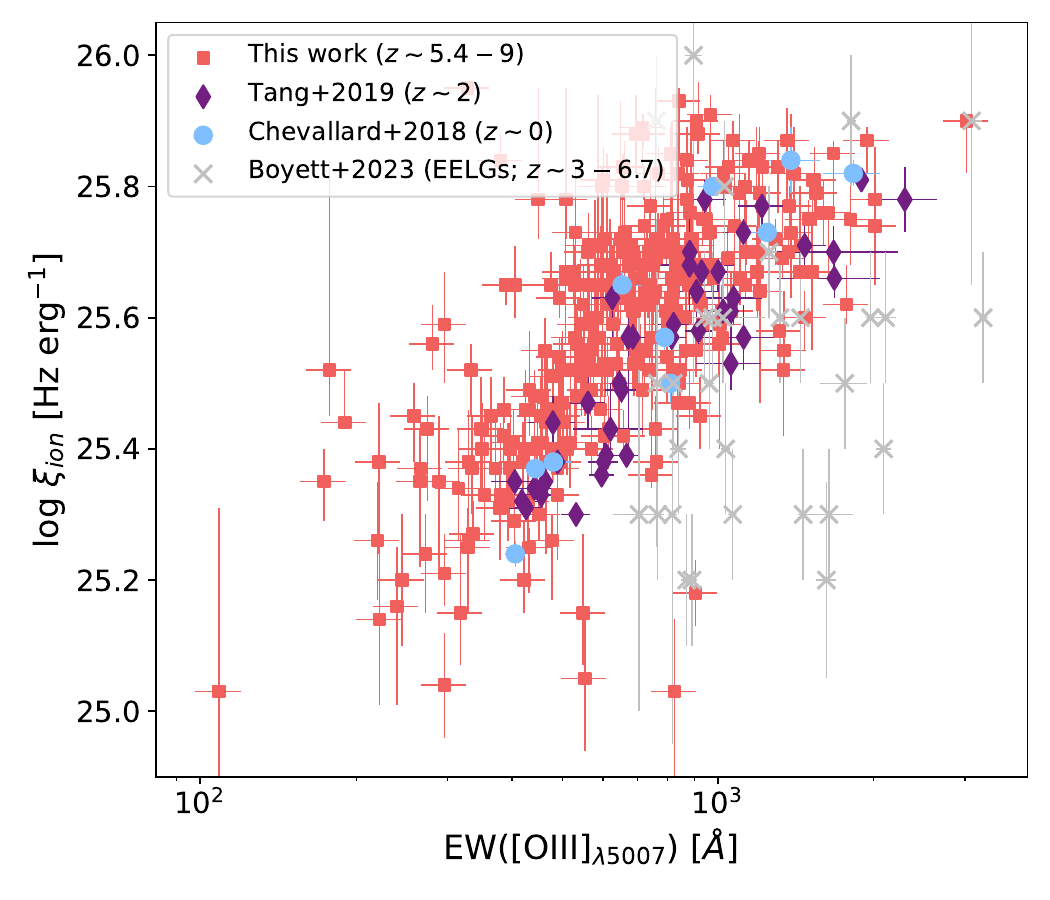}
   \caption{\xion\ versus \oiii\/$_{\lambda 5007}$ equivalent widths (figure adapted from \citet{Tang2019}). The red squares represent our sample, described by 
   log \xion\/ $= (0.57 \pm 0.09)\times$ log(EW[\oiii\/$_{\lambda 5007}$]) + $ (23.97 \pm 0.25)$ , while the purple diamonds show the results from \citet{Tang2019}, and the light blue circles those from \citet{Chevallard2016}. We also include the EELGs from Boyett et al. (in prep.) as grey crosses. The \xion\ for our sample is provided by \texttt{Prospector}. Except for a few outliers, our sample follows the same trend as the previous works, confirming that \oiii\ strength is also a reliable tracer of \xion\ in the early Universe.}
              \label{fig:Tang}%
    \end{figure}

\section{Constraints on \xion} 
\label{CONSTRAINTS}

After confirming the overall consistency of our flux measurements, we estimate \xion\ following the methods described in Section~\ref{OBS_METHODS}, and compare them to those inferred by \texttt{Prospector}. We provide an excerpt of the results in Table~\ref{table:excerpt}
, and present them visually in Figure~\ref{fig:xion_redshift}. We find a good agreement between the \xion\ obtained through \ha\/, \oiii\/, and \texttt{Prospector}, this agreement is highlighted in Figure~\ref{fig:xi_comparison}. In addition, we include seven LAEs studied in \citep{Saxena2023} using NIRSpec spectra, for which \xion\ was measured directly from Balmer recombination lines (\ha\ and \hb\/). These seven LAEs overlap with our sample and our results are consistent with theirs (see Table~\ref{table:LAE_comparison}). Moreover, our \xion\ values agree with those found in literature. For example, \cite{Stefanon2022} compiled \xion\ measurements up to $z \sim 8$ \citep[using data points from ][]{Stark2015,Stark2017,Marmol-Queralto2016,Nakajima2016,Bouwens2016,Matthee2017,Harikane2018,Shivaei2018,DeBarros2019,Lam2019,Faisst2019,Tang2019,Nanayakkara2020,Emami2020,Endsley2021,Naidu2022,Atek2022}. With this extensive compilation, they provided a best fit to the slope of \xion\/ as a function of redshift (given by dlog \xionnofesc / dz $= 0.09 \pm 0.01$), which is consistent within errors with this work (dlog \xion / dz $= 0.07 \pm 0.02$). More recently, JWST has been used to estimate \xion\ for individual galaxies up to $z \sim 8$ \citep[see ][]{Ning2023,Prieto-Lyon2023,Simmonds2023,Rinaldi2023arXiv}, and  this  work is also consistent with those. In the bottom panel of Figure~\ref{fig:xion_redshift}, \xion\ is shown as a function of redshift but colour-coded by M$_{\rm{UV}}$. The horizontal dashed lines show the intercepts of the best-fit relations between \xion\ and M$_{\rm{UV}}$ per redshift bin, discussed in the next paragraph, for a constant M$_{\rm{UV}}$ of -18. Their increase demonstrates that for a fixed M$_{\rm{UV}}$, \xion\ evolves with redshift.

\xion\ has been shown to vary due to the metallicity, age and dust content of galaxies \citep{Shivaei2018}, as well as due to their UV luminosities \citep{Duncan2015}, where fainter galaxies are more efficient at producing ionising radiation. This is clearly illustrated in Figure 3 from \cite{Maseda2020}, which consists of a compilation of measurements from literature \citep[specifically; ][]{Bouwens2016, Matthee2017,Harikane2018,Lam2019}. We check for this relation in our data and find a similar trend, shown in Figure~\ref{fig:xion_MUV}. For clarity, the sample is separated into redshift bins. As expected, there are less galaxies in the higher redshift bins, however, we consistently find the fainter galaxies in our sample have increased \xion\/. In addition, the higher redshift bins in our sample ($z > 7$) are populated by fainter galaxies than the other bins. This is potentially a result of our selection function, and will be discussed later. We note that an opposite trend is seen with \ndot\/, namely, that \ndot\ decreases for fainter galaxies (see Appendix~\ref{Appendix:nion}). Given the observed trends of \xion\ with redshift and with M$_{\rm{UV}}$, and the reliability of the \texttt{Prospector}-inferred \xion\ 
 (and M$_{\rm{UV}}$) measurements for our sample, we perform a 2-dimensional line fit combining these parameters and find:
\begin{equation}
    \log \xi_{\rm{ion}} (z,\text{M}_{\rm{UV}}) = (0.05 \pm 0.02)z + (0.11 \pm 0.02) \text{M}_{\rm{UV}} + (27.33 \pm 0.37)
\end{equation}
where \xion\ is in units of Hz erg$^{-1}$. This equation simultaneously describes the positive evolution of \xion\ with $z$ and M$_{\rm{UV}}$, which is shown in Figure~\ref{fig:xion_redshift}.

Regarding the use of \oiii\ EWs to estimate \xion\/, in Figure~\ref{fig:Tang} we present an update to a figure from \cite{Tang2019}, showing how \xion\ and \oiii\ EWs correlate. We plot their $z \sim 2$ results, along with the local ones from \cite{Chevallard2018}, and the ones estimated for a sample of Extreme Emission Line Galaxies (EELGs) at $z \sim 3 - 6.7$ in Boyett et al. (in prep.). We include our \oiii\ EWs obtained from photometry and the \xion\ from \texttt{Prospector}. We remind the reader that the use of a fixed circular aperture in the photometry might result in an underestimation of fluxes when sources are extended. Additionally, differences in dust treatments affect the EW measurements, for example, adopting a higher ratio between nebular and continuum dust attenuation would increase the measurements of this work, making them more compatible with those from Boyet et al. (in prep.). Independent of these slight discrepancies, for every sample, \xion\ is seen to increase with \oiii\ EWs. This  connection between \xion\ and \oiii\ EWs has also been seen in some simulations, for example, \cite{Seeyave2023} report a positive correlation between \oiii\ EWs and \xion\ in the First Light And Reionisation Epoch Simulations \citep[FLARES; ][]{Lovell2021,Vijayan2021}. Our results broadly follow the expected relation, and, by comparing the results derived by emission line fluxes to those inferred by \texttt{Prospector}, we corroborate that \oiii\ strengths are good tracers of \xion\/. This is particularly useful in the high redshift Universe, where Hydrogen recombination lines are not easily accessible.

    \begin{figure}
        \centering
   \includegraphics[width=0.45\textwidth]{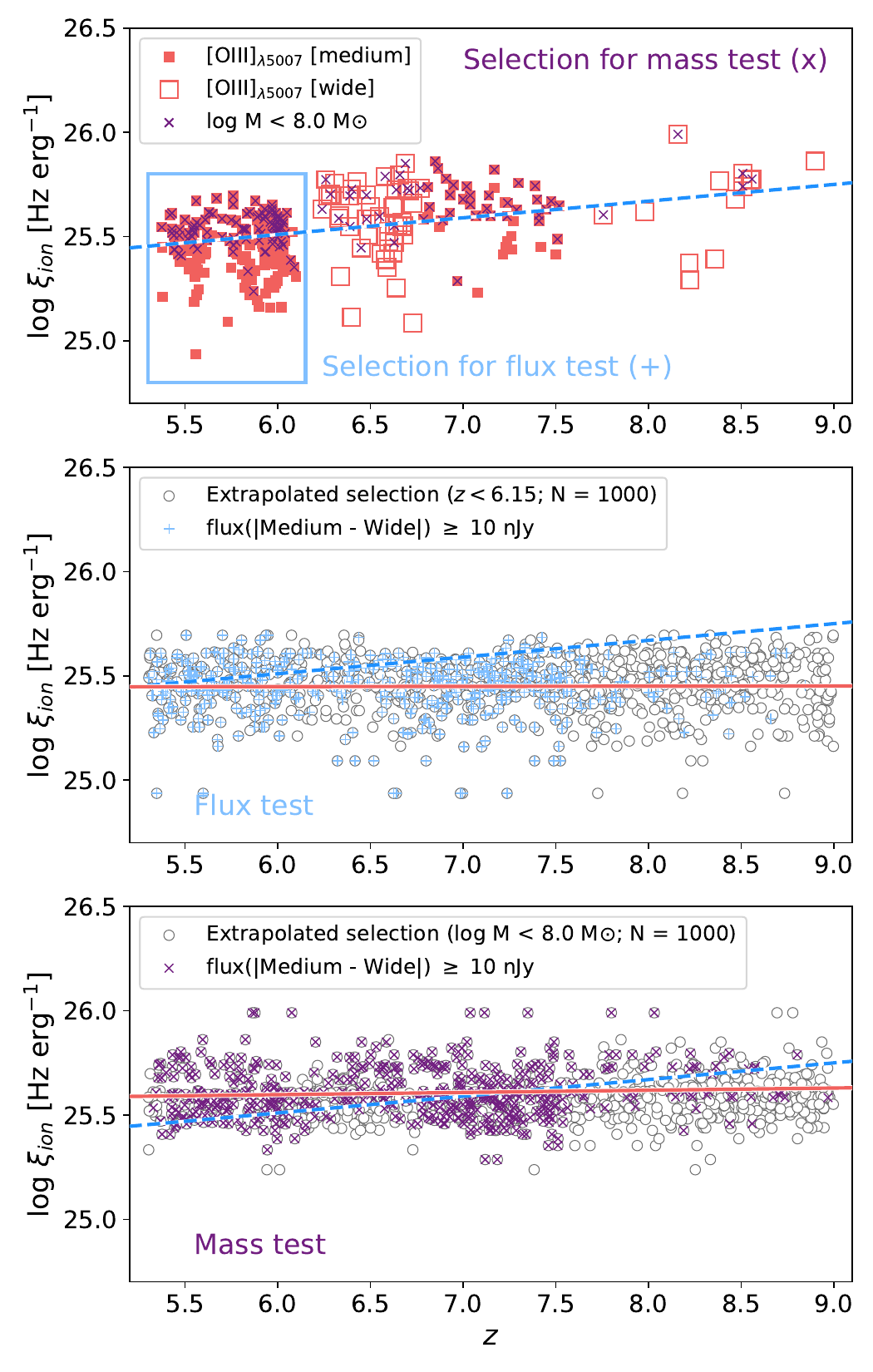}
   \caption{Null hypothesis test. \xion\ versus redshift, the blue dashed line (in all panels) is the best fit to the \xion\ inferred via emission lines, as seen in Figure~\ref{fig:xion_redshift}. For simplicity, errors have been omitted. \textsl{Top panel}: \xion\ values from this work, obtained via \oiii\ EWs. The blue rectangle shows the galaxies selected as seeds in order to simulate 1000 galaxies covering the whole redshift range shown ($z \sim 5.5 - 9$) in the middle panel. While the purple crosses mark the galaxies that have low stellar masses (log M $< 8.0$ M$_{\odot}$), and are used as seeds to simulate 1000 galaxies in the bottom panel. \textsl{Middle panel:} simulated galaxies (white circles with grey edges) and ones that would be observable according to our sample selection criteria (blue plus signs; flux difference between filter pairs of at least 10 nJy). \textsl{Bottom panel:} simulated galaxies (white circles with grey edges) and ones that would be observable according to our sample selection criteria (purple crosses; flux difference between filter pairs of at least 10 nJy). The red solid line in the middle and bottom panels is the best fit to the blue plus signs and purple crosses, respectively. The slope of the red line does not match the slope of the blue dashed one, indicating that the null hypothesis is wrong in both cases, and that the increase of \xion\ with redshift is not due to a luminosity or mass bias in our selection criteria.}
              \label{fig:bias}%
    \end{figure}

    \begin{figure*}
        \centering
   \includegraphics[width=0.75\textwidth]{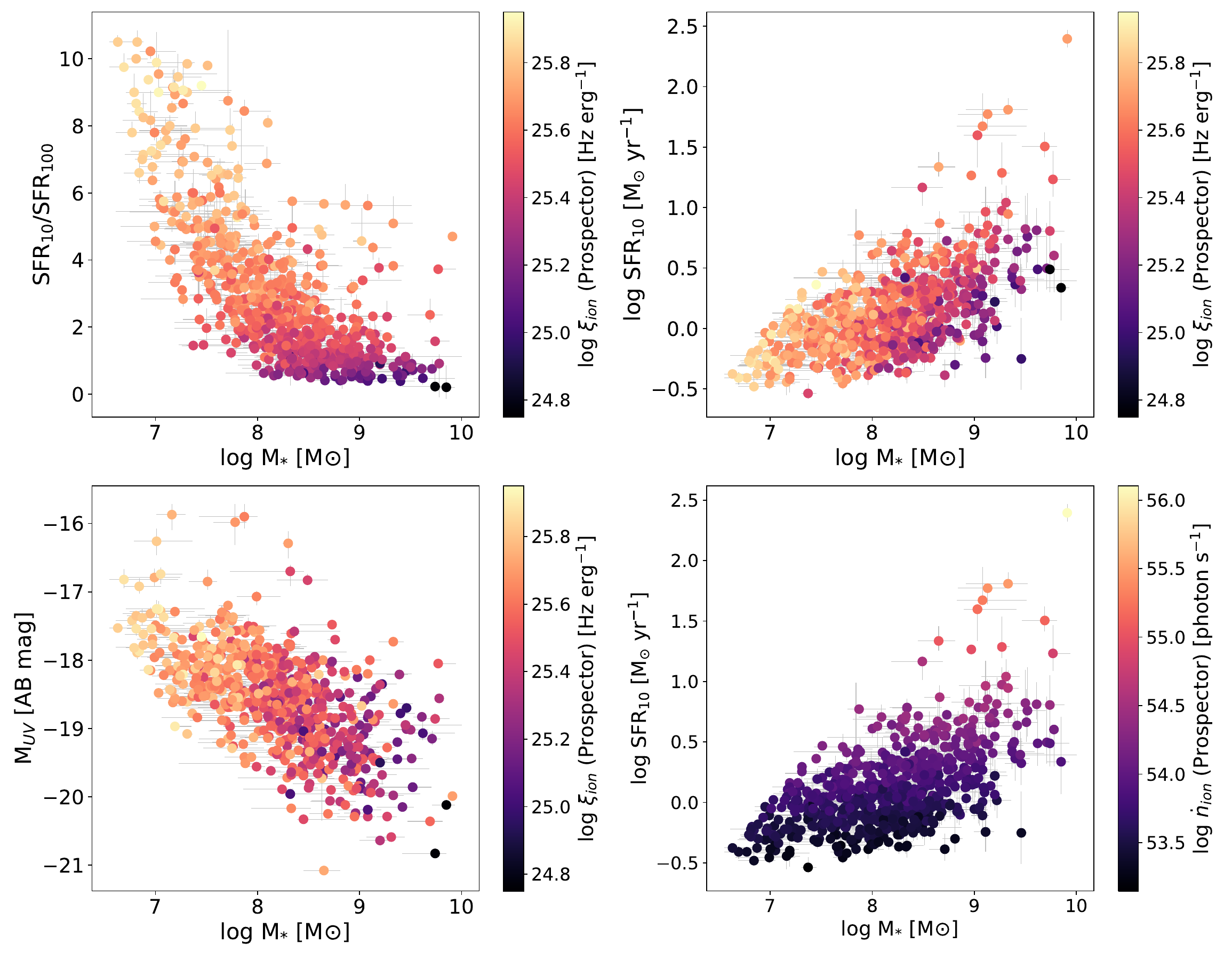}
   \caption{Relations between relevant galactic properties, as inferred by \texttt{Prospector}. \textsl{Top left panel:} ratio between recent (within 10 Myr) and sustained over 100 Myr star formation, SFR$_{10}$/SFR$_{100}$ (i.e. burstiness of star formation), over stellar mass, and colour-coded by \xion\/. The Spearman's rank correlation coefficient value is 0.914, while the p-value is consistent with zero. Lower mass galaxies with bursty SFHs have an increased \xion\ with respect to non-bursty higher-mass galaxies. \textsl{Bottom left panel:} correlation between UV luminosity and stellar mass, colour-coded by \xion\/. There is a strong trend of decreasing stellar mass with increasing M$_{\rm{UV}}$. \textsl{Top and bottom right panels:} relation between recent star formation, stellar mass and \xion\ (top) or \ndot\ (bottom).}
              \label{fig:burstiness}%
    \end{figure*}

\begin{table*}
    \small
    \centering
    \begin{tabular}{cccccc}
    \hline
    \noalign{\smallskip}
    Name & $z$ & $\log$ \xionnofesc\ (\ha\/) & $\log$ \xion\ (\oiii\/) & $\log$ \xion\ (Prospector) & $\log$ \xion\ (NIRSpec)\\
        &    &  [Hz erg$^{-1}$]  & [Hz erg$^{-1}$]  & [Hz erg$^{-1}$] & [Hz erg$^{-1}$] \\
    \noalign{\smallskip}
    \hline
    \noalign{\smallskip}
    JADES-GS+53.17657-27.77113 & $5.89\pm 0.07$ & - & 25.58$^{+0.11}_{-0.15}$ & 25.52$^{+0.04}_{-0.03}$ & 25.66$^{+0.05}_{-0.05}$ (\ha\/)\\
     JADES-GS+53.12176-27.79764 & $5.93\pm 0.14$ & 25.57$^{+0.20}_{-0.40}$ & 25.54$^{+0.11}_{-0.15}$ & 25.52$^{+0.26}_{-0.00}$ & 25.70$^{+0.01}_{-0.01}$ (\ha\/)\\
     JADES-GS+53.11042-27.80892 & $5.94\pm 0.06$ & 25.10$^{+0.20}_{-0.40}$ & 25.43$^{+0.11}_{-0.15}$ & 25.40$^{+0.07}_{-0.04}$ & 25.30$^{+0.02}_{-0.02}$ (\ha\/)\\
     JADES-GS+53.16062-27.77161 & $5.97\pm 0.07$ & - & 25.46$^{+0.11}_{-0.15}$ &  25.52$^{+0.04}_{-0.03}$ & 25.51$^{+0.03}_{-0.03}$ (\ha\/)\\
     JADES-GS+53.13492-27.77271 & $6.33\pm 0.09$ & 25.19$^{+0.20}_{-0.40}$ & 25.60$^{+0.20}_{-0.40}$ & 25.47$^{+0.05}_{-0.04}$ & 25.40$^{+0.02}_{-0.02}$ (\ha\/)\\
     JADES-GS+53.16905-27.77883 & $6.63\pm 0.08$ & - &  25.47$^{+0.20}_{-0.40}$& 25.71$^{+0.23}_{-0.09}$& 25.69$^{+0.03}_{-0.04}$ (\ha\/)\\
     JADES-GS+53.15683-27.76716 & $7.98\pm 0.10$ & - & 25.62$^{+0.20}_{-0.40}$ & 25.59$^{+0.05}_{-0.04}$ & 25.59$^{+0.05}_{-0.05}$ (\hb\/)\\
    \end{tabular}
    \caption{Comparison between the \xion\ values from this work to those presented in \citet{Saxena2023}. The values estimated from emission line fluxes and those derived from \texttt{Prospector}, are in broad agreement with those found through NIRSpec spectra. The columns are as in Table~\ref{table:excerpt}, but now include NIRSpec measurements. \textsl{Column 1:} JADES identifier, composed of the coordinates of the centroid rounded to the fifth decimal place, in units of degrees. \textsl{Column 2:} photometric redshift  inferred using the template-fitting code \texttt{EAZY}. \textsl{Columns 3 - 6:} logarithm of the ionising photon production efficiency estimations in units of Hz erg$^{-1}$. For the NIRSpec estimations, we specify whether \ha\ or \hb\ fluxes were used to infer \xion\/.}
    \label{table:LAE_comparison}
\end{table*}

        \begin{figure}
        \centering
   \includegraphics[width=0.45\textwidth]{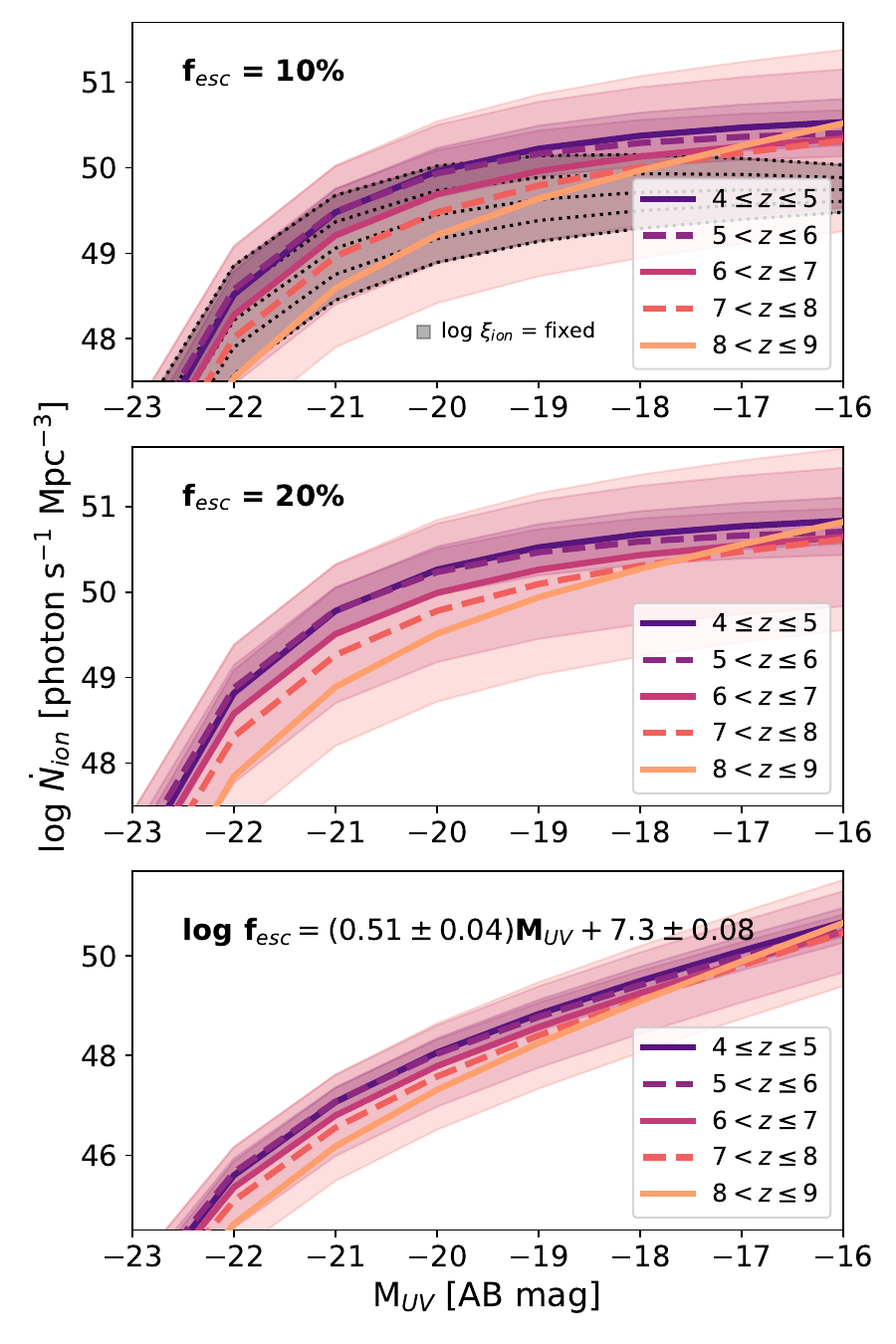}
   \caption{\Ndot\ as a function of M$_{\rm{UV}}$, by redshift bins, assuming a \fesc\ indicated in each panel and a \xion\ described by our data. The prescription of \fesc\ that varies with M$_{\rm{UV}}$ was taken from \citet{Anderson2017}, while the UV luminosity functions were adopted from \citet{Bouwens2021}, for the redshifts relevant to this work. For comparison, in the top panel, results adopting a constant log \xion\ = 25.2 Hz erg$^{-1}$ are shown in black. At every redshift and for every \fesc\/, galaxies fainter than M$_{\rm{UV}} \sim -19$ dominate the ionisation budget, indicating that faint (but not necessarily extremely faint) galaxies contribute significantly to reionisation.}
              \label{fig:cosmic_nion_UV}%
    \end{figure}

   \begin{figure}
        \centering
   \includegraphics[width=0.5\textwidth]{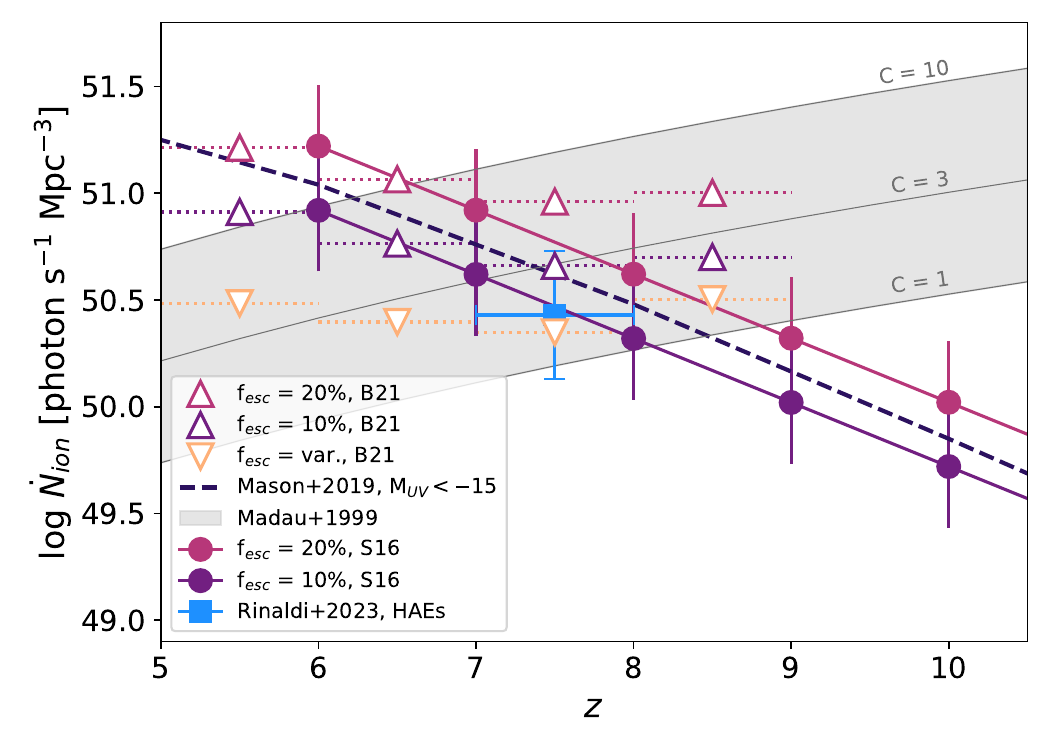}
   \caption{Cosmic rate of ionising photons being emitted per second and per unit of volume, \Ndot\/, as a function of redshift, assuming a \fesc\ as indicated, and a \xion\ described by our data (i.e. log \xion\/ $= (0.07\pm 0.02)z + 25.05\pm 0.11$). We show the results obtained when adopting the UV luminosity density from \citet{Sun2016} (circles labelled "S16"),  as well as results obtained by integrating the UV luminosity density curves from \citet{Bouwens2021} down to M$_{\rm{UV}}$ = -16 (triangles labelled "B21", curves shown in Figure~\ref{fig:cosmic_nion_UV}). As comparison, we include the curve from \citet{Mason2015} assuming constant \xion\ and \fesc\/, integrated down to a M$_{\rm{UV}}$ of -15 \citep[as in ][]{Mason2019}, as well as the \Ndot\ reported in \citet{Rinaldi2023arXiv} for \ha\ emitters at $z \sim 7-8$ (blue square). Finally, we include the estimated \Ndot\ needed to maintain Hydrogen ionisation in the IGM \citep{Madau1999}, adopting clumping factors of 1, 3 and 10. We find a cosmic \Ndot\ consistent with literature up to $z \sim 8$, but that starts to rise at the highest redshift bin due to the dependence of \xion\ with  M$_{\rm{UV}}$.}
              \label{fig:cosmic_nion}%
    \end{figure}

\section{Discussion}
\label{DISCUSSION} 
We begin by first addressing the biases that could potentially affect the results from this work. By construction, only galaxies with emission lines that can be measured from photometry were selected. Therefore, we are mostly focusing on star-forming galaxies. This was a necessary step in order to estimate \xion\ from either \ha\ or \oiii\/. As an experiment, we used \texttt{Prospector} to fit a small subsample of galaxies with no obvious emission lines in the filter pairs of interest (F335M-F356W and F410M-F444W). As previously stated, \texttt{Prospector} does not rely on emission lines for the measurement of \xion\/. We find that in these cases \xion\ is consistently below $10^{25}$ Hz erg$^{-1}$, with values as low as $10^{23}$ Hz erg$^{-1}$, suggesting that there is a population of galaxies for which \xion\ falls below the relation shown in Figure~\ref{fig:xion_redshift}, possibly explaining the origin of the trend of \xion\ with M$_{\rm{UV}}$. This work is not representative of those galaxies, rather, it sheds light on the galaxies and mechanisms most likely responsible for reionising the Universe. It must be noted that recent work by \cite{Looser2023a} and \cite{Looser2023b}, among others, show that galaxies with no emission lines might be only temporarily quiescent, as a result of extremely bursty SFHs \citep{Tibor2023}. Therefore, these kinds of galaxies are interesting to study \citep{Katz2023}, and are potentially important in the context of the EoR. The contribution of each galaxy population to the cosmic reionisation budget is beyond the scope of this work, and will be presented in a future work, where the full capacity of JADES photometry will be combined with the power of \texttt{Prospector} to quantify the relative importance of different populations.

Using our sample, we now investigate the nature of the positive slope of \xion\ with redshift, aiming to answer two questions: (1) Is it physical or is it a result of our selection? and (2) If it is real, what is driving it? 

\subsection{Does \xion\ evolve with redshift?}

In order to answer this question we conduct a simple null-hypothesis test, shown in Figure~\ref{fig:bias} (for simplicity, errors have been ignored in this test). We first select a subsample of galaxies at $z \sim 5.5 - 6$ from the galaxies used in this work, for which \xion\ has been inferred through the \oiii\ EW (framed with a blue rectangle in the top panel). We assume that there is no evolution of \xion\ with redshift, and that any observational study that says the contrary suffers from a luminosity bias (i.e. that at higher redshift we can only see the fainter galaxies with stronger emission lines). Under that assumption, we use our selected galaxies as seeds to produce 1000 simulated galaxies located randomly between $z \sim 5.5 - 9$, and that have been dimmed according to their luminosity distance (white circles with grey edges in middle panel). For these galaxies the rest-frame \oiii\ EWs estimated originally are used to obtain \xion\/ (see equation 3), so \xion\ does not change with redshift for a specific seed. Finally, we apply the same selection criteria we did when constructing our sample, i.e. a difference of 10 nJy between filter pairs (F335M-F356W or F410M-F444W). The galaxies deemed observable and that would be selected in our sample are shown as blue plus signs. The \xion\/ slope derived through emission line fluxes is shown in all panels as a dashed blue line, whereas the slope obtained after this test is shown in red. It is clear that these slopes do not match and that the red slope is flat. From this exercise we can conclude that the increase of \xion\ with redshift is not mainly due to our selection criteria. Furthermore, we investigate the possibility of a stellar mass bias driving the observed \xion\ evolution with redshift. In the bottom panel we conduct a similar experiment as the one just described, but now using as seeds the galaxies in our sample with low stellar masses (log M/M$_{\odot}$ $< 8.0$; shown as purple crosses in the top panel), it is important to note that in our sample, this is equivalent to studying galaxies fainter than M$_{\rm{UV}} \simeq -19$. We find that lower mass (fainter) galaxies have higher \xion\/, but that this property alone is insufficient in explaining the increase of \xion\ with redshift. Therefore, we go forward under the assumption that, even if there is a degree of observational bias, there is a physical cause driving the observed \xion\ evolution with redshift. 

\subsection{What drives the \xion\ evolution?}

We now aim to investigate the main driver of the observed increase of \xion\ with redshift. Throughout this paper we have demonstrated that our simple prescription to measure line fluxes from photometry is adequate, agreeing with both NIRSpec and FRESCO grism flux measurements (when available). We have also shown that our \xion\ values agree with those found in the literature, and finally, with those inferred with \texttt{Prospector}. We now focus particularly on this last point, and exploit the synergy between observations and SED fitting to find which galactic property (or properties) is  (are) driving the \xion\ evolution. For this purpose we calculate a Spearman's rank correlation coefficient for \xion\ against the following properties: redshift, stellar mass, UV magnitude (both observed and intrinsic), recent SFR (SFR$_{10}$; in the past 10 Myr), SFR in the past 100 Myr (SFR$_{100}$), stellar metallicity, ionisation parameter, dust index \citep[dust2 in the prescription of ][]{Conroy2009}, half-mass assembly time (t50), and ionising photons emitted per second (\ndot\/). All of the results are shown in Appendix~\ref{Appendix:Prospector}, with their correlation and p-value shown in the title of each panel. An excerpt of the table containing all the values can be found in Table~\ref{table:excerpt_prospector}.
We find that in our sample \xion\ correlates with M$_{\rm{UV}}$, half-mass stellar age and metallicity, however, the strongest correlations are those of \xion\ with stellar mass, where lower masses lead to higher \xion\/ values, and with SFR. Motivated by these findings, we explore the correlation of \xion\ with the SFH burstiness, which translates into the ratio between both recent and older SFR (SFR$_{10}$/SFR$_{100}$). Figure~\ref{fig:burstiness} shows how burstiness becomes increasingly important at lower stellar masses, and that low-mass galaxies with bursty star formation have the highest \xion\ values in the sample. Also shown is the correlation between stellar mass and M$_{\rm{UV}}$, and recent star formation versus stellar mass (colour-coded by \xion\ and \ndot\/, respectively). The Spearman's correlation coefficient value for SFR$_{10}$/SFR$_{100}$ is 0.914, with a p-value consistent with zero, indicating a strong positive correlation between \xion\ and burstiness of the SFH. Therefore, from \texttt{Prospector} we conclude that low mass and burstiness in a galaxy are the most important properties driving \xion\/.

Burstiness in star formation is usually associated with low stellar masses \citep{Weisz2012,Guo2016}, mainly due to stellar feedback. In brief, supernovae occurring after intense star formation heat up and expel gas. This leads to star formation being temporarily quenched \citep{Stinson2007,Tibor2023}, followed by new gas accretion, which results in new star forming episodes. Burstiness in high redshift galaxies can also be explained by their dynamical timescale, which becomes too short for supernovae feedback to respond to gravitational collapse \citep{Faucher-Giguere2018,Tacchella2020}. At high redshift, galaxies with low stellar masses are expected to be more numerous \citep{Bouwens2015,Austin2023,Bouwens2023,Harikane2023}. Additionally, these types of galaxies are thought to be the main sources responsible for reionising the Universe \citep{Hassan2018,Rosdahl2018,Trebitsch2020}. In a recent work, \cite{Atek2023} present spectroscopic observations of extremely low mass lensed galaxies (log M/M$_{\odot}$ $\sim 6 - 7$) with high \xion\/ (log \xion\//Hz erg$^{-1}$ $\sim$ 25.8, measured through the \ha\ recombination line). These kinds of galaxies are likely key in the reionisation of the Universe. Our results support the scenario of low-mass galaxies being efficient producers of ionising radiation, in agreement with previous findings.

\subsection{The impact of \xion\ on the cosmic ionisation budget}

We first study how the number of ionising photons produced per volume unit, \Ndot\/, varies with M$_{\rm{UV}}$ and redshift. We adopt the UV luminosity functions from \cite{Bouwens2021}, and two different prescriptions for \fesc\/: constant \citep[of 10 and 20\%; ][]{Ouchi2009,Robertson2013,Robertson2015}, and varying with M$_{\rm{UV}}$. For the variable prescription we follow the work of \cite{Anderson2017}, who estimate \fesc\ over a large range of galaxy masses, using the high-resolution, uniform volume simulation \texttt{Vulcan}. This simulation provides detailed distributions of gas and stars in resolved galaxies, allowing precise measurements of \fesc\/. \cite{Anderson2017} find a dependence of \fesc\ with M$_{\rm{UV}}$ given by: $\log \text{f}_{\rm{esc}} = (0.51\pm 0.4)\text{M}_{\rm{UV}} + 7.3\pm 0.08$.
For \xion\/, we assume the best fit lines to our observations given in Figure~\ref{fig:xion_MUV}. It is important to mention that by following this prescription we are assuming that the \xion\ evolution is representative for all low-mass faint galaxies, when in fact, it does not represent galaxies in quiescent phases (without detectable emission lines). Therefore, the cosmic ionising photon budgets here derived should be taken as upper limits. In a future work we will quantify the contribution of different galaxy populations to reionisation, and these calculations will be further constrained. 

The \Ndot\ results as a function of M$_{\rm{UV}}$ are presented in Figure~\ref{fig:cosmic_nion_UV}, where each panel shows a different escape fraction. At every redshift bin, the fainter galaxies dominate the budget of cosmic reionisation. In particular, that galaxies fainter than M$_{\rm{UV}} \sim -19$ account for at least 90\% of the total \Ndot\/. This is especially true for the case with the variable \fesc\/ from \citep{Anderson2017}, where \Ndot\ has a steeper dependence with M$_{\rm{UV}}$, and galaxies fainter than M$_{\rm{UV}}$ = -18 account for more than 90\% of the total ionising budget at all redshift bins. It is important to mention that the curves start to flatten at M$_{\rm{UV}} \sim -20$ for the constant \fesc\/ cases, which means that faint (but not necessarily extremely faint) galaxies are significant contributors to reionisation. If we use the same luminosity functions but instead adopt a constant \xion\ value of log \xion\/ = 25.2 Hz erg$^{-1}$ \citep[motivated by stellar populations, as in ][]{Robertson2013}, and a constant \fesc\ of 10\% (seen as black shaded area in top panel), we find that at the highest redshift bin investigated ($z = 8 - 9$) our results are significantly higher, and can translate to a reduction in the average \fesc\ from 10 to $\sim 2$\%. This is a natural result of a \xion\ dependant on both galaxy mass and redshift \citep{Finkelstein2019}.\\

Using the curves derived in the previous step, we now investigate the effect of our \xion\ estimation in the evolution of the cosmic ionisation budget, \Ndot\ with redshift, given by:
\begin{equation}
    \dot{N}_{\rm{ion}}(z) = \text{f}_{\rm{esc}} \times \xi_{\rm{ion}}(z) \times \rho_{\rm{UV}}(z)
\end{equation}
where \Ndot\ is in units of photon s$^{-1}$ Mpc$^{-3}$, \xion\ is in units of Hz erg$^{-1}$, and $\rho_{\rm{UV}}$ in units of erg s$^{-1}$ Mpc$^{-3}$. The escape fraction is dimensionless and can be assumed constant. In addition to the \citep{Bouwens2021} luminosity functions, we now include the red solid curve provided in Figure 7 of \cite{Sun2016}, which fits a power law to the low-mass end. To estimate \xion\ we use the equation that describes the best fit to our data (see Figure~\ref{fig:xion_redshift}), and assume \fesc\ values of 10 and 20\%, in accordance to the canonical average \fesc\ values needed for galaxies to be capable of ionising the Universe \citep{Ouchi2009,Robertson2013,Robertson2015}. In addition, we integrate the curves from Figure~\ref{fig:cosmic_nion_UV} down to M$_{\rm{UV}} = -16$, and show the results as open triangles in Figure~\ref{fig:cosmic_nion}. The values adopting the variable \xion\ from this work (triangles) are consistent with those from literature up to $z \sim 8$ \citep[e.g. ][]{Bouwens2015,Mason2015,Mason2019,Naidu2020,Rinaldi2023arXiv}. However, there is an upturn in the last redshift bin, where faint low-mass galaxies dominate and the \xion\ dependence with M$_{\rm{UV}}$ and redshift becomes more important. As comparison, we add the estimated \Ndot\ that is required to maintain the ionisation of Hydrogen according to the models of \citep{Madau1999}, adopting clumping factors of 1, 3 and 10. A clumping factor of unity represents a uniform IGM, whereas larger clumping factors imply that an increased number of recombinations are taking place in the IGM. This leads to the need for a higher number of ionising photons to be produced, in order to reach a balance between ionisation and recombination rates. If the \xion\ derived in this work is representative of the faint low-mass galaxy population, then these kind of galaxies would produce an ionising photon budget sufficient to ionise the Universe by the end of the EoR. 

\subsection{Implications for reionisation}
The connection between the cosmic \Ndot\ estimations and our previous conclusions comes through the stellar mass of galaxies. Stellar mass has been seen to decrease as galaxies become fainter, for example, \cite{Bhatawdekar2019} analyse this relation at $z = 6 - 9$ using data from the Hubble Frontier Fields. They notice that despite seeing a few high-mass galaxies with faint UV luminosities, there is a clear trend (with a large scatter) of stellar mass decreasing as galaxies become fainter in M$_{\rm{UV}}$ \citep[see also; ][]{Song2016}. In particular, galaxies fainter than M$_{\rm{UV}} \sim -18$ have stellar masses below $\sim 10^8$ M$_{\odot}$. In our sample, galaxies with comparable mass have the highest \xion\/, which is illustrated in Figure~\ref{fig:burstiness}, where we also show the correlation between stellar mass and M$_{\rm{UV}}$. Therefore, the conclusions made from estimating the cosmic ionising photon budget agree with the ones drawn from combining our emission line estimations with \texttt{Prospector}. In particular, that low-mass galaxies in the fainter end of luminosity functions are more efficient in producing ionising radiation, and might be the main drivers of reionisation. As mentioned previously, this conclusion depends on how representative our sample is of the general galaxy population, and how common low-mass bursty galaxies in a quiescent phase are, both topics to be presented in a future study. Promisingly, \cite{Rinaldi2023arXiv} find that \ha\ emitters (HAEs) contribute significantly more to \Ndot\ than their non-\ha\ emitting counterparts, for a sample of galaxies at $z \sim 7 - 8$.

In brief, based on our sample of galaxies with detectable \ha\ and/or \oiii\ emission lines, we conclude that the increase of \xion\ with redshift in this population is likely physical in origin. The main driver of the observed evolution is the stellar mass of galaxies, which leads to bursty SFHs and result in higher \xion\/ (and possibly higher \fesc\/).  Additionally, we convolve our \xion\ estimations with UV luminosity functions from literature, and find that if our findings are representative of the faint low-mass galaxy population, then these galaxies can produce enough ionising photons to ionise the Universe by the end of the EoR. In particular, we find that the \xion\ relations found in this work can reduce the requirement of average escape fractions, if assumed constant, to $< 10$\%.  The effect is more significant at higher redshifts where faint low-mass galaxies dominate luminosity functions.

\section{Conclusions}
\label{CONLCUSIONS}

In summary, we use NIRCam Deep imaging to build a sample of 677 galaxies at $z = 3.9-8.9$, for which \ha\ and/or \oiii\/$_{\lambda 5007}$ fluxes can be estimated from photometry. By construction, this sample does not include galaxies in quiescent phases. Depending on the redshift, we estimate \xion\ through \ha\ and/or EW(\oiii\/$_{\lambda 5007}$), measured from photometry in the filter pairs: F335M-F356W and F410M-F444W. We adopt an SMC dust attenuation curve, proven to be adequate at high redshifts. Simultaneously, we fit all the photometry with \texttt{Prospector} and derive \xion\/, in addition to relevant galaxy properties. The \xion\ measurements inferred through emission line fluxes agree with the values derived by \texttt{Prospector}. We find that \xion\ evolves with both redshift and M$_{\rm{UV}}$, and this evolution is not only due to observational biases. To place our results on a cosmic scale, we combine our relations of \xion\ with redshift and M$_{\rm{UV}}$, along with two different \fesc\ treatments: constant (10 and 20\%), and variable as a function of M$_{\rm{UV}}$, to constrain the cosmic budget of reionisation, \Ndot\/, and make conclusions about which kind of galaxies dominate this budget.
The main conclusions of this work are the following:

\begin{itemize}
    \item By comparing the resulting \xion\ using \oiii\ EWs with those inferred by \texttt{Prospector}, we confirm the effectiveness of EW(\oiii\/) to estimate \xion\ in the high redshift Universe
    \item For our sample, \xion\ evolves positively with redshift as: $\log \xi_{\rm{ion}}  = (0.07\pm 0.02)z + (25.05\pm 0.11)$ 
    \item We perform a 2-dimensional fit to account for the evolution of \xion\ with both redshift and M$_{\rm{UV}}$, and find: $\log \xi_{\rm{ion}} (z,\text{M}_{\rm{UV}}) = (0.05 \pm 0.02)z + (0.11 \pm 0.02) \text{M}_{\rm{UV}} + (27.33 \pm 0.37)$ 
    \item The observed evolution of \xion\ is likely has a physical origin, and is driven by specific star formation rate of galaxies. Specifically, lower mass leads to burstier SFHs, which we find is the property that has the strongest correlation with \xion\
    \item By comparing \Ndot\ obtained by adopting a constant \fesc\ of 10\% and a constant ionising photon production efficiency of log \xion\//[Hz erg$^{-1}$] = 25.2, with our evolving \xion\ prescriptions, we conclude that the average \fesc\ requirement can be reduced to $< 10$\%, an effect that increases with redshift (as low as $\sim 2$\% for our highest redshift bin)
    \item If our sample is representative of faint-low mass galaxies, then these kind of galaxies can account for the budget of ionising photons required to ionise the Universe by the end of the EoR
\end{itemize}

In this study, we conclude that low-mass faint galaxies with bursty SFHs are efficient enough in producing ionising photons to be the main sources responsible for ionising the Universe. We note that the sample used in this work was constructed to have detectable emission lines, particularly, \ha\ and/or \oiii\/$_{\lambda 5007}$, and is therefore not representative of every galaxy population. However, the population here studied is likely representative of the galaxies responsible for ionising the Universe. In a future study, we will use the full potential of JADES photometry to shed light on the contribution different galaxy populations have to the total cosmic ionising budget.

\section*{Data Availability}
The data underlying this article will be shared on reasonable request to the corresponding author. 


\section*{Acknowledgements}
The JADES Collaboration thanks the Instrument Development Teams and the instrument teams at the European Space Agency and the Space Telescope Science
Institute for the support that made this program possible. We also thank our program coordinators at STScI
for their help in planning complicated parallel observations.

\noindent CS thanks James Leftley for insightful discussions and IT support. RM, CS, WB, WC, JS and JW acknowledge support by the Science and Technology Facilities Council (STFC) and by the ERC through Advanced Grant number 695671 ‘QUENCH’, and by the UKRI Frontier Research grant RISEandFALL. RM also acknowledges funding from a research professorship from the Royal Society.
ECL acknowledges support of an STFC Webb Fellowship (ST/W001438/1). 
AJB and AS acknowledge funding from the "FirstGalaxies" Advanced Grant from the
European Research Council (ERC) under the European Union’s Horizon 2020 research and innovation
programme (Grant agreement No. 789056).
DJE is supported as a Simons Investigator and by JWST/NIRCam contract to the University of Arizona,
NAS5-02015. BDJ, BER, EE and FS acknowledge support by the JWST/NIRCam contract to the University of Arizona NAS5-02015. WM thanks the Science and Technology Facilities Council (STFC) Center for Doctoral Training (CDT) in Data intensive Science at the University of Cambridge (STFC grant number 2742968) for a PhD studentship.
CW thanks the Science and Technology Facilities Council (STFC) for a PhD studentship, funded by UKRI grant 2602262.
The research of CCW is supported by NOIRLab, which is managed by the Association of Universities for
Research in Astronomy (AURA) under a cooperative agreement with the National Science Foundation.
This research is supported in part by the Australian Research Council Centre of Excellence for All Sky
Astrophysics in 3 Dimensions (ASTRO 3D), through project number CE170100013.
Funding for this research was provided by the Johns Hopkins University, Institute for Data Intensive
Engineering and Science (IDIES).



\bibliographystyle{mnras}
\bibliography{bib} 

\begin{thebibliography}{}
\makeatletter
\relax
\def\mn@urlcharsother{\let\do\@makeother \do\$\do\&\do\#\do\^\do\_\do\%\do\~}
\def\mn@doi{\begingroup\mn@urlcharsother \@ifnextchar [ {\mn@doi@}
  {\mn@doi@[]}}
\def\mn@doi@[#1]#2{\def\@tempa{#1}\ifx\@tempa\@empty \href
  {http://dx.doi.org/#2} {doi:#2}\else \href {http://dx.doi.org/#2} {#1}\fi
  \endgroup}
\def\mn@eprint#1#2{\mn@eprint@#1:#2::\@nil}
\def\mn@eprint@arXiv#1{\href {http://arxiv.org/abs/#1} {{\tt arXiv:#1}}}
\def\mn@eprint@dblp#1{\href {http://dblp.uni-trier.de/rec/bibtex/#1.xml}
  {dblp:#1}}
\def\mn@eprint@#1:#2:#3:#4\@nil{\def\@tempa {#1}\def\@tempb {#2}\def\@tempc
  {#3}\ifx \@tempc \@empty \let \@tempc \@tempb \let \@tempb \@tempa \fi \ifx
  \@tempb \@empty \def\@tempb {arXiv}\fi \@ifundefined
  {mn@eprint@\@tempb}{\@tempb:\@tempc}{\expandafter \expandafter \csname
  mn@eprint@\@tempb\endcsname \expandafter{\@tempc}}}

\bibitem[\protect\citeauthoryear{{Anderson}, {Governato}, {Karcher}, {Quinn}
  \& {Wadsley}}{{Anderson} et~al.}{2017}]{Anderson2017}
{Anderson} L.,  {Governato} F.,  {Karcher} M.,  {Quinn} T.,   {Wadsley} J.,
  2017, \mn@doi [\mnras] {10.1093/mnras/stx709}, \href
  {https://ui.adsabs.harvard.edu/abs/2017MNRAS.468.4077A} {468, 4077}

\bibitem[\protect\citeauthoryear{{Atek}, {Furtak}, {Oesch}, {van Dokkum},
  {Reddy}, {Contini}, {Illingworth}  \& {Wilkins}}{{Atek}
  et~al.}{2022}]{Atek2022}
{Atek} H.,  {Furtak} L.~J.,  {Oesch} P.,  {van Dokkum} P.,  {Reddy} N.,
  {Contini} T.,  {Illingworth} G.,   {Wilkins} S.,  2022, \mn@doi [\mnras]
  {10.1093/mnras/stac360}, \href
  {https://ui.adsabs.harvard.edu/abs/2022MNRAS.511.4464A} {511, 4464}

\bibitem[\protect\citeauthoryear{{Atek} et~al.,}{{Atek}
  et~al.}{2023}]{Atek2023}
{Atek} H.,  et~al., 2023, \mn@doi [arXiv e-prints] {10.48550/arXiv.2308.08540},
  \href {https://ui.adsabs.harvard.edu/abs/2023arXiv230808540A} {p.
  arXiv:2308.08540}

\bibitem[\protect\citeauthoryear{{Austin} et~al.,}{{Austin}
  et~al.}{2023}]{Austin2023}
{Austin} D.,  et~al., 2023, \mn@doi [\apjl] {10.3847/2041-8213/ace18d}, \href
  {https://ui.adsabs.harvard.edu/abs/2023ApJ...952L...7A} {952, L7}

\bibitem[\protect\citeauthoryear{{Becker} et~al.,}{{Becker}
  et~al.}{2001}]{Becker2001}
{Becker} R.~H.,  et~al., 2001, \mn@doi [\aj] {10.1086/324231}, \href
  {https://ui.adsabs.harvard.edu/abs/2001AJ....122.2850B} {122, 2850}

\bibitem[\protect\citeauthoryear{{Beckwith} et~al.,}{{Beckwith}
  et~al.}{2006}]{Beckwith2006}
{Beckwith} S. V.~W.,  et~al., 2006, \mn@doi [\aj] {10.1086/507302}, \href
  {https://ui.adsabs.harvard.edu/abs/2006AJ....132.1729B} {132, 1729}

\bibitem[\protect\citeauthoryear{{Bhatawdekar}, {Conselice},
  {Margalef-Bentabol}  \& {Duncan}}{{Bhatawdekar}
  et~al.}{2019}]{Bhatawdekar2019}
{Bhatawdekar} R.,  {Conselice} C.~J.,  {Margalef-Bentabol} B.,   {Duncan} K.,
  2019, \mn@doi [\mnras] {10.1093/mnras/stz866}, \href
  {https://ui.adsabs.harvard.edu/abs/2019MNRAS.486.3805B} {486, 3805}

\bibitem[\protect\citeauthoryear{{Bian}, {Fan}, {McGreer}, {Cai}  \&
  {Jiang}}{{Bian} et~al.}{2017}]{Bian2017}
{Bian} F.,  {Fan} X.,  {McGreer} I.,  {Cai} Z.,   {Jiang} L.,  2017, \mn@doi
  [\apjl] {10.3847/2041-8213/aa5ff7}, \href
  {https://ui.adsabs.harvard.edu/abs/2017ApJ...837L..12B} {837, L12}

\bibitem[\protect\citeauthoryear{{Borthakur}, {Heckman}, {Leitherer}  \&
  {Overzier}}{{Borthakur} et~al.}{2014}]{Borthakur2014}
{Borthakur} S.,  {Heckman} T.~M.,  {Leitherer} C.,   {Overzier} R.~A.,  2014,
  \mn@doi [Science] {10.1126/science.1254214}, \href
  {https://ui.adsabs.harvard.edu/abs/2014Sci...346..216B} {346, 216}

\bibitem[\protect\citeauthoryear{{Bosman} et~al.,}{{Bosman}
  et~al.}{2022}]{Bosman2022}
{Bosman} S. E.~I.,  et~al., 2022, \mn@doi [\mnras] {10.1093/mnras/stac1046},
  \href {https://ui.adsabs.harvard.edu/abs/2022MNRAS.514...55B} {514, 55}

\bibitem[\protect\citeauthoryear{{Bouwens} et~al.,}{{Bouwens}
  et~al.}{2015}]{Bouwens2015}
{Bouwens} R.~J.,  et~al., 2015, \mn@doi [\apj] {10.1088/0004-637X/803/1/34},
  \href {https://ui.adsabs.harvard.edu/abs/2015ApJ...803...34B} {803, 34}

\bibitem[\protect\citeauthoryear{{Bouwens}, {Smit}, {Labb{\'e}}, {Franx},
  {Caruana}, {Oesch}, {Stefanon}  \& {Rasappu}}{{Bouwens}
  et~al.}{2016}]{Bouwens2016}
{Bouwens} R.~J.,  {Smit} R.,  {Labb{\'e}} I.,  {Franx} M.,  {Caruana} J.,
  {Oesch} P.,  {Stefanon} M.,   {Rasappu} N.,  2016, \mn@doi [\apj]
  {10.3847/0004-637X/831/2/176}, \href
  {https://ui.adsabs.harvard.edu/abs/2016ApJ...831..176B} {831, 176}

\bibitem[\protect\citeauthoryear{{Bouwens} et~al.,}{{Bouwens}
  et~al.}{2021}]{Bouwens2021}
{Bouwens} R.~J.,  et~al., 2021, \mn@doi [\aj] {10.3847/1538-3881/abf83e}, \href
  {https://ui.adsabs.harvard.edu/abs/2021AJ....162...47B} {162, 47}

\bibitem[\protect\citeauthoryear{{Bouwens} et~al.,}{{Bouwens}
  et~al.}{2023}]{Bouwens2023}
{Bouwens} R.~J.,  et~al., 2023, \mn@doi [\mnras] {10.1093/mnras/stad1145},
  \href {https://ui.adsabs.harvard.edu/abs/2023MNRAS.523.1036B} {523, 1036}

\bibitem[\protect\citeauthoryear{{Bowler}, {Bourne}, {Dunlop}, {McLure}  \&
  {McLeod}}{{Bowler} et~al.}{2018}]{Bowler2018}
{Bowler} R.~A.~A.,  {Bourne} N.,  {Dunlop} J.~S.,  {McLure} R.~J.,   {McLeod}
  D.~J.,  2018, \mn@doi [\mnras] {10.1093/mnras/sty2368}, \href
  {https://ui.adsabs.harvard.edu/abs/2018MNRAS.481.1631B} {481, 1631}

\bibitem[\protect\citeauthoryear{{Bowler}, {Cullen}, {McLure}, {Dunlop}  \&
  {Avison}}{{Bowler} et~al.}{2022}]{Bowler2022}
{Bowler} R.~A.~A.,  {Cullen} F.,  {McLure} R.~J.,  {Dunlop} J.~S.,   {Avison}
  A.,  2022, \mn@doi [\mnras] {10.1093/mnras/stab3744}, \href
  {https://ui.adsabs.harvard.edu/abs/2022MNRAS.510.5088B} {510, 5088}

\bibitem[\protect\citeauthoryear{{Boyett}, {Stark}, {Bunker}, {Tang}  \&
  {Maseda}}{{Boyett} et~al.}{2022}]{Boyett2022}
{Boyett} K. N.~K.,  {Stark} D.~P.,  {Bunker} A.~J.,  {Tang} M.,   {Maseda}
  M.~V.,  2022, \mn@doi [\mnras] {10.1093/mnras/stac1109}, \href
  {https://ui.adsabs.harvard.edu/abs/2022MNRAS.513.4451B} {513, 4451}

\bibitem[\protect\citeauthoryear{Bradley et~al.,}{Bradley
  et~al.}{2022}]{bradley2022a}
Bradley L.,  et~al., 2022, astropy/photutils: 1.5.0,
  \mn@doi{10.5281/zenodo.6825092}, \url
  {https://doi.org/10.5281/zenodo.6825092}

\bibitem[\protect\citeauthoryear{{Brammer}, {van Dokkum}  \& {Coppi}}{{Brammer}
  et~al.}{2008}]{Brammer2008}
{Brammer} G.~B.,  {van Dokkum} P.~G.,   {Coppi} P.,  2008, \mn@doi [\apj]
  {10.1086/591786}, \href
  {https://ui.adsabs.harvard.edu/abs/2008ApJ...686.1503B} {686, 1503}

\bibitem[\protect\citeauthoryear{{Bunker}, {Warren}, {Hewett}  \&
  {Clements}}{{Bunker} et~al.}{1995}]{Bunker1995}
{Bunker} A.~J.,  {Warren} S.~J.,  {Hewett} P.~C.,   {Clements} D.~L.,  1995,
  \mn@doi [\mnras] {10.1093/mnras/273.2.513}, \href
  {https://ui.adsabs.harvard.edu/abs/1995MNRAS.273..513B} {273, 513}

\bibitem[\protect\citeauthoryear{{Byler}, {Dalcanton}, {Conroy}  \&
  {Johnson}}{{Byler} et~al.}{2017}]{Byler2017}
{Byler} N.,  {Dalcanton} J.~J.,  {Conroy} C.,   {Johnson} B.~D.,  2017, \mn@doi
  [\apj] {10.3847/1538-4357/aa6c66}, \href
  {https://ui.adsabs.harvard.edu/abs/2017ApJ...840...44B} {840, 44}

\bibitem[\protect\citeauthoryear{{Calzetti}, {Kinney}  \&
  {Storchi-Bergmann}}{{Calzetti} et~al.}{1994}]{Calzetti1994}
{Calzetti} D.,  {Kinney} A.~L.,   {Storchi-Bergmann} T.,  1994, \mn@doi [\apj]
  {10.1086/174346}, \href
  {https://ui.adsabs.harvard.edu/abs/1994ApJ...429..582C} {429, 582}

\bibitem[\protect\citeauthoryear{{Cameron} et~al.,}{{Cameron}
  et~al.}{2023}]{Cameron2023}
{Cameron} A.~J.,  et~al., 2023, \mn@doi [arXiv e-prints]
  {10.48550/arXiv.2302.04298}, \href
  {https://ui.adsabs.harvard.edu/abs/2023arXiv230204298C} {p. arXiv:2302.04298}

\bibitem[\protect\citeauthoryear{{Carnall}, {McLure}, {Dunlop}  \&
  {Dav{\'e}}}{{Carnall} et~al.}{2018}]{Carnall2018}
{Carnall} A.~C.,  {McLure} R.~J.,  {Dunlop} J.~S.,   {Dav{\'e}} R.,  2018,
  \mn@doi [\mnras] {10.1093/mnras/sty2169}, \href
  {https://ui.adsabs.harvard.edu/abs/2018MNRAS.480.4379C} {480, 4379}

\bibitem[\protect\citeauthoryear{{Chabrier}}{{Chabrier}}{2003}]{Chabrier2003}
{Chabrier} G.,  2003, \mn@doi [\pasp] {10.1086/376392}, \href
  {https://ui.adsabs.harvard.edu/abs/2003PASP..115..763C} {115, 763}

\bibitem[\protect\citeauthoryear{{Charlot} \& {Longhetti}}{{Charlot} \&
  {Longhetti}}{2001}]{Charlot2001}
{Charlot} S.,  {Longhetti} M.,  2001, \mn@doi [\mnras]
  {10.1046/j.1365-8711.2001.04260.x}, \href
  {https://ui.adsabs.harvard.edu/abs/2001MNRAS.323..887C} {323, 887}

\bibitem[\protect\citeauthoryear{{Chevallard} \& {Charlot}}{{Chevallard} \&
  {Charlot}}{2016}]{Chevallard2016}
{Chevallard} J.,  {Charlot} S.,  2016, \mn@doi [\mnras]
  {10.1093/mnras/stw1756}, \href
  {https://ui.adsabs.harvard.edu/abs/2016MNRAS.462.1415C} {462, 1415}

\bibitem[\protect\citeauthoryear{{Chevallard} et~al.,}{{Chevallard}
  et~al.}{2018}]{Chevallard2018}
{Chevallard} J.,  et~al., 2018, \mn@doi [\mnras] {10.1093/mnras/sty1461}, \href
  {https://ui.adsabs.harvard.edu/abs/2018MNRAS.479.3264C} {479, 3264}

\bibitem[\protect\citeauthoryear{{Conroy}, {Gunn}  \& {White}}{{Conroy}
  et~al.}{2009}]{Conroy2009}
{Conroy} C.,  {Gunn} J.~E.,   {White} M.,  2009, \mn@doi [\apj]
  {10.1088/0004-637X/699/1/486}, \href
  {https://ui.adsabs.harvard.edu/abs/2009ApJ...699..486C} {699, 486}

\bibitem[\protect\citeauthoryear{{De Barros}, {Oesch}, {Labb{\'e}}, {Stefanon},
  {Gonz{\'a}lez}, {Smit}, {Bouwens}  \& {Illingworth}}{{De Barros}
  et~al.}{2019}]{DeBarros2019}
{De Barros} S.,  {Oesch} P.~A.,  {Labb{\'e}} I.,  {Stefanon} M.,
  {Gonz{\'a}lez} V.,  {Smit} R.,  {Bouwens} R.~J.,   {Illingworth} G.~D.,
  2019, \mn@doi [\mnras] {10.1093/mnras/stz940}, \href
  {https://ui.adsabs.harvard.edu/abs/2019MNRAS.489.2355D} {489, 2355}

\bibitem[\protect\citeauthoryear{{Dome}, {Tacchella}, {Fialkov}, {Dekel},
  {Ginzburg}, {Lapiner}  \& {Looser}}{{Dome} et~al.}{2023}]{Tibor2023}
{Dome} T.,  {Tacchella} S.,  {Fialkov} A.,  {Dekel} A.,  {Ginzburg} O.,
  {Lapiner} S.,   {Looser} T.~J.,  2023, \mn@doi [arXiv e-prints]
  {10.48550/arXiv.2305.07066}, \href
  {https://ui.adsabs.harvard.edu/abs/2023arXiv230507066D} {p. arXiv:2305.07066}

\bibitem[\protect\citeauthoryear{{Duncan} \& {Conselice}}{{Duncan} \&
  {Conselice}}{2015}]{Duncan2015}
{Duncan} K.,  {Conselice} C.~J.,  2015, \mn@doi [\mnras]
  {10.1093/mnras/stv1049}, \href
  {https://ui.adsabs.harvard.edu/abs/2015MNRAS.451.2030D} {451, 2030}

\bibitem[\protect\citeauthoryear{{Eisenstein} et~al.,}{{Eisenstein}
  et~al.}{2023}]{Eisenstein2023}
{Eisenstein} D.~J.,  et~al., 2023, \mn@doi [arXiv e-prints]
  {10.48550/arXiv.2306.02465}, \href
  {https://ui.adsabs.harvard.edu/abs/2023arXiv230602465E} {p. arXiv:2306.02465}

\bibitem[\protect\citeauthoryear{{Eldridge}, {Stanway}, {Xiao}, {McClelland},
  {Taylor}, {Ng}, {Greis}  \& {Bray}}{{Eldridge} et~al.}{2017}]{Eldridge2017}
{Eldridge} J.~J.,  {Stanway} E.~R.,  {Xiao} L.,  {McClelland} L.~A.~S.,
  {Taylor} G.,  {Ng} M.,  {Greis} S.~M.~L.,   {Bray} J.~C.,  2017, \mn@doi
  [\pasa] {10.1017/pasa.2017.51}, \href
  {https://ui.adsabs.harvard.edu/abs/2017PASA...34...58E} {34, e058}

\bibitem[\protect\citeauthoryear{{Emami}, {Siana}, {Alavi}, {Gburek},
  {Freeman}, {Richard}, {Weisz}  \& {Stark}}{{Emami} et~al.}{2020}]{Emami2020}
{Emami} N.,  {Siana} B.,  {Alavi} A.,  {Gburek} T.,  {Freeman} W.~R.,
  {Richard} J.,  {Weisz} D.~R.,   {Stark} D.~P.,  2020, \mn@doi [\apj]
  {10.3847/1538-4357/ab8f97}, \href
  {https://ui.adsabs.harvard.edu/abs/2020ApJ...895..116E} {895, 116}

\bibitem[\protect\citeauthoryear{{Endsley}, {Stark}, {Chevallard}  \&
  {Charlot}}{{Endsley} et~al.}{2021}]{Endsley2021}
{Endsley} R.,  {Stark} D.~P.,  {Chevallard} J.,   {Charlot} S.,  2021, \mn@doi
  [\mnras] {10.1093/mnras/staa3370}, \href
  {https://ui.adsabs.harvard.edu/abs/2021MNRAS.500.5229E} {500, 5229}

\bibitem[\protect\citeauthoryear{{Faisst} et~al.,}{{Faisst}
  et~al.}{2016}]{Faisst2016}
{Faisst} A.~L.,  et~al., 2016, \mn@doi [\apj] {10.3847/0004-637X/821/2/122},
  \href {https://ui.adsabs.harvard.edu/abs/2016ApJ...821..122F} {821, 122}

\bibitem[\protect\citeauthoryear{{Faisst}, {Capak}, {Emami}, {Tacchella}  \&
  {Larson}}{{Faisst} et~al.}{2019}]{Faisst2019}
{Faisst} A.~L.,  {Capak} P.~L.,  {Emami} N.,  {Tacchella} S.,   {Larson} K.~L.,
   2019, \mn@doi [\apj] {10.3847/1538-4357/ab425b}, \href
  {https://ui.adsabs.harvard.edu/abs/2019ApJ...884..133F} {884, 133}

\bibitem[\protect\citeauthoryear{{Fan} et~al.,}{{Fan} et~al.}{2006}]{Fan2006}
{Fan} X.,  et~al., 2006, \mn@doi [\aj] {10.1086/500296}, \href
  {https://ui.adsabs.harvard.edu/abs/2006AJ....131.1203F} {131, 1203}

\bibitem[\protect\citeauthoryear{{Faucher-Gigu{\`e}re}}{{Faucher-Gigu{\`e}re}}{2018}]{Faucher-Giguere2018}
{Faucher-Gigu{\`e}re} C.-A.,  2018, \mn@doi [\mnras] {10.1093/mnras/stx2595},
  \href {https://ui.adsabs.harvard.edu/abs/2018MNRAS.473.3717F} {473, 3717}

\bibitem[\protect\citeauthoryear{{Ferland} et~al.,}{{Ferland}
  et~al.}{2013}]{Ferland2013}
{Ferland} G.~J.,  et~al., 2013, \mn@doi [\rmxaa] {10.48550/arXiv.1302.4485},
  \href {https://ui.adsabs.harvard.edu/abs/2013RMxAA..49..137F} {49, 137}

\bibitem[\protect\citeauthoryear{{Ferland} et~al.,}{{Ferland}
  et~al.}{2017}]{Ferland2017}
{Ferland} G.~J.,  et~al., 2017, \rmxaa, \href
  {https://ui.adsabs.harvard.edu/abs/2017RMxAA..53..385F} {53, 385}

\bibitem[\protect\citeauthoryear{{Finkelstein} et~al.,}{{Finkelstein}
  et~al.}{2019}]{Finkelstein2019}
{Finkelstein} S.~L.,  et~al., 2019, \mn@doi [\apj] {10.3847/1538-4357/ab1ea8},
  \href {https://ui.adsabs.harvard.edu/abs/2019ApJ...879...36F} {879, 36}

\bibitem[\protect\citeauthoryear{{Flury} et~al.,}{{Flury}
  et~al.}{2022}]{Flury2022a}
{Flury} S.~R.,  et~al., 2022, \mn@doi [\apjs] {10.3847/1538-4365/ac5331}, \href
  {https://ui.adsabs.harvard.edu/abs/2022ApJS..260....1F} {260, 1}

\bibitem[\protect\citeauthoryear{{Gallerani} et~al.,}{{Gallerani}
  et~al.}{2010}]{Gallerani2010}
{Gallerani} S.,  et~al., 2010, \mn@doi [\aap] {10.1051/0004-6361/201014721},
  \href {https://ui.adsabs.harvard.edu/abs/2010A&A...523A..85G} {523, A85}

\bibitem[\protect\citeauthoryear{{Gardner} et~al.,}{{Gardner}
  et~al.}{2023}]{Gardner2023}
{Gardner} J.~P.,  et~al., 2023, \mn@doi [\pasp] {10.1088/1538-3873/acd1b5},
  \href {https://ui.adsabs.harvard.edu/abs/2023PASP..135f8001G} {135, 068001}

\bibitem[\protect\citeauthoryear{{Giavalisco} et~al.,}{{Giavalisco}
  et~al.}{2004}]{Giavalisco2004}
{Giavalisco} M.,  et~al., 2004, \mn@doi [\apjl] {10.1086/379232}, \href
  {https://ui.adsabs.harvard.edu/abs/2004ApJ...600L..93G} {600, L93}

\bibitem[\protect\citeauthoryear{{Gordon} \& {Clayton}}{{Gordon} \&
  {Clayton}}{1998}]{Gordon1998}
{Gordon} K.~D.,  {Clayton} G.~C.,  1998, \mn@doi [\apj] {10.1086/305774}, \href
  {https://ui.adsabs.harvard.edu/abs/1998ApJ...500..816G} {500, 816}

\bibitem[\protect\citeauthoryear{{Gordon}, {Clayton}, {Misselt}, {Landolt}  \&
  {Wolff}}{{Gordon} et~al.}{2003}]{Gordon2003}
{Gordon} K.~D.,  {Clayton} G.~C.,  {Misselt} K.~A.,  {Landolt} A.~U.,   {Wolff}
  M.~J.,  2003, \mn@doi [\apj] {10.1086/376774}, \href
  {https://ui.adsabs.harvard.edu/abs/2003ApJ...594..279G} {594, 279}

\bibitem[\protect\citeauthoryear{{Guo} et~al.,}{{Guo} et~al.}{2016}]{Guo2016}
{Guo} Y.,  et~al., 2016, \mn@doi [\apj] {10.3847/1538-4357/833/1/37}, \href
  {https://ui.adsabs.harvard.edu/abs/2016ApJ...833...37G} {833, 37}

\bibitem[\protect\citeauthoryear{{Hainline} et~al.,}{{Hainline}
  et~al.}{2023}]{Hainline2023}
{Hainline} K.~N.,  et~al., 2023, \mn@doi [arXiv e-prints]
  {10.48550/arXiv.2306.02468}, \href
  {https://ui.adsabs.harvard.edu/abs/2023arXiv230602468H} {p. arXiv:2306.02468}

\bibitem[\protect\citeauthoryear{{Harikane} et~al.,}{{Harikane}
  et~al.}{2018}]{Harikane2018}
{Harikane} Y.,  et~al., 2018, \mn@doi [\apj] {10.3847/1538-4357/aabd80}, \href
  {https://ui.adsabs.harvard.edu/abs/2018ApJ...859...84H} {859, 84}

\bibitem[\protect\citeauthoryear{{Harikane} et~al.,}{{Harikane}
  et~al.}{2023}]{Harikane2023}
{Harikane} Y.,  et~al., 2023, \mn@doi [\apjs] {10.3847/1538-4365/acaaa9}, \href
  {https://ui.adsabs.harvard.edu/abs/2023ApJS..265....5H} {265, 5}

\bibitem[\protect\citeauthoryear{{Hassan}, {Dav{\'e}}, {Mitra}, {Finlator},
  {Ciardi}  \& {Santos}}{{Hassan} et~al.}{2018}]{Hassan2018}
{Hassan} S.,  {Dav{\'e}} R.,  {Mitra} S.,  {Finlator} K.,  {Ciardi} B.,
  {Santos} M.~G.,  2018, \mn@doi [\mnras] {10.1093/mnras/stx2194}, \href
  {https://ui.adsabs.harvard.edu/abs/2018MNRAS.473..227H} {473, 227}

\bibitem[\protect\citeauthoryear{{Izotov}, {Worseck}, {Schaerer}, {Guseva},
  {Chisholm}, {Thuan}, {Fricke}  \& {Verhamme}}{{Izotov}
  et~al.}{2021}]{Izotov2021}
{Izotov} Y.~I.,  {Worseck} G.,  {Schaerer} D.,  {Guseva} N.~G.,  {Chisholm} J.,
   {Thuan} T.~X.,  {Fricke} K.~J.,   {Verhamme} A.,  2021, \mn@doi [\mnras]
  {10.1093/mnras/stab612}, \href
  {https://ui.adsabs.harvard.edu/abs/2021MNRAS.503.1734I} {503, 1734}

\bibitem[\protect\citeauthoryear{{Johnson}, {Leja}, {Conroy}  \&
  {Speagle}}{{Johnson} et~al.}{2019}]{Johnson2019}
{Johnson} B.~D.,  {Leja} J.~L.,  {Conroy} C.,   {Speagle} J.~S.,  2019,
  {Prospector: Stellar population inference from spectra and SEDs},
  Astrophysics Source Code Library, record ascl:1905.025 (\mn@eprint {ascl}
  {1905.025})

\bibitem[\protect\citeauthoryear{{Johnson}, {Leja}, {Conroy}  \&
  {Speagle}}{{Johnson} et~al.}{2021}]{Johnson2021}
{Johnson} B.~D.,  {Leja} J.,  {Conroy} C.,   {Speagle} J.~S.,  2021, \mn@doi
  [\apjs] {10.3847/1538-4365/abef67}, \href
  {https://ui.adsabs.harvard.edu/abs/2021ApJS..254...22J} {254, 22}

\bibitem[\protect\citeauthoryear{{Katz} et~al.,}{{Katz}
  et~al.}{2023}]{Katz2023}
{Katz} H.,  et~al., 2023, \mn@doi [\mnras] {10.1093/mnras/stac3019}, \href
  {https://ui.adsabs.harvard.edu/abs/2023MNRAS.518..270K} {518, 270}

\bibitem[\protect\citeauthoryear{{Keating}, {Weinberger}, {Kulkarni},
  {Haehnelt}, {Chardin}  \& {Aubert}}{{Keating} et~al.}{2020}]{Keating2020}
{Keating} L.~C.,  {Weinberger} L.~H.,  {Kulkarni} G.,  {Haehnelt} M.~G.,
  {Chardin} J.,   {Aubert} D.,  2020, \mn@doi [\mnras] {10.1093/mnras/stz3083},
  \href {https://ui.adsabs.harvard.edu/abs/2020MNRAS.491.1736K} {491, 1736}

\bibitem[\protect\citeauthoryear{{Lam} et~al.,}{{Lam} et~al.}{2019}]{Lam2019}
{Lam} D.,  et~al., 2019, \mn@doi [\aap] {10.1051/0004-6361/201935227}, \href
  {https://ui.adsabs.harvard.edu/abs/2019A&A...627A.164L} {627, A164}

\bibitem[\protect\citeauthoryear{{Leitet}, {Bergvall}, {Hayes}, {Linn{\'e}}  \&
  {Zackrisson}}{{Leitet} et~al.}{2013}]{Leitet2013}
{Leitet} E.,  {Bergvall} N.,  {Hayes} M.,  {Linn{\'e}} S.,   {Zackrisson} E.,
  2013, \mn@doi [\aap] {10.1051/0004-6361/201118370}, \href
  {https://ui.adsabs.harvard.edu/abs/2013A&A...553A.106L} {553, A106}

\bibitem[\protect\citeauthoryear{{Leitherer}, {Hernandez}, {Lee}  \&
  {Oey}}{{Leitherer} et~al.}{2016}]{Leitherer2016}
{Leitherer} C.,  {Hernandez} S.,  {Lee} J.~C.,   {Oey} M.~S.,  2016, \mn@doi
  [\apj] {10.3847/0004-637X/823/1/64}, \href
  {https://ui.adsabs.harvard.edu/abs/2016ApJ...823...64L} {823, 64}

\bibitem[\protect\citeauthoryear{{Leja}, {Carnall}, {Johnson}, {Conroy}  \&
  {Speagle}}{{Leja} et~al.}{2019}]{Leja2019}
{Leja} J.,  {Carnall} A.~C.,  {Johnson} B.~D.,  {Conroy} C.,   {Speagle} J.~S.,
   2019, \mn@doi [\apj] {10.3847/1538-4357/ab133c}, \href
  {https://ui.adsabs.harvard.edu/abs/2019ApJ...876....3L} {876, 3}

\bibitem[\protect\citeauthoryear{{Looser} et~al.,}{{Looser}
  et~al.}{2023a}]{Looser2023a}
{Looser} T.~J.,  et~al., 2023a, \mn@doi [arXiv e-prints]
  {10.48550/arXiv.2302.14155}, \href
  {https://ui.adsabs.harvard.edu/abs/2023arXiv230214155L} {p. arXiv:2302.14155}

\bibitem[\protect\citeauthoryear{{Looser} et~al.,}{{Looser}
  et~al.}{2023b}]{Looser2023b}
{Looser} T.~J.,  et~al., 2023b, \mn@doi [arXiv e-prints]
  {10.48550/arXiv.2306.02470}, \href
  {https://ui.adsabs.harvard.edu/abs/2023arXiv230602470L} {p. arXiv:2306.02470}

\bibitem[\protect\citeauthoryear{{Lovell}, {Vijayan}, {Thomas}, {Wilkins},
  {Barnes}, {Irodotou}  \& {Roper}}{{Lovell} et~al.}{2021}]{Lovell2021}
{Lovell} C.~C.,  {Vijayan} A.~P.,  {Thomas} P.~A.,  {Wilkins} S.~M.,  {Barnes}
  D.~J.,  {Irodotou} D.,   {Roper} W.,  2021, \mn@doi [\mnras]
  {10.1093/mnras/staa3360}, \href
  {https://ui.adsabs.harvard.edu/abs/2021MNRAS.500.2127L} {500, 2127}

\bibitem[\protect\citeauthoryear{{Ma} et~al.,}{{Ma} et~al.}{2019}]{Ma2019}
{Ma} X.,  et~al., 2019, \mn@doi [\mnras] {10.1093/mnras/stz1324}, \href
  {https://ui.adsabs.harvard.edu/abs/2019MNRAS.487.1844M} {487, 1844}

\bibitem[\protect\citeauthoryear{{Madau}}{{Madau}}{1995}]{Madau1995}
{Madau} P.,  1995, \mn@doi [\apj] {10.1086/175332}, \href
  {https://ui.adsabs.harvard.edu/abs/1995ApJ...441...18M} {441, 18}

\bibitem[\protect\citeauthoryear{{Madau}, {Haardt}  \& {Rees}}{{Madau}
  et~al.}{1999}]{Madau1999}
{Madau} P.,  {Haardt} F.,   {Rees} M.~J.,  1999, \mn@doi [\apj]
  {10.1086/306975}, \href
  {https://ui.adsabs.harvard.edu/abs/1999ApJ...514..648M} {514, 648}

\bibitem[\protect\citeauthoryear{{Maiolino} \& {Mannucci}}{{Maiolino} \&
  {Mannucci}}{2019}]{Maiolino2019}
{Maiolino} R.,  {Mannucci} F.,  2019, \mn@doi [\aapr]
  {10.1007/s00159-018-0112-2}, \href
  {https://ui.adsabs.harvard.edu/abs/2019A&ARv..27....3M} {27, 3}

\bibitem[\protect\citeauthoryear{{Maiolino} et~al.,}{{Maiolino}
  et~al.}{2023}]{Maiolino2023}
{Maiolino} R.,  et~al., 2023, \mn@doi [arXiv e-prints]
  {10.48550/arXiv.2308.01230}, \href
  {https://ui.adsabs.harvard.edu/abs/2023arXiv230801230M} {p. arXiv:2308.01230}

\bibitem[\protect\citeauthoryear{{M{\'a}rmol-Queralt{\'o}}, {McLure}, {Cullen},
  {Dunlop}, {Fontana}  \& {McLeod}}{{M{\'a}rmol-Queralt{\'o}}
  et~al.}{2016}]{Marmol-Queralto2016}
{M{\'a}rmol-Queralt{\'o}} E.,  {McLure} R.~J.,  {Cullen} F.,  {Dunlop} J.~S.,
  {Fontana} A.,   {McLeod} D.~J.,  2016, \mn@doi [\mnras]
  {10.1093/mnras/stw1212}, \href
  {https://ui.adsabs.harvard.edu/abs/2016MNRAS.460.3587M} {460, 3587}

\bibitem[\protect\citeauthoryear{{Maseda} et~al.,}{{Maseda}
  et~al.}{2020}]{Maseda2020}
{Maseda} M.~V.,  et~al., 2020, \mn@doi [\mnras] {10.1093/mnras/staa622}, \href
  {https://ui.adsabs.harvard.edu/abs/2020MNRAS.493.5120M} {493, 5120}

\bibitem[\protect\citeauthoryear{{Mason}, {Trenti}  \& {Treu}}{{Mason}
  et~al.}{2015}]{Mason2015}
{Mason} C.~A.,  {Trenti} M.,   {Treu} T.,  2015, \mn@doi [\apj]
  {10.1088/0004-637X/813/1/21}, \href
  {https://ui.adsabs.harvard.edu/abs/2015ApJ...813...21M} {813, 21}

\bibitem[\protect\citeauthoryear{{Mason}, {Naidu}, {Tacchella}  \&
  {Leja}}{{Mason} et~al.}{2019}]{Mason2019}
{Mason} C.~A.,  {Naidu} R.~P.,  {Tacchella} S.,   {Leja} J.,  2019, \mn@doi
  [\mnras] {10.1093/mnras/stz2291}, \href
  {https://ui.adsabs.harvard.edu/abs/2019MNRAS.489.2669M} {489, 2669}

\bibitem[\protect\citeauthoryear{{Matthee}, {Sobral}, {Darvish}, {Santos},
  {Mobasher}, {Paulino-Afonso}, {R{\"o}ttgering}  \& {Alegre}}{{Matthee}
  et~al.}{2017}]{Matthee2017}
{Matthee} J.,  {Sobral} D.,  {Darvish} B.,  {Santos} S.,  {Mobasher} B.,
  {Paulino-Afonso} A.,  {R{\"o}ttgering} H.,   {Alegre} L.,  2017, \mn@doi
  [\mnras] {10.1093/mnras/stx2061}, \href
  {https://ui.adsabs.harvard.edu/abs/2017MNRAS.472..772M} {472, 772}

\bibitem[\protect\citeauthoryear{{Naidu}, {Tacchella}, {Mason}, {Bose}, {Oesch}
   \& {Conroy}}{{Naidu} et~al.}{2020}]{Naidu2020}
{Naidu} R.~P.,  {Tacchella} S.,  {Mason} C.~A.,  {Bose} S.,  {Oesch} P.~A.,
  {Conroy} C.,  2020, \mn@doi [\apj] {10.3847/1538-4357/ab7cc9}, \href
  {https://ui.adsabs.harvard.edu/abs/2020ApJ...892..109N} {892, 109}

\bibitem[\protect\citeauthoryear{{Naidu} et~al.,}{{Naidu}
  et~al.}{2022}]{Naidu2022}
{Naidu} R.~P.,  et~al., 2022, \mn@doi [\mnras] {10.1093/mnras/stab3601}, \href
  {https://ui.adsabs.harvard.edu/abs/2022MNRAS.510.4582N} {510, 4582}

\bibitem[\protect\citeauthoryear{{Nakajima}, {Ellis}, {Iwata}, {Inoue},
  {Kusakabe}, {Ouchi}  \& {Robertson}}{{Nakajima} et~al.}{2016}]{Nakajima2016}
{Nakajima} K.,  {Ellis} R.~S.,  {Iwata} I.,  {Inoue} A.~K.,  {Kusakabe} H.,
  {Ouchi} M.,   {Robertson} B.~E.,  2016, \mn@doi [\apjl]
  {10.3847/2041-8205/831/1/L9}, \href
  {https://ui.adsabs.harvard.edu/abs/2016ApJ...831L...9N} {831, L9}

\bibitem[\protect\citeauthoryear{{Nanayakkara} et~al.,}{{Nanayakkara}
  et~al.}{2020}]{Nanayakkara2020}
{Nanayakkara} T.,  et~al., 2020, \mn@doi [\apj] {10.3847/1538-4357/ab65eb},
  \href {https://ui.adsabs.harvard.edu/abs/2020ApJ...889..180N} {889, 180}

\bibitem[\protect\citeauthoryear{{Ning}, {Cai}, {Jiang}, {Lin}, {Fu}  \&
  {Spinoso}}{{Ning} et~al.}{2023}]{Ning2023}
{Ning} Y.,  {Cai} Z.,  {Jiang} L.,  {Lin} X.,  {Fu} S.,   {Spinoso} D.,  2023,
  \mn@doi [\apjl] {10.3847/2041-8213/acb26b}, \href
  {https://ui.adsabs.harvard.edu/abs/2023ApJ...944L...1N} {944, L1}

\bibitem[\protect\citeauthoryear{{Oesch} et~al.,}{{Oesch}
  et~al.}{2023}]{Oesch2023}
{Oesch} P.~A.,  et~al., 2023, \mn@doi [arXiv e-prints]
  {10.48550/arXiv.2304.02026}, \href
  {https://ui.adsabs.harvard.edu/abs/2023arXiv230402026O} {p. arXiv:2304.02026}

\bibitem[\protect\citeauthoryear{{Onodera} et~al.,}{{Onodera}
  et~al.}{2020}]{Onodera2020}
{Onodera} M.,  et~al., 2020, \mn@doi [\apj] {10.3847/1538-4357/abc174}, \href
  {https://ui.adsabs.harvard.edu/abs/2020ApJ...904..180O} {904, 180}

\bibitem[\protect\citeauthoryear{{Osterbrock} \& {Ferland}}{{Osterbrock} \&
  {Ferland}}{2006}]{Osterbrock2006}
{Osterbrock} D.~E.,  {Ferland} G.~J.,  2006, {Astrophysics of gaseous nebulae
  and active galactic nuclei}

\bibitem[\protect\citeauthoryear{{Ouchi} et~al.,}{{Ouchi}
  et~al.}{2009}]{Ouchi2009}
{Ouchi} M.,  et~al., 2009, \mn@doi [\apj] {10.1088/0004-637X/706/2/1136}, \href
  {https://ui.adsabs.harvard.edu/abs/2009ApJ...706.1136O} {706, 1136}

\bibitem[\protect\citeauthoryear{{Paardekooper}, {Khochfar}  \& {Dalla
  Vecchia}}{{Paardekooper} et~al.}{2015}]{Paardekooper2015}
{Paardekooper} J.-P.,  {Khochfar} S.,   {Dalla Vecchia} C.,  2015, \mn@doi
  [\mnras] {10.1093/mnras/stv1114}, \href
  {https://ui.adsabs.harvard.edu/abs/2015MNRAS.451.2544P} {451, 2544}

\bibitem[\protect\citeauthoryear{{Pannella} et~al.,}{{Pannella}
  et~al.}{2015}]{Pannella2015}
{Pannella} M.,  et~al., 2015, \mn@doi [\apj] {10.1088/0004-637X/807/2/141},
  \href {https://ui.adsabs.harvard.edu/abs/2015ApJ...807..141P} {807, 141}

\bibitem[\protect\citeauthoryear{{Planck Collaboration} et~al.,}{{Planck
  Collaboration} et~al.}{2016}]{Planck2016}
{Planck Collaboration} et~al., 2016, \mn@doi [\aap]
  {10.1051/0004-6361/201628897}, \href
  {https://ui.adsabs.harvard.edu/abs/2016A&A...596A.108P} {596, A108}

\bibitem[\protect\citeauthoryear{{Planck Collaboration} et~al.,}{{Planck
  Collaboration} et~al.}{2020}]{Planck2020}
{Planck Collaboration} et~al., 2020, \mn@doi [\aap]
  {10.1051/0004-6361/201833910}, \href
  {https://ui.adsabs.harvard.edu/abs/2020A&A...641A...6P} {641, A6}

\bibitem[\protect\citeauthoryear{{Prevot}, {Lequeux}, {Maurice}, {Prevot}  \&
  {Rocca-Volmerange}}{{Prevot} et~al.}{1984}]{Prevot1984}
{Prevot} M.~L.,  {Lequeux} J.,  {Maurice} E.,  {Prevot} L.,
  {Rocca-Volmerange} B.,  1984, \aap, \href
  {https://ui.adsabs.harvard.edu/abs/1984A&A...132..389P} {132, 389}

\bibitem[\protect\citeauthoryear{{Prieto-Lyon} et~al.,}{{Prieto-Lyon}
  et~al.}{2023}]{Prieto-Lyon2023}
{Prieto-Lyon} G.,  et~al., 2023, \mn@doi [\aap] {10.1051/0004-6361/202245532},
  \href {https://ui.adsabs.harvard.edu/abs/2023A&A...672A.186P} {672, A186}

\bibitem[\protect\citeauthoryear{{Reddy} et~al.,}{{Reddy}
  et~al.}{2018}]{Reddy2018}
{Reddy} N.~A.,  et~al., 2018, \mn@doi [\apj] {10.3847/1538-4357/aaa3e7}, \href
  {https://ui.adsabs.harvard.edu/abs/2018ApJ...853...56R} {853, 56}

\bibitem[\protect\citeauthoryear{{Rieke} et~al.,}{{Rieke}
  et~al.}{2023a}]{Rieke2023}
{Rieke} M.~J.,  et~al., 2023a, \mn@doi [arXiv e-prints]
  {10.48550/arXiv.2306.02466}, \href
  {https://ui.adsabs.harvard.edu/abs/2023arXiv230602466R} {p. arXiv:2306.02466}

\bibitem[\protect\citeauthoryear{{Rieke} et~al.,}{{Rieke}
  et~al.}{2023b}]{Rieke2023instrument}
{Rieke} M.~J.,  et~al., 2023b, \mn@doi [\pasp] {10.1088/1538-3873/acac53},
  \href {https://ui.adsabs.harvard.edu/abs/2023PASP..135b8001R} {135, 028001}

\bibitem[\protect\citeauthoryear{{Rinaldi} et~al.,}{{Rinaldi}
  et~al.}{2023a}]{Rinaldi2023arXiv}
{Rinaldi} P.,  et~al., 2023a, arXiv e-prints, \href
  {https://ui.adsabs.harvard.edu/abs/2023arXiv230915671R} {p. arXiv:2309.15671}

\bibitem[\protect\citeauthoryear{{Rinaldi} et~al.,}{{Rinaldi}
  et~al.}{2023b}]{Rinaldi2023}
{Rinaldi} P.,  et~al., 2023b, \mn@doi [\apj] {10.3847/1538-4357/acdc27}, \href
  {https://ui.adsabs.harvard.edu/abs/2023ApJ...952..143R} {952, 143}

\bibitem[\protect\citeauthoryear{{Robertson}}{{Robertson}}{2022}]{Robertson2022}
{Robertson} B.~E.,  2022, \mn@doi [\araa]
  {10.1146/annurev-astro-120221-044656}, \href
  {https://ui.adsabs.harvard.edu/abs/2022ARA&A..60..121R} {60, 121}

\bibitem[\protect\citeauthoryear{{Robertson} et~al.,}{{Robertson}
  et~al.}{2013}]{Robertson2013}
{Robertson} B.~E.,  et~al., 2013, \mn@doi [\apj] {10.1088/0004-637X/768/1/71},
  \href {https://ui.adsabs.harvard.edu/abs/2013ApJ...768...71R} {768, 71}

\bibitem[\protect\citeauthoryear{{Robertson}, {Ellis}, {Furlanetto}  \&
  {Dunlop}}{{Robertson} et~al.}{2015}]{Robertson2015}
{Robertson} B.~E.,  {Ellis} R.~S.,  {Furlanetto} S.~R.,   {Dunlop} J.~S.,
  2015, \mn@doi [\apjl] {10.1088/2041-8205/802/2/L19}, \href
  {https://ui.adsabs.harvard.edu/abs/2015ApJ...802L..19R} {802, L19}

\bibitem[\protect\citeauthoryear{{Rosdahl} et~al.,}{{Rosdahl}
  et~al.}{2018}]{Rosdahl2018}
{Rosdahl} J.,  et~al., 2018, \mn@doi [\mnras] {10.1093/mnras/sty1655}, \href
  {https://ui.adsabs.harvard.edu/abs/2018MNRAS.479..994R} {479, 994}

\bibitem[\protect\citeauthoryear{{Saxena} et~al.,}{{Saxena}
  et~al.}{2023}]{Saxena2023}
{Saxena} A.,  et~al., 2023, \mn@doi [arXiv e-prints]
  {10.48550/arXiv.2306.04536}, \href
  {https://ui.adsabs.harvard.edu/abs/2023arXiv230604536S} {p. arXiv:2306.04536}

\bibitem[\protect\citeauthoryear{{Seeyave} et~al.,}{{Seeyave}
  et~al.}{2023}]{Seeyave2023}
{Seeyave} L. T.~C.,  et~al., 2023, \mn@doi [\mnras] {10.1093/mnras/stad2487},
  \href {https://ui.adsabs.harvard.edu/abs/2023MNRAS.tmp.2360S} {}

\bibitem[\protect\citeauthoryear{{Shivaei} et~al.,}{{Shivaei}
  et~al.}{2018}]{Shivaei2018}
{Shivaei} I.,  et~al., 2018, \mn@doi [\apj] {10.3847/1538-4357/aaad62}, \href
  {https://ui.adsabs.harvard.edu/abs/2018ApJ...855...42S} {855, 42}

\bibitem[\protect\citeauthoryear{{Shivaei} et~al.,}{{Shivaei}
  et~al.}{2020}]{Shivaei2020}
{Shivaei} I.,  et~al., 2020, \mn@doi [\apj] {10.3847/1538-4357/aba35e}, \href
  {https://ui.adsabs.harvard.edu/abs/2020ApJ...899..117S} {899, 117}

\bibitem[\protect\citeauthoryear{{Simmonds} et~al.,}{{Simmonds}
  et~al.}{2023}]{Simmonds2023}
{Simmonds} C.,  et~al., 2023, \mn@doi [\mnras] {10.1093/mnras/stad1749}, \href
  {https://ui.adsabs.harvard.edu/abs/2023MNRAS.523.5468S} {523, 5468}

\bibitem[\protect\citeauthoryear{{Song} et~al.,}{{Song}
  et~al.}{2016}]{Song2016}
{Song} M.,  et~al., 2016, \mn@doi [\apj] {10.3847/0004-637X/825/1/5}, \href
  {https://ui.adsabs.harvard.edu/abs/2016ApJ...825....5S} {825, 5}

\bibitem[\protect\citeauthoryear{{Stark}, {Schenker}, {Ellis}, {Robertson},
  {McLure}  \& {Dunlop}}{{Stark} et~al.}{2013}]{Stark2013}
{Stark} D.~P.,  {Schenker} M.~A.,  {Ellis} R.,  {Robertson} B.,  {McLure} R.,
  {Dunlop} J.,  2013, \mn@doi [\apj] {10.1088/0004-637X/763/2/129}, \href
  {https://ui.adsabs.harvard.edu/abs/2013ApJ...763..129S} {763, 129}

\bibitem[\protect\citeauthoryear{{Stark} et~al.,}{{Stark}
  et~al.}{2015}]{Stark2015}
{Stark} D.~P.,  et~al., 2015, \mn@doi [\mnras] {10.1093/mnras/stv1907}, \href
  {https://ui.adsabs.harvard.edu/abs/2015MNRAS.454.1393S} {454, 1393}

\bibitem[\protect\citeauthoryear{{Stark} et~al.,}{{Stark}
  et~al.}{2017}]{Stark2017}
{Stark} D.~P.,  et~al., 2017, \mn@doi [\mnras] {10.1093/mnras/stw2233}, \href
  {https://ui.adsabs.harvard.edu/abs/2017MNRAS.464..469S} {464, 469}

\bibitem[\protect\citeauthoryear{{Stefanon}, {Bouwens}, {Illingworth},
  {Labb{\'e}}, {Oesch}  \& {Gonzalez}}{{Stefanon} et~al.}{2022}]{Stefanon2022}
{Stefanon} M.,  {Bouwens} R.~J.,  {Illingworth} G.~D.,  {Labb{\'e}} I.,
  {Oesch} P.~A.,   {Gonzalez} V.,  2022, \mn@doi [\apj]
  {10.3847/1538-4357/ac7e44}, \href
  {https://ui.adsabs.harvard.edu/abs/2022ApJ...935...94S} {935, 94}

\bibitem[\protect\citeauthoryear{Steidel, Bogosavljevi{\'c}, Shapley, Reddy,
  Rudie, Pettini, Trainor  \& Strom}{Steidel et~al.}{2018}]{Steidel2018}
Steidel C.~C.,  Bogosavljevi{\'c} M.,  Shapley A.~E.,  Reddy N.~A.,  Rudie
  G.~C.,  Pettini M.,  Trainor R.~F.,   Strom A.~L.,  2018, The Astrophysical
  Journal, 869, 123

\bibitem[\protect\citeauthoryear{{Stinson}, {Dalcanton}, {Quinn}, {Kaufmann}
  \& {Wadsley}}{{Stinson} et~al.}{2007}]{Stinson2007}
{Stinson} G.~S.,  {Dalcanton} J.~J.,  {Quinn} T.,  {Kaufmann} T.,   {Wadsley}
  J.,  2007, \mn@doi [\apj] {10.1086/520504}, \href
  {https://ui.adsabs.harvard.edu/abs/2007ApJ...667..170S} {667, 170}

\bibitem[\protect\citeauthoryear{{Sugahara}, {Inoue}, {Fudamoto}, {Hashimoto},
  {Harikane}  \& {Yamanaka}}{{Sugahara} et~al.}{2022}]{Sugahara2022}
{Sugahara} Y.,  {Inoue} A.~K.,  {Fudamoto} Y.,  {Hashimoto} T.,  {Harikane} Y.,
    {Yamanaka} S.,  2022, \mn@doi [\apj] {10.3847/1538-4357/ac7fed}, \href
  {https://ui.adsabs.harvard.edu/abs/2022ApJ...935..119S} {935, 119}

\bibitem[\protect\citeauthoryear{{Sun} \& {Furlanetto}}{{Sun} \&
  {Furlanetto}}{2016}]{Sun2016}
{Sun} G.,  {Furlanetto} S.~R.,  2016, \mn@doi [\mnras] {10.1093/mnras/stw980},
  \href {https://ui.adsabs.harvard.edu/abs/2016MNRAS.460..417S} {460, 417}

\bibitem[\protect\citeauthoryear{{Tacchella}, {Forbes}  \&
  {Caplar}}{{Tacchella} et~al.}{2020}]{Tacchella2020}
{Tacchella} S.,  {Forbes} J.~C.,   {Caplar} N.,  2020, \mn@doi [\mnras]
  {10.1093/mnras/staa1838}, \href
  {https://ui.adsabs.harvard.edu/abs/2020MNRAS.497..698T} {497, 698}

\bibitem[\protect\citeauthoryear{{Tacchella} et~al.,}{{Tacchella}
  et~al.}{2022a}]{Tacchella2022stellar_pops}
{Tacchella} S.,  et~al., 2022a, arXiv e-prints, \href
  {https://ui.adsabs.harvard.edu/abs/2022arXiv220803281T} {p. arXiv:2208.03281}

\bibitem[\protect\citeauthoryear{{Tacchella} et~al.,}{{Tacchella}
  et~al.}{2022b}]{Tacchella2022}
{Tacchella} S.,  et~al., 2022b, \mn@doi [\mnras] {10.1093/mnras/stac818}, \href
  {https://ui.adsabs.harvard.edu/abs/2022MNRAS.513.2904T} {513, 2904}

\bibitem[\protect\citeauthoryear{{Tacchella} et~al.,}{{Tacchella}
  et~al.}{2022c}]{Tacchella2022bursty}
{Tacchella} S.,  et~al., 2022c, \mn@doi [\apj] {10.3847/1538-4357/ac4cad},
  \href {https://ui.adsabs.harvard.edu/abs/2022ApJ...927..170T} {927, 170}

\bibitem[\protect\citeauthoryear{{Tang}, {Stark}, {Chevallard}  \&
  {Charlot}}{{Tang} et~al.}{2019}]{Tang2019}
{Tang} M.,  {Stark} D.~P.,  {Chevallard} J.,   {Charlot} S.,  2019, \mn@doi
  [\mnras] {10.1093/mnras/stz2236}, \href
  {https://ui.adsabs.harvard.edu/abs/2019MNRAS.489.2572T} {489, 2572}

\bibitem[\protect\citeauthoryear{{Tang} et~al.,}{{Tang}
  et~al.}{2023}]{Tang2023}
{Tang} M.,  et~al., 2023, \mn@doi [arXiv e-prints] {10.48550/arXiv.2301.07072},
  \href {https://ui.adsabs.harvard.edu/abs/2023arXiv230107072T} {p.
  arXiv:2301.07072}

\bibitem[\protect\citeauthoryear{{Trebitsch}, {Volonteri}  \&
  {Dubois}}{{Trebitsch} et~al.}{2020}]{Trebitsch2020}
{Trebitsch} M.,  {Volonteri} M.,   {Dubois} Y.,  2020, \mn@doi [\mnras]
  {10.1093/mnras/staa1012}, \href
  {https://ui.adsabs.harvard.edu/abs/2020MNRAS.494.3453T} {494, 3453}

\bibitem[\protect\citeauthoryear{{Vanzella} et~al.,}{{Vanzella}
  et~al.}{2018}]{Vanzella2018}
{Vanzella} E.,  et~al., 2018, \mn@doi [\mnras] {10.1093/mnrasl/sly023}, \href
  {https://ui.adsabs.harvard.edu/abs/2018MNRAS.476L..15V} {476, L15}

\bibitem[\protect\citeauthoryear{{Vijayan}, {Lovell}, {Wilkins}, {Thomas},
  {Barnes}, {Irodotou}, {Kuusisto}  \& {Roper}}{{Vijayan}
  et~al.}{2021}]{Vijayan2021}
{Vijayan} A.~P.,  {Lovell} C.~C.,  {Wilkins} S.~M.,  {Thomas} P.~A.,  {Barnes}
  D.~J.,  {Irodotou} D.,  {Kuusisto} J.,   {Roper} W.~J.,  2021, \mn@doi
  [\mnras] {10.1093/mnras/staa3715}, \href
  {https://ui.adsabs.harvard.edu/abs/2021MNRAS.501.3289V} {501, 3289}

\bibitem[\protect\citeauthoryear{Virtanen et~al.,}{Virtanen
  et~al.}{2020}]{Virtanen2020}
Virtanen P.,  et~al., 2020, \mn@doi [Nature Methods]
  {10.1038/s41592-019-0686-2}, \href {https://rdcu.be/b08Wh} {17, 261}

\bibitem[\protect\citeauthoryear{{Weisz} et~al.,}{{Weisz}
  et~al.}{2012}]{Weisz2012}
{Weisz} D.~R.,  et~al., 2012, \mn@doi [\apj] {10.1088/0004-637X/744/1/44},
  \href {https://ui.adsabs.harvard.edu/abs/2012ApJ...744...44W} {744, 44}

\bibitem[\protect\citeauthoryear{{Williams} et~al.,}{{Williams}
  et~al.}{2023}]{Williams2023}
{Williams} C.~C.,  et~al., 2023, \mn@doi [arXiv e-prints]
  {10.48550/arXiv.2301.09780}, \href
  {https://ui.adsabs.harvard.edu/abs/2023arXiv230109780W} {p. arXiv:2301.09780}

\bibitem[\protect\citeauthoryear{{Yang} et~al.,}{{Yang}
  et~al.}{2020}]{Yang2020}
{Yang} J.,  et~al., 2020, \mn@doi [\apj] {10.3847/1538-4357/abbc1b}, \href
  {https://ui.adsabs.harvard.edu/abs/2020ApJ...904...26Y} {904, 26}

\makeatother
\end{thebibliography}



\appendix
\section{\texttt{Prospector} results}
\label{Appendix:Prospector}
\begin{figure*}
        \centering
   \includegraphics[width=0.85\textwidth]{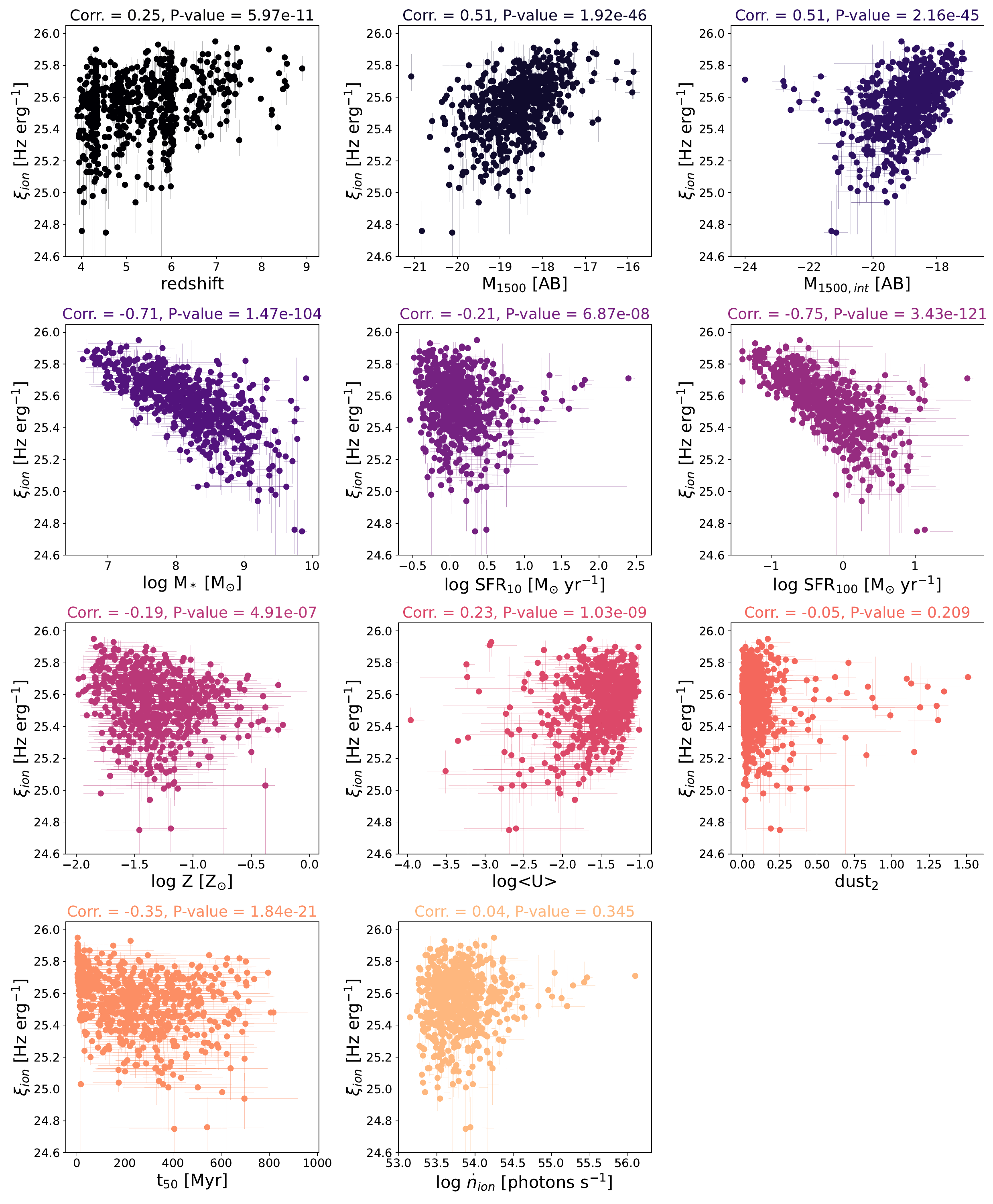}
   \caption{Exploring tentative correlations between \xion\ and different properties. The vertical axis is \xion\ in all panels, while the name of each property is given in the x-label. The title of the panels show the Spearman's rank coefficients for each parameter, indicating how strong the correlation is with \xion\/. The strongest correlations are found for SFR$_{100}$ (right panel of second row) and stellar mass (left panel of second row). From top to bottom and left to right, the parameters are: redshift, observed UV magnitude (M$_{1500}$), intrinsic UV magnitude (M$_{1500,int}$), stellar mass (M$_*$), SFR in the past 10 Myr (SFR$_{10}$), SFR in the past 100 Myr (SFR$_{100}$), metallicity ($Z$), ionisation parameter (log$\langle U \rangle$), dust2, half-mass assembly time (t50), and rate of ionising photons being emitted (\ndot\/).}
              \label{fig:spearman}%
    \end{figure*}

Here we present the galaxy properties inferred by \texttt{Prospector}. They are shown visually in Figure~\ref{fig:spearman}, and as a table in Table~\ref{table:excerpt_prospector}. 

\begin{landscape}
    \begin{table}
    \small
    \centering
    \begin{tabular}{cccccccccccc}
    \hline
    \noalign{\smallskip}
    Name & $z$ & M$_{1500}$ & M$_{1500,\rm{int}}$ & log M  & log SFR$_{10}$  & log SFR$_{100}$ & log Z & log$\langle U \rangle$ & dust$_{2}$ & $\dot{n}_{ion}$  \\
      & & [AB]] & [AB] & [M$_{\odot}$]  & [M$_{\odot}$ yr$^{-1}$]  & [M$_{\odot}$ yr$^{-1}$] & [Z$_{\odot}$] &  &  & [s$^{-1}$]  \\
    \noalign{\smallskip}
    \hline
    \noalign{\smallskip}
    JADES-GS+53.11634-27.81272 & 3.91$\pm$0.05 & -19.52$_{-0.03}^{0.05}$ & -19.52$_{-0.07}^{0.06}$ & 8.98$_{-0.19}^{0.05}$ & 0.34$_{-0.06}^{0.06}$ & 0.09$_{-0.11}^{0.19}$ & -1.14$_{-0.22}^{0.32}$ & -1.34$_{-0.33}^{0.27}$ & 0.02$_{-0.01}^{0.03}$ & 817.19$_{-381.59}^{141.22}$ & 53.94$_{-0.05}^{0.02}$ \\ 
JADES-GS+53.20925-27.75711 & 3.92$\pm$0.04 & -17.51$_{-0.07}^{0.07}$ & -17.69$_{-0.19}^{0.10}$ & 7.96$_{-0.13}^{0.15}$ & -0.39$_{-0.08}^{0.11}$ & -0.68$_{-0.09}^{0.18}$ & -1.13$_{-0.50}^{0.31}$ & -1.25$_{-0.53}^{0.15}$ & 0.05$_{-0.03}^{0.06}$ & 502.82$_{-298.10}^{116.09}$ & 53.25$_{-0.04}^{0.07}$ \\ 
JADES-GS+53.12549-27.78044 & 3.94$\pm$0.05 & -19.55$_{-0.04}^{0.04}$ & -19.51$_{-0.11}^{0.05}$ & 8.44$_{-0.06}^{0.09}$ & 0.23$_{-0.06}^{0.10}$ & 0.13$_{-0.04}^{0.08}$ & -1.59$_{-0.14}^{0.23}$ & -1.66$_{-0.26}^{0.36}$ & 0.01$_{-0.01}^{0.03}$ & 302.96$_{-177.69}^{66.06}$ & 53.82$_{-0.02}^{0.04}$ \\ 
JADES-GS+53.18436-27.80581 & 3.95$\pm$0.04 & -19.15$_{-0.09}^{0.11}$ & -20.51$_{-0.66}^{0.46}$ & 9.71$_{-0.13}^{0.08}$ & 0.50$_{-0.18}^{0.00}$ & 0.62$_{-0.15}^{0.00}$ & -1.24$_{-0.26}^{0.47}$ & -2.52$_{-0.90}^{1.01}$ & 0.32$_{-0.15}^{0.30}$ & 697.12$_{-224.10}^{132.32}$ & 54.06$_{-0.15}^{0.29}$ \\ 
JADES-GS+53.16268-27.73611 & 3.95$\pm$0.04 & -18.06$_{-0.05}^{0.06}$ & -18.26$_{-0.15}^{0.16}$ & 8.71$_{-0.15}^{0.04}$ & -0.39$_{-0.09}^{0.18}$ & -0.40$_{-0.14}^{0.40}$ & -1.69$_{-0.21}^{0.55}$ & -1.43$_{-0.42}^{0.26}$ & 0.07$_{-0.05}^{0.05}$ & 560.40$_{-171.13}^{174.10}$ & 53.32$_{-0.06}^{0.09}$ \\ 
JADES-GS+53.15123-27.79826 & 3.96$\pm$0.04 & -18.35$_{-0.07}^{0.07}$ & -19.63$_{-0.53}^{0.43}$ & 9.22$_{-0.17}^{0.14}$ & 0.02$_{-0.14}^{0.69}$ & 0.35$_{-0.14}^{0.69}$ & -1.25$_{-0.41}^{0.36}$ & -2.05$_{-0.84}^{0.60}$ & 0.32$_{-0.17}^{0.26}$ & 502.74$_{-220.83}^{290.13}$ & 53.51$_{-0.13}^{0.26}$ \\ 
JADES-GS+53.15282-27.79549 & 3.97$\pm$0.04 & -18.09$_{-0.06}^{0.00}$ & -18.06$_{-0.01}^{0.10}$ & 7.59$_{-0.01}^{0.17}$ & -0.11$_{-0.04}^{0.00}$ & -0.92$_{-0.00}^{0.08}$ & -1.03$_{-0.04}^{0.11}$ & -1.20$_{-0.02}^{0.11}$ & 0.04$_{-0.03}^{0.01}$ & 557.16$_{-51.46}^{40.65}$ & 53.57$_{-0.05}^{0.05}$ \\ 
JADES-GS+53.19804-27.76002 & 3.97$\pm$0.04 & -18.48$_{-0.05}^{0.06}$ & -18.69$_{-0.19}^{0.15}$ & 8.68$_{-0.23}^{0.12}$ & -0.08$_{-0.10}^{1.22}$ & -0.17$_{-0.12}^{0.30}$ & -1.04$_{-0.46}^{0.53}$ & -1.26$_{-0.25}^{0.21}$ & 0.05$_{-0.03}^{0.06}$ & 586.50$_{-213.14}^{149.05}$ & 53.53$_{-0.05}^{0.05}$ \\ 
JADES-GS+53.13905-27.75893 & 3.97$\pm$0.05 & -19.04$_{-0.03}^{0.06}$ & -18.93$_{-0.08}^{0.09}$ & 8.01$_{-0.08}^{0.05}$ & 0.17$_{-0.03}^{0.06}$ & -0.16$_{-0.21}^{0.09}$ & -1.84$_{-0.09}^{0.61}$ & -1.39$_{-0.10}^{0.22}$ & 0.02$_{-0.01}^{0.01}$ & 97.09$_{-15.53}^{337.60}$ & 53.88$_{-0.03}^{0.02}$ \\ 
JADES-GS+53.15186-27.75258 & 4.00$\pm$0.06 & -17.80$_{-0.05}^{0.04}$ & -17.85$_{-0.21}^{0.15}$ & 7.43$_{-0.16}^{0.12}$ & -0.31$_{-0.04}^{0.09}$ & -0.66$_{-0.10}^{0.17}$ & -1.22$_{-0.16}^{0.17}$ & -1.23$_{-0.16}^{0.12}$ & 0.03$_{-0.02}^{0.04}$ & 41.59$_{-24.23}^{79.71}$ & 53.39$_{-0.03}^{0.05}$ \\ 

JADES-GS+53.12644-27.79200 & 5.37$\pm$0.06 & -18.49$_{-0.03}^{0.03}$ & -18.35$_{-0.17}^{0.06}$ & 7.37$_{-0.09}^{0.21}$ & -0.00$_{-0.03}^{0.05}$ & -0.77$_{-0.09}^{0.15}$ & -1.61$_{-0.29}^{0.61}$ & -1.10$_{-0.20}^{0.08}$ & 0.01$_{-0.01}^{0.03}$ & 43.43$_{-33.08}^{183.82}$ & 53.78$_{-0.03}^{0.02}$ \\ 
JADES-GS+53.12775-27.78098 & 5.38$\pm$0.06 & -18.48$_{-0.03}^{0.06}$ & -19.03$_{-0.14}^{0.20}$ & 8.29$_{-0.48}^{0.24}$ & 0.17$_{-0.09}^{0.22}$ & -0.22$_{-0.11}^{0.74}$ & -1.12$_{-0.58}^{0.50}$ & -1.26$_{-0.26}^{0.19}$ & 0.10$_{-0.05}^{0.07}$ & 320.79$_{-300.27}^{173.58}$ & 53.79$_{-0.07}^{0.05}$ \\ 
JADES-GS+53.16729-27.75273 & 5.38$\pm$0.06 & -18.88$_{-0.03}^{0.03}$ & -18.91$_{-0.15}^{0.09}$ & 8.48$_{-0.16}^{0.13}$ & -0.02$_{-0.06}^{0.15}$ & -0.00$_{-0.11}^{0.31}$ & -1.33$_{-0.33}^{0.31}$ & -1.36$_{-0.37}^{0.23}$ & 0.03$_{-0.02}^{0.05}$ & 303.50$_{-151.42}^{119.74}$ & 53.66$_{-0.05}^{0.08}$ \\ 
JADES-GS+53.14381-27.80835 & 5.42$\pm$0.08 & -17.93$_{-0.03}^{0.03}$ & -17.80$_{-0.08}^{0.06}$ & 7.24$_{-0.30}^{0.16}$ & -0.21$_{-0.09}^{0.05}$ & -0.92$_{-0.12}^{0.78}$ & -1.46$_{-0.31}^{0.19}$ & -1.30$_{-0.32}^{0.16}$ & 0.01$_{-0.01}^{0.01}$ & 25.68$_{-19.30}^{86.95}$ & 53.49$_{-0.03}^{0.03}$ \\ 
JADES-GS+53.10726-27.81102 & 5.43$\pm$0.07 & -16.82$_{-0.12}^{0.15}$ & -17.21$_{-0.06}^{0.14}$ & 6.69$_{-0.14}^{0.53}$ & -0.41$_{-0.05}^{0.05}$ & -1.40$_{-0.10}^{0.30}$ & -1.76$_{-0.16}^{0.09}$ & -1.83$_{-0.50}^{0.30}$ & 0.10$_{-0.03}^{0.05}$ & 4.34$_{-1.81}^{247.58}$ & 53.41$_{-0.02}^{0.06}$ \\ 
JADES-GS+53.14676-27.79738 & 5.43$\pm$0.06 & -17.92$_{-0.04}^{0.03}$ & -17.81$_{-0.13}^{0.07}$ & 7.30$_{-0.21}^{0.12}$ & -0.23$_{-0.04}^{0.02}$ & -0.92$_{-0.07}^{0.23}$ & -1.27$_{-0.51}^{0.15}$ & -1.70$_{-0.38}^{0.52}$ & 0.01$_{-0.01}^{0.01}$ & 42.70$_{-28.41}^{234.59}$ & 53.48$_{-0.02}^{0.03}$ \\ 
JADES-GS+53.12301-27.79661 & 5.44$\pm$0.06 & -18.04$_{-0.04}^{0.03}$ & -18.22$_{-0.05}^{0.06}$ & 7.11$_{-0.06}^{0.07}$ & -0.04$_{-0.02}^{0.03}$ & -0.92$_{-0.07}^{0.04}$ & -1.25$_{-0.08}^{0.10}$ & -1.45$_{-0.15}^{0.32}$ & 0.04$_{-0.01}^{0.01}$ & 4.92$_{-0.92}^{3.41}$ & 53.74$_{-0.02}^{0.02}$ \\ 
JADES-GS+53.12247-27.79653 & 5.44$\pm$0.07 & -18.11$_{-0.03}^{0.04}$ & -18.52$_{-0.27}^{0.06}$ & 7.75$_{-0.59}^{0.17}$ & 0.06$_{-0.05}^{0.08}$ & -0.59$_{-0.18}^{0.00}$ & -1.51$_{-0.05}^{0.59}$ & -1.67$_{-0.61}^{0.23}$ & 0.09$_{-0.02}^{0.07}$ & 31.12$_{-27.09}^{423.39}$ & 53.75$_{-0.04}^{0.05}$ \\ 
JADES-GS+53.16407-27.79972 & 5.44$\pm$0.08 & -20.19$_{-0.03}^{0.03}$ & -20.21$_{-0.13}^{0.06}$ & 9.08$_{-0.29}^{0.13}$ & 0.27$_{-0.13}^{0.32}$ & 0.67$_{-0.11}^{0.18}$ & -1.16$_{-0.35}^{0.33}$ & -2.19$_{-0.79}^{0.45}$ & 0.02$_{-0.01}^{0.05}$ & 341.47$_{-286.55}^{174.28}$ & 53.79$_{-0.08}^{0.10}$ \\ 
JADES-GS+53.12874-27.79788 & 5.44$\pm$0.06 & -17.92$_{-0.04}^{0.03}$ & -18.10$_{-0.25}^{0.12}$ & 7.95$_{-0.35}^{0.11}$ & -0.12$_{-0.07}^{0.08}$ & -0.64$_{-0.12}^{0.22}$ & -1.28$_{-0.38}^{0.49}$ & -1.36$_{-0.23}^{0.29}$ & 0.06$_{-0.03}^{0.06}$ & 376.88$_{-299.49}^{131.65}$ & 53.56$_{-0.06}^{0.07}$ \\ 

JADES-GS+53.19106-27.79732 & 7.26$\pm$0.07 & -19.13$_{-0.04}^{0.04}$ & -19.96$_{-0.32}^{0.31}$ & 8.51$_{-0.30}^{0.15}$ & 0.54$_{-0.08}^{0.14}$ & 0.19$_{-0.16}^{0.91}$ & -1.62$_{-0.20}^{0.43}$ & -1.85$_{-0.36}^{0.33}$ & 0.18$_{-0.09}^{0.11}$ & 103.35$_{-74.78}^{157.81}$ & 54.20$_{-0.10}^{0.13}$ \\ 
JADES-GS+53.17976-27.77465 & 7.27$\pm$0.08 & -19.15$_{-0.03}^{0.05}$ & -20.70$_{-0.32}^{0.60}$ & 8.66$_{-0.15}^{0.10}$ & 0.87$_{-0.16}^{0.13}$ & 0.47$_{-0.12}^{0.12}$ & -1.55$_{-0.22}^{0.28}$ & -1.67$_{-0.14}^{0.51}$ & 0.49$_{-0.28}^{0.26}$ & 90.19$_{-43.73}^{102.14}$ & 54.55$_{-0.15}^{0.22}$ \\ 
JADES-GS+53.16579-27.82179 & 7.29$\pm$0.09 & -17.91$_{-0.04}^{0.04}$ & -18.41$_{-0.26}^{0.15}$ & 7.43$_{-0.36}^{0.35}$ & 0.04$_{-0.06}^{0.06}$ & -0.72$_{-0.17}^{0.28}$ & -1.48$_{-0.42}^{0.33}$ & -1.67$_{-0.70}^{0.33}$ & 0.10$_{-0.03}^{0.09}$ & 53.84$_{-50.66}^{241.98}$ & 53.79$_{-0.07}^{0.05}$ \\ 
JADES-GS+53.18334-27.79050 & 7.30$\pm$0.07 & -17.82$_{-0.10}^{0.09}$ & -18.76$_{-0.28}^{0.18}$ & 7.73$_{-0.28}^{0.22}$ & 0.15$_{-0.08}^{0.19}$ & -0.49$_{-0.13}^{0.30}$ & -1.59$_{-0.26}^{0.54}$ & -1.95$_{-0.36}^{0.60}$ & 0.15$_{-0.06}^{0.13}$ & 89.85$_{-73.40}^{92.18}$ & 53.89$_{-0.08}^{0.11}$ \\ 
JADES-GS+53.13219-27.78578 & 7.39$\pm$0.09 & -18.07$_{-0.05}^{0.05}$ & -18.25$_{-0.33}^{0.16}$ & 7.42$_{-0.21}^{0.26}$ & -0.05$_{-0.04}^{0.46}$ & -0.74$_{-0.12}^{0.30}$ & -1.52$_{-0.32}^{0.27}$ & -1.32$_{-0.34}^{0.22}$ & 0.07$_{-0.03}^{0.20}$ & 87.39$_{-79.55}^{91.48}$ & 53.75$_{-0.10}^{0.06}$ \\ 
JADES-GS+53.16638-27.81237 & 7.40$\pm$0.08 & -17.78$_{-0.06}^{0.03}$ & -17.73$_{-0.10}^{0.07}$ & 6.88$_{-0.05}^{0.34}$ & -0.18$_{-0.04}^{0.02}$ & -1.10$_{-0.05}^{0.60}$ & -1.78$_{-0.16}^{0.21}$ & -1.55$_{-0.30}^{0.30}$ & 0.04$_{-0.02}^{0.02}$ & 5.32$_{-0.67}^{18.39}$ & 53.53$_{-0.03}^{0.02}$ \\ 
JADES-GS+53.18405-27.79783 & 7.41$\pm$0.08 & -18.15$_{-0.04}^{0.09}$ & -18.34$_{-0.34}^{0.16}$ & 7.25$_{-0.26}^{0.23}$ & 0.02$_{-0.08}^{0.06}$ & -0.85$_{-0.11}^{0.00}$ & -1.39$_{-0.12}^{0.51}$ & -1.49$_{-0.32}^{0.36}$ & 0.07$_{-0.02}^{0.15}$ & 8.38$_{-5.31}^{49.78}$ & 53.71$_{-0.03}^{0.09}$ \\ 
JADES-GS+53.18536-27.77319 & 7.41$\pm$0.06 & -18.84$_{-0.04}^{0.03}$ & -18.69$_{-0.16}^{0.23}$ & 7.41$_{-0.35}^{0.39}$ & 0.06$_{-0.07}^{0.14}$ & -0.59$_{-0.19}^{0.00}$ & -1.60$_{-0.24}^{0.45}$ & -1.55$_{-0.47}^{0.35}$ & 0.01$_{-0.00}^{0.01}$ & 8.97$_{-6.42}^{32.63}$ & 53.77$_{-0.09}^{0.16}$ \\ 
JADES-GS+53.18393-27.79999 & 7.42$\pm$0.07 & -19.32$_{-0.04}^{0.04}$ & -19.35$_{-0.11}^{0.17}$ & 8.19$_{-0.48}^{0.23}$ & 0.33$_{-0.08}^{0.09}$ & -0.19$_{-0.09}^{0.31}$ & -1.46$_{-0.28}^{0.34}$ & -2.41$_{-0.24}^{0.86}$ & 0.02$_{-0.01}^{0.02}$ & 114.51$_{-102.80}^{185.35}$ & 54.00$_{-0.07}^{0.04}$ \\ 
JADES-GS+53.18301-27.78946 & 7.45$\pm$0.07 & -18.21$_{-0.05}^{0.04}$ & -18.63$_{-0.23}^{0.09}$ & 7.27$_{-0.12}^{0.47}$ & 0.16$_{-0.04}^{0.04}$ & -0.80$_{-0.05}^{1.20}$ & -1.79$_{-0.10}^{0.13}$ & -2.94$_{-0.06}^{0.35}$ & 0.13$_{-0.02}^{0.03}$ & 3.94$_{-1.33}^{132.92}$ & 54.01$_{-0.06}^{0.03}$ \\ 

    \end{tabular}
    \caption{Table excerpt showing a selection of galaxies in our sample, for clarity, we have chosen to display the results for the same galaxies shown in Table~\ref{table:excerpt}. \textsl{Column 1:} JADES identifier, composed of the coordinates of the centroid rounded to the fifth decimal place, in units of degrees. \textsl{Column 2:} photometric redshift inferred using the template-fitting code \texttt{EAZY}. \textsl{Column 3,4:} observed and intrinsic UV magnitudes at rest-frame 1500 \AA\/. \textsl{Column 5:} logarithm of the stellar mass in units of solar masses (M$_{\odot}$). \textsl{Column 6:} logarithm of the recent star formation rate (last 10 Myr) in units of M$_{\odot}$ yr$^{-1}$. \textsl{Column 7:} logarithm of the star formation rate in the past 100 Myr in units of M$_{\odot}$ yr$^{-1}$. \textsl{Column 8:} logarithm of the stellar metallicity in units of solar metallicities (Z$_{\odot}$). \textsl{Column 9:} dimensionless ionisation parameter, log$\langle U \rangle$. \textsl{Column 10:} one of the governing dust parameters in the dust model adopted (see \citet{Conroy2009}). \textsl{Column 11:} lookback time at which half of the mass was assembled in units of Myr. \textsl{Column 11:} ionising photons emitted per second.}
    \label{table:excerpt_prospector}
    \end{table}
\end{landscape}

\section{Trends with ionising photon production}
\label{Appendix:nion}
\begin{figure*}
        \centering
   \includegraphics[width=0.75\textwidth]{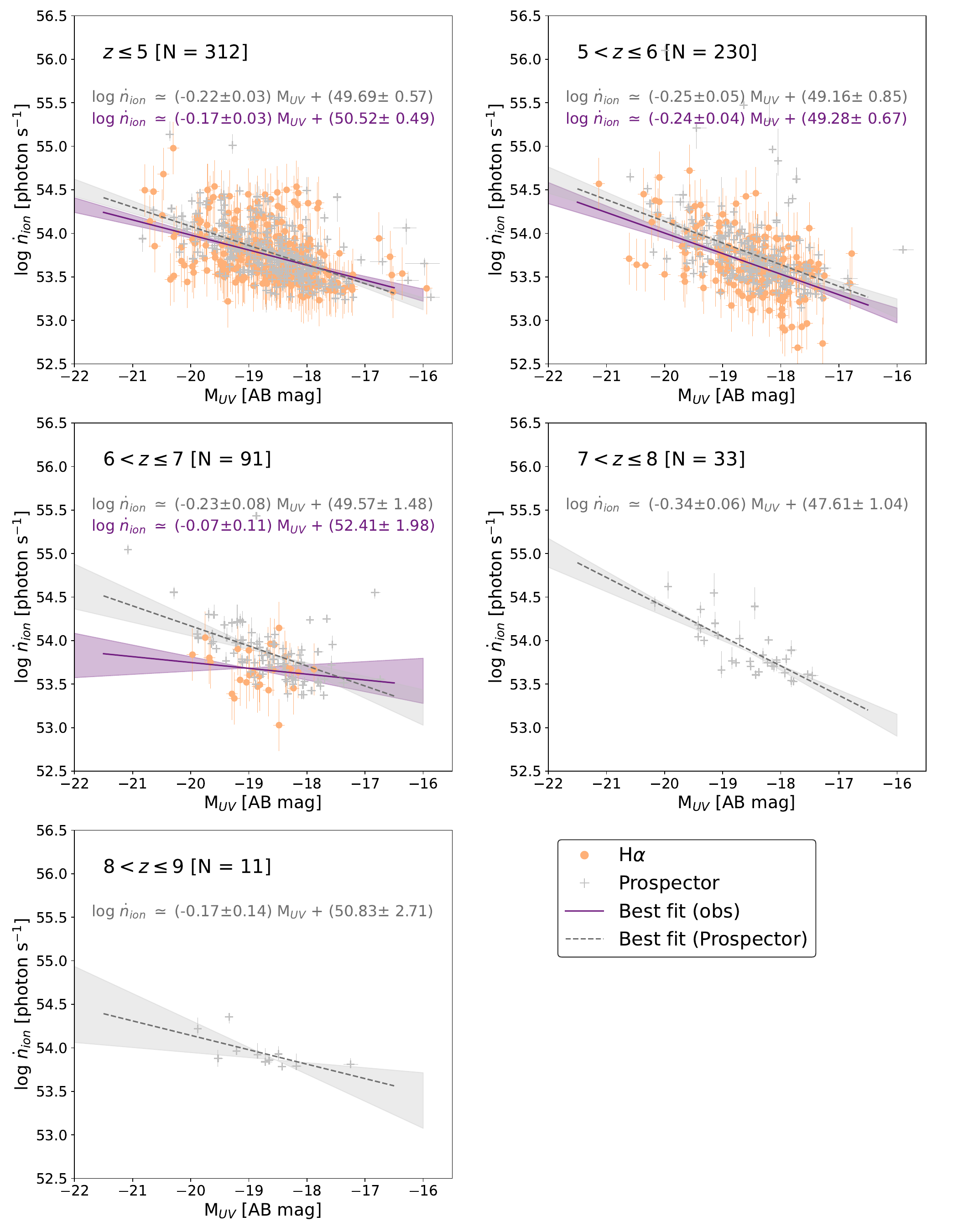}
   \caption{Evolution of \ndot\ with UV magnitude, separated in redshift bins, analogous to Figure~\ref{fig:xion_MUV}. The coloured circles are values estimated from the dust-corrected \ha\ luminosities, as: \ndot\/ = $7.35 \times 10^{11}$ L(\ha\/). The number of galaxies in each redshift bin is indicated in the top left corner of each panel. The filled (dashed) line is the best fit to the data obtained via photometry (\texttt{Prospector}). Contrary to \xion\/, \ndot\ decreases as galaxies become fainter.}
              \label{fig:nion_MUV}%
    \end{figure*}

Figure~\ref{fig:nion_MUV} shows the evolution of \ndot\ as a function of M$_{\rm{UV}}$ for different redshift bins. \ndot\ correlates negatively with UV magnitude, with a large scatter, indicating fainter galaxies produce in average a smaller amount of ionising photons.    

\bsp	
\label{lastpage}
\end{document}